\documentclass[11pt, prd, aps, showpacs, superscriptaddress, floatfix, nofootinbib]{revtex4-1}
\usepackage{amsmath,amssymb}
\pdfoutput=1 

\usepackage[latin1]{inputenc}
\usepackage{graphicx}
\usepackage{textcomp}
\usepackage{enumitem}
\usepackage{graphicx}
\usepackage{dsfont}
\usepackage{bm}
\usepackage{xcolor}
\usepackage{subfig}
\usepackage{caption}
\usepackage{comment}
\usepackage{bbold}
\usepackage{cancel}

\newcommand{\be}{\begin{equation}}
\newcommand{\ee}{\end{equation}}
\newcommand{\beq}{\begin{equation}}
\newcommand{\eeq}{\end{equation}}
\newcommand{\bea}{\begin{eqnarray}}
\newcommand{\eea}{\end{eqnarray}}
\newcommand{\kket}[1]{\ensuremath{ {#1} \rangle }}
\newcommand{\bbra}[1]{\ensuremath{\langle {#1} }}
\newcommand{\ket}[1]{\ensuremath{| {#1} \rangle }}
\newcommand{\bra}[1]{\ensuremath{\langle {#1} |}}

\newcommand{\beqn}{\begin{eqnarray}}   
\newcommand{\eeqn}{\end{eqnarray}}

\newcommand{\RomatreINFN}{Istituto Nazionale di Fisica Nucleare, Sezione di Roma Tre,\\ Via della Vasca Navale 84, I-00146 Rome, Italy}

\newcommand{\LaSapienza}{Physics Department and INFN Sezione di Roma La Sapienza,\\ Piazzale Aldo Moro 5, 00185 Roma, Italy}
\newcommand{\SNS}{Scuola Normale Superiore, Piazza dei Cavalieri 7, I-56126, Pisa, Italy}
\newcommand{\INFNpisa}{Istituto Nazionale di Fisica Nucleare, Sezione di Pisa,\\ Largo Bruno Pontecorvo 3, I-56127 Pisa, Italy} 
\newcommand{\pisa}{Dipartimento di Fisica dell'Università di Pisa and INFN - Sezione di Pisa,\\ Largo Bruno Pontecorvo 3, I-56127 Pisa, Italy} 
\newcommand{\Edin}{Higgs Centre for Theoretical Physics, School of Physics and Astronomy,\\ The University of Edinburgh, Edinburgh EH9 3FD, UK}

\begin{document}

\title{Unitarity Bounds for Semileptonic Decays in Lattice QCD}

\author{M.~Di Carlo}\affiliation{\Edin}\affiliation{\LaSapienza}
\author{G.~Martinelli}\affiliation{\LaSapienza}
\author{M.~Naviglio}\affiliation{\pisa}
\author{F.~Sanfilippo}\affiliation{\RomatreINFN}
\author{S.~Simula}\affiliation{\RomatreINFN}
\author{L.~Vittorio}\affiliation{\SNS},\affiliation{\INFNpisa}


\begin{abstract}
In this work we discuss in  detail the non-perturbative  determination of the momentum dependence of the form factors entering in  semileptonic decays using unitarity and analyticity constraints. The method contains several new elements with respect to previous proposals and allows to extract, using suitable two-point functions computed non-perturbatively, the form factors at low momentum transfer $q^2$ from those computed explicitly on the lattice at large $q^2$, without any assumption about their $q^2$-dependence. The approach will be very useful for exclusive semileptonic $B$-meson decays, where the direct calculation of the form factors at low $q^2$ is particularly difficult due to large statistical fluctuations and discretisation effects. As a testing ground we apply our approach to the semileptonic $D \to K \ell \nu_\ell$ decay, where we can compare the results of the unitarity approach to the explicit direct lattice calculation of the form factors in the full $q^2$-range. We show that the method is very effective and that it allows to compute the form factors with rather good precision.
\end{abstract}

\maketitle

\section{Introduction}
\label{sec:intro}

In this work we present an extended study of  two- and three-point lattice correlation functions which are  used, together with  dispersive techniques~\cite{Okubo:1971jf}-\cite{Bourrely:1980gp}, to constrain the lattice predictions for the form factors (FFs)  relevant to  exclusive semileptonic decays.  The form factors are then obtained in a substantially non-perturbative way and without any specific assumption on their momentum dependence. To achieve this goal a complex strategy and several improvements with respect to the original proposal of Ref.~\cite{Lellouch:1995yv} are introduced and applied to several calculations. 
This strategy will be described in detail in the following.

The measurement of the weak charged-current  $b\to c$ and $b\to u$ transitions, more specifically  the semileptonic $B \to D^{(*)} \ell \nu_\ell$ decays, received an increasing attention in the recent past. The first reason is the precise determination of two of the fundamental parameters of the Standard Model, namely the Cabibbo-Kobayashi-Maskawa (CKM) matrix elements $|V_{cb}|$ and $|V_{ub}|$~\cite{Amhis:2019ckw}.
The second reason is the apparent tension between inclusive~\cite{Gambino:2013rza,Alberti:2014yda} and exclusive determinations of  $|V_{cb}|$,
which may be related to other lepton anomalies which have been observed experimentally, see for example~\cite{Alpigiani:2017lpj} and references therein.  Among the others let us mention the deviations from lepton-flavour universality (LFU) in the measurements of   $R_{D^{(*)}}$~\cite{Amhis:2016xyh}, the ratios of the branching fractions $B \to D^{(*)} \tau \nu$ over $B \to D^{(*)} \ell \nu$, $\ell = e, \mu$, made by Belle, BABAR  and LHCb~\cite{Lees:2012xj,Lees:2013uzd,Aaij:2015yra,Huschle:2015rga,Sato:2016svk,Hirose:2016wfn,Aaij:2017uff,Hirose:2017dxl,Aaij:2017deq}. 
These deviations may be interpreted as a  hint  of the presence of  New Physics (NP)~\cite{Crivellin:2014zpa,Bernlochner:2017jka,Bernlochner:2017xyx,Jung:2018lfu,Colangelo:2018cnj,Azatov:2018knx,Feruglio:2018fxo, Bordone:2019vic,Bordone:2019guc,Bobeth:2021lya}.
Anomalous values of the ratios of the branching fractions $BR(B\to K^{(*)} \mu^+\mu^-)$ to $BR(B\to K^{(*)} e^+e^-)$ have also been interpreted as further signals of a possible violation of LFU, see~\cite{Ciuchini:2020gvn}  and references therein.  
The tension between inclusive and exclusive determinations of $|V_{ub}|$ is  even larger and without a satisfactory explanation so far~\cite{Alpigiani:2017lpj,Aoki:2019cca,Zyla:2020zbs}. 
From the experimental point of view new data and a  better understanding of the experimental systematics could still change  the present scenario. 

An improvement of the theory, mainly if not uniquely, for exclusive decays is expected from progress in lattice QCD calculations of the relevant form factors.
Indeed, most of the reliable  information about the  form factors relevant in semileptonic $B$ decays is given by first principles calculations made in lattice QCD. 
One is however limited by the cutoff effects induced by the presence of a quark as heavy as  the $b$-quark in calculations done at a finite lattice spacing $a$.  
Actually,  most of the numerical simulations with heavy quarks are performed  at $a$ larger than about $0.05$~fm, so that  for the physical $b$-quark mass we have $m_b a \gtrsim 1$ and an extrapolation in $m_b$ from unphysical values is necessary. 
 In this context discretisation errors affect the value of the form factors at zero recoil and make it difficult to study the momentum  dependence of the form factors at large recoil, namely at  small $q^2$, where $q=p_B-p_{D^{(*)}}$ is the lepton pair momentum in the decay\footnote{In some cases the dependence of the form factors on $q^2$ in the whole allowed kinematical region is supplemented by using also the results of QCD sum rules calculations at small $q^2$.}. 
For this reason, for example, results for the $B \to D$ form factors from lattice QCD~\cite{Na:2015kha,Lattice:2015rga}, together with their uncertainties and correlations, are only available in the range $9.3~{\rm GeV}^2 \lesssim q^2 \lesssim 11.7~{\rm GeV}^2$,  much smaller than the physical range, $0 \lesssim q^2 \lesssim 11.7~{\rm GeV}^2$. 
More recently the first, preliminary results for the momentum-dependence of the form factors in $B \to D^*  \ell \nu$ decays appeared, but  the kinematical region is always restricted at small recoil and  not all the form factors are determined with good accuracy~\cite{Kaneko:2019vkx,Aviles-Casco:2019zop}. 
There exist also calculations of the form factors relevant for $B_s \to D^{(*)}_s$ decays, as well as recent updates of these quantities~\cite{Harrison:2017fmw,McLean:2019qcx,Flynn:2020nmk}. 

In order to supply the lack of information from explicit calculations of the form factors in the full kinematical range, both the experimental analyses  (in order to account for efficiencies and response functions) and the theoretical studies have to assume some parameterisation of the form factors. It is well possible then that the extraction of $|V_{cb}|$ from experiments is biased by the theoretical model adopted in the fits of the data. 
In the years most of the analyses used two popular parameterisations called Boyd-Grinstein-Lebed (BGL)~\cite{ Boyd:1995cf,Boyd:1995sq,Boyd:1997kz} or Caprini-Lellouch-Neubert (CLN)~\cite{Caprini:1995wq,Caprini:1997mu} after the name of the authors.
 

An important step in constraining in a model independent way the $q^2$-dependence of the semileptonic FFs, extracted on the lattice, along the full kinematical range was proposed by L.~Lellouch in a pioneering work~\cite{Lellouch:1995yv}. 
This proposal uses the dispersive techniques mentioned before~\cite{ Boyd:1995cf,Boyd:1995sq,Boyd:1997kz}, see also the original proposal in Refs.~\cite{Okubo:1971jf,Okubo:1971my,Okubo:1972ih,Bourrely:1980gp}, applied to lattice data and introduces a formalism to take into account the errors of the lattice results. 
In spite of the use of the form factors derived from first principles in lattice QCD, the proposal of Ref.~\cite{Lellouch:1995yv} relies, for the unitarity constraints, on the perturbative calculation of the two-point current correlation functions. 
To our knowledge, in spite of the large use of the dispersive techniques discussed above, no one has systematically used the lattice two-point correlators computed non perturbatively in numerical simulations to constraint the FFs in semileptonic decays, not even in the original work~\cite{Lellouch:1995yv}. 

In this work we present an extended study of the two- and three-point lattice correlation functions which are used, together with the dispersive techniques, to constrain the lattice predictions for the form factors. This approach will require a detailed understanding of dispersion relations on the lattice, the combination of perturbative and non perturbative calculations, the renormalisation of  lattice T-products, the control of lattice  artefacts and the peculiar treatment of the statistical errors to be discussed in the following. At the end we will be able to  constrain FFs in a substantially non-perturbative way.

With respect to the proposal by L.~Lellouch~\cite{Lellouch:1995yv} and other previous studies, the main novelties in this work are as follows:
\begin{enumerate}
 \item  The non perturbative determination of the relevant two-point current correlation functions on the lattice which are then used to implement the dispersive bounds;
 \item The possibility of implementing the constraints from the two-point correlation function  computed non-perturbatively also in regions where   the perturbative calculations used in previous analyses are non reliable and could not be used;
 \item The reduction of lattice artefacts in the two-point  correlation functions using fixed-order perturbation theory on the lattice and in the continuum;
 \item A quite  simpler treatment of the lattice uncertainties with respect to the method proposed in Ref.~\cite{Lellouch:1995yv}; 
 \item A new approach to a realistic estimate of the systematic errors present at small values of $q^2$, namely at large momenta of the final meson, both at finite lattice spacing and in the continuum limit, based on the results of Refs.~\cite{DAgostini:2020vsk,DAgostini:2020pim}.
 \end{enumerate}
 
 We stress an important feature of the method that will be presented in this work.
 Consider a set of lattice data for a form factor evaluated at a series of values $q_j^2$ of the squared 4-momentum transfer ($ j = 1, ..., N$).
 The data are distributed according to a multivariate distribution with given uncertainties and correlations.
 The (non-perturbative) unitarity bounds act as a filter by selecting only those combinations of the data that satisfy unitarity and analyticity. 
Since the two-point correlation functions provide an extra information, in general a new multivariate distribution is obtained, which may even correspond to a more precise (and differently correlated) data set.
Using the new distribution the method of Ref.~\cite{Lellouch:1995yv} reproduces exactly each of the data point when $q^2 \to q_j^2$.
It behaves like a fitting procedure passing exactly for the given data set.
This is at variance with what may happen adopting the BGL or CLN parameterisations.
Indeed, in these cases there is no guarantee that the parameterization reproduces exactly the data set and, therefore, the impact of the unitarity filter may be different.
 
We have applied the method described in this work to the analysis of the lattice data of the semileptonic $D \to K \ell \nu_\ell$ decays obtained in Ref.~\cite{Lubicz:2017syv}. We use this process as a training ground for the dispersive approach to show that 
starting from a limited set of data at large $q^2$ it is possible to determine quite precisely the form factors in a model independent way in the full kinematical range, obtaining a remarkable agreement with the direct calculations from Ref.~\cite{Lubicz:2017syv}.
This finding opens the possibility to obtain non-perturbatively the form factors entering the semileptonic $B$ decays in their full kinematical range.
The extension of the non-perturbative dispersive approach described here to $B$ decays requires a remarkable  effort in the reduction of  discretisation effects due to the large masses and momenta involved and it will be the subject of a forthcoming  publication.     

The plan of the paper is as follows. The definition of  the relevant FFs entering in semileptonic  decays is introduced in Section~\ref{sec:GBgk}, where we also give their expressions in the Heavy Quark Effective Theory (HQET)~\cite{Caprini:1997mu}. The definition of the hadronic tensors from the two-point Green functions of suitable bilinear currents and their expression in terms of the form factors is given in Section~\ref{sec:dispbound}, where we also recall the dispersive bounds that can be extracted from these two-point Green functions. The material in Sections~\ref{sec:GBgk}-\ref{sec:dispbound} can be found in several papers. It is, however, useful to have it collected before its use in the present analysis. In particular, we recall the basic features of the dispersive matrix approach of Ref.~\cite{Lellouch:1995yv} and in Appendix~\ref{sec:detine} we also provide new analytical expressions for the numerical evaluation of the unitarity bands of the form factors. The Euclidean lattice correlation functions corresponding to the Minkowskian Green functions that we used in our numerical simulation are presented in Section~\ref{sec:lattice}. In Section~\ref{sec:errors} we discuss the treatment of the statistical and systematic errors with the dispersive bounds and in the presence of kinematical constraints among different form factors. This Section will give us the opportunity of discussing briefly the treatment of the statistical and systematic errors using the Bayesian approach of Refs.~\cite{DAgostini:2020vsk,DAgostini:2020pim}, which is different from the treatment of the errors suggested in the original proposal by L.~Lellouch in Ref.~\cite{Lellouch:1995yv}. 
In Section~\ref{sec:pertlatt} we give the results of the calculation of the two-point correlation functions in perturbation theory both in the continuum and on the lattice, while in Section~\ref{sec:susceptibilities} the two-point correlation functions are evaluated making use of the gauge ensembles produced by the Extended Twisted Mass Collaboration (ETMC) with $N_f = 2+1+1$ dynamical quarks~\cite{Baron:2010bv,Baron:2011sf} (see Appendix~\ref{sec:simulations}). The perturbative calculations are used to reduce discretisation errors and to improve the extrapolation of the numerical non-perturbative results to the continuum limit. 
In Section~\ref{sec:proto} the dispersive matrix approach is applied to the $D \to K$ form factors obtained in Ref.~\cite{Lubicz:2017syv} both in the continuum limit and at finite lattice spacing (using the same gauge ensembles adopted for the evaluation of the two-point correlation functions). Finally, Section~\ref{sec:conclu} contains our conclusion and outlooks for future developments.

\section{Form Factors in  Semileptonic  Decays}
\label{sec:GBgk}

In this Section we introduce the relevant FFs for semileptonic  decays and convert them in the form used in the HQET, which is particularly suitable for the expansion in inverse powers of the heavy quark mass. For definiteness, since this is our final goal, the formulae will refer to  $B \to D$ and $B  \to D^*$ decays. With trivial modifications, the same formalism can be applied to $D \to K$ and $D \to K^*$ semileptonic decays, which is the case study for which we present a complete numerical analysis here, as  well as also to $K \to \pi$ and other semileptonic decays. 

We use the following  form factor classification: 
\begin{itemize}
\item vector current matrix elements
\begin{eqnarray}
\sqrt{m_{B} m_{D}}\kket{ \bbra{D(p_D)| V^{\mu} |\bar{B}(p_B)}} &=& f_{+} (p_B + p_D)^{\mu} + f_{-} (p_B - p_D)^{\mu}\,  ; \nonumber\\
\sqrt{m_{B} m_{D^*}}\kket{ \bbra{D^{*}(p_D,\epsilon_D)| V^{\mu} |\bar{B}(p_B)} }&=& i f_V \epsilon^{\mu \nu \alpha \beta} \epsilon_{D \nu}^{*} p_{D \alpha} p_{B\beta} \,  ; \nonumber \\
\sqrt{m_{B^*} m_{D}}\kket{ \bbra{D(p_D)| V^{\mu} |\bar{B}^{*}(p_B,\epsilon_B)} }&=& i f_{\bar{V}} \epsilon^{\mu \nu \alpha \beta} \epsilon_{B \nu} p_{D \alpha} p_{B\beta}\,  ; \label{eq:VV}\\
\sqrt{m_{B^*} m_{D^*}}\kket{ \bbra{D^{*}(p_D,\epsilon_D)| V^{\mu} |\bar{B}^{*}(p_B,\epsilon_B)}}&=&-[f_{1} (p_B + p_D)^{\mu} + f_{2} (p_B - p_D)^{\mu}] \,  (\epsilon_{D}^{*} \cdot \epsilon_B) \nonumber \\
&\times& + f_{3} (\epsilon_{D}^{*} \cdot p_B) \epsilon_B^{\mu} + f_{4} ( \epsilon_{B} \cdot p_D )\epsilon_{D}^{* \mu}  \nonumber\\
&-& [f_{5} p_B^{\mu} + f_{6} p_D^{\mu}] (\epsilon_{D}^{*} \cdot p_B )(\epsilon_B \cdot p_D) \, ;  \nonumber \end{eqnarray}
\item axial current matrix elements
\begin{eqnarray}
\sqrt{m_{B} m_{D^*}}\kket{ \bbra{D^{*}(p_D,\epsilon_D)| A^{\mu} |\bar{B}(p_B)}} &=& f_{A_1} \epsilon_{D}^{* \mu} - [f_{A_2} p_B^{\mu} + f_{A_3} p_D^{\mu}] (\epsilon_{D}^{*} \cdot p_B)  \,  ; \nonumber\\ 
\sqrt{m_{B^*} m_{D}}\kket{ \bbra{D(p_D)| A^{\mu} |\bar{B}^{*}(p_B,\epsilon_B)} }&=& f_{\bar{A}_1} \epsilon_{B}^{\mu} - [f_{\bar{A}_2} p_B^{\mu} + f_{\bar{A}_3} p_D^{\mu}] (\epsilon_{B} \cdot p_D ) \,  ;  \label{eq:AA} \\
\sqrt{m_{B^*} m_{D^*}}\kket{ \bbra{D^{*}(p_D,\epsilon_D)| A^{\mu} |\bar{B}^{*}(p_B,\epsilon_B)}} &=& i\epsilon^{\mu \nu \alpha \beta} \{ [f_7 (p_B + p_D)_{\nu}  \nonumber\\
&+& f_8 (p_B - p_D)_{\nu}] \epsilon_{B \alpha} \epsilon_{D \beta}^{*}+ [f_9( \epsilon_{D}^{*} \cdot p_B ) \epsilon_{B\nu}\nonumber \\
 &+& f_{10} (\epsilon_{B} \cdot p_D )\epsilon_{D \nu}^{*}] p_{D \alpha} p_{B \beta}   \}\,  , \nonumber
\end{eqnarray}
\end{itemize}
where  the weak currents are given by 
\begin{equation}
\label{corr}
V^{\mu}= \bar{c} \gamma^{\mu} b, \,\,\,\,\, A^{\mu}= \bar{c} \gamma^{\mu} \gamma^5 b\, , 
\end{equation}
 $f_i=f_i(q^2)$ is the generic hadronic FF that depends only on the squared 4-momentum transfer $q^2 =( p_{B^{(*)}} - p_{D^{(*)}})^2$, which is the only non-trivial Lorentz-invariant quantity, and $\epsilon_{D(B)}$ is a polarization 4-vector. Although the initial state is always a $\bar{B}^{(*)}$ meson containing a $b$ quark,  rather than a $B^{(*)}$ meson, to simplify the notation in the following we will omit the bar. The normalisation of the states differs from the Feynman one by a factor equal to the square root of the meson mass, $\ket{p_M}_{\rm Feynman} = \sqrt{M_M}\,  \ket{p_M}$, because this is more convenient in the framework of the HQET.
 
One can define the recoil variable $w$, given by  the scalar product of the meson four-velocities, namely $w = v_{B^{(*)}} \cdot v_{D^{(*)}}$, related to squared 4-momentum transfer $q^2$ by
\begin{equation}
w = \frac{m_{B^{(*)}}^2 + m_{D^{(*)}}^2 - q^2}{2m_{B^{(*)}}m_{D^{(*)}}} \, ,  \label{eq:w}
\end{equation}
and  a new dimensionless variable $r\equiv m_{D^{(*)}}/m_{B^{(*)}}$.
In Ref.~\cite{Caprini:1997mu} Caprini, Lellouch and Neubert introduced the quantities $h_i$, which are linear combinations of the FFs  $f_i$  previously defined, expressed as functions  of the recoil variable  $w$  rather than of $q^2$ (the index $i$ labels a generic form factor). The $h_i$ describe the decompositions (\ref{eq:VV})-(\ref{eq:AA}) in terms of the meson 4-velocities instead of the meson 4-momenta, according to the following classification:
\begin{itemize}
\item scalar FFs
\begin{eqnarray}
S_1^{BD} &=& h_{+} - \frac{1+r}{1-r} \frac{w-1}{w+1} h_{-}; \nonumber \\
S_2^{B^{*}D^{*}} &=& h_1 - \frac{1+r}{1-r} \frac{w-1}{w+1} h_2 \, ; \nonumber \\ 
S_3^{B^{*}D^{*}} &=& w \left[ h_1 - \frac{1+r}{1-r} \frac{w-1}{w+1} h_2 \right] \\ 
&+& \frac{w-1}{1-r} \left[ rh_3-h_4+(1-wr)h_5 + (w-r) h_6 \right] \, ; \nonumber 
\end{eqnarray}
\item vector FFs
\begin{eqnarray}
V_1^{BD} &=& h_{+} - \frac{1-r}{1+r} h_{-}; \nonumber \\
V_2^{B^{*}D^{*}} &=& h_1 - \frac{1-r}{1+r} h_2\, ; \nonumber \\ 
V_3^{B^{*}D^{*}} &=& w \left[ h_1 - \frac{1-r}{1+r} h_2 \right] \, \nonumber \\
&+& \frac{1}{1+r} \left[ (1-wr)h_3 + (r-w) h_4 + (w^2 -1)(rh_5+h_6) \right] \\  
V_4^{BD^{*}} &=& h_V; \nonumber \\
V_5^{B^{*}D} &=& h_{\bar{V}}\, ; \nonumber \\  
V_6^{B^{*}D^{*}} &=& h_3,; \nonumber \\
V_7^{B^{*}D^{*}} &=& h_4\, ; \nonumber  
\end{eqnarray}
\item pseudoscalar FFs
\begin{eqnarray}
P_1^{BD^{*}} &=& \frac{1}{1+r} [(w+1) h_{A_1} - (1-rw) h_{A_2} - (w-r) h_{A_3}]\, ; \nonumber \\ 
P_2^{B^{*}D} &=& \frac{1}{1+r} [r(w+1) h_{\bar{A}_1} - (r-w) h_{\bar{A}_2} - (rw-1) h_{\bar{A}_3}]\, ; \\ 
P_3^{B^{*}D^{*}} &=& h_7 - \frac{1-r}{1+r} h_8\, ; \nonumber  
\end{eqnarray}
\item axial FFs
\begin{eqnarray}
A_1^{BD^{*}} &=& h_{A_1}; \nonumber \\
A_2^{B^{*}D} &=& h_{\bar{A}_1}\, ; \nonumber \\  
A_3^{B^{*}D^{*}} &=& h_7 - \frac{w-1}{w+1} h_8 + (w-1) h_{10}\, ; \nonumber \\
A_4^{B^{*}D^{*}} &=& h_7 + \frac{w-1}{w+1} h_8 + (w-1) h_{9}\, ; \\ 
A_5^{BD^{*}} &=& \frac{1}{1-r} [(w-r) h_{A_1} - (w-1)(rh_{A_2} + h_{A_3})]\, ; \nonumber \\ 
A_6^{B^{*}D}& =& \frac{1}{1-r} [(1-wr) h_{\bar{A}_1} + (w-1)(h_{\bar{A}_2} + rh_{\bar{A}_3})]\, ;\nonumber \\ 
A_7^{B^{*}D^{*}} &=& h_7 - \frac{1+r}{1-r} \frac{w-1}{w+1} h_8 \, . \nonumber  
\end{eqnarray}
\end{itemize}

The last classification of the FFs allows to separate different  values of spin and parity quantum numbers from each other. For simplicity, all scalar, pseudoscalar, vector and axial quantities have been presented indicating initial and final states as superscripts, so that we can easily remember to which process each form factor refers to.

\section{Two-point correlation functions}
\label{sec:dispbound}

The bounds on the different FFs are derived from the two-point functions of suitable currents. 
The starting point is the Fourier transform of the T-product of two hadronic currents, which generalizes the definition of the hadronic vacuum polarization (HVP) tensor. 
Assuming  $x^0>0$ we have 
\begin{equation}
\begin{aligned}
\label{eq:cf}
&\int d^4x \, e^{iq\cdot x}\,  \bra{0} T\{J^{\mu\dagger}(x) J^{\nu}(0)\} \ket{0} =\int d^4x \, e^{iq\cdot x} \sum_n \bra{0} J^{\mu\dagger}(x) \ket{n} \bra{n} J^{\nu}(0) \ket{0}\\
&\hskip 5.52truecm=\sum_n\, \int d^4x \, e^{iq\cdot x}  e^{-ip_n\cdot x}\, \bra{0} J^{\mu\dagger}(0) \ket{n} \bra{n} J^{\nu}(0) \ket{0}\\
&\hskip 5.52truecm=\sum_{n}\, (2 \pi)^4 \delta^{(4)} (q - p_n)\,  \bra{0} J^{\mu\dagger}(0) \ket{n} \bra{n} J^{\nu}(0) \ket{0}\, , \\
\end{aligned}
\end{equation}
where $q$ is the current 4-momentum transfer and $p_n$ the 4-momentum of the intermediate $n$-particle state. 
A similar result can be derived for the case  $x^0<0$. 
The completeness sum runs over all possible intermediate hadronic states and in particular we will focus our attention onto either a single-particle $B_c^{(*)}$-meson state or two-particle states composed by a $B^{(*)}$-meson and a $D^{(*)}$-meson. 
Note that pseudoscalar (scalar) particles will be characterized only by their four-momentum $p^{\mu}$ while the vector (axial) mesons will be distinguished also by their polarization four-vector $\epsilon^{\mu}$.
The link between the two-particle states appearing in the completeness sum (\ref{eq:cf}) and the classification of the FFs introduced in Eqs.\,(\ref{eq:VV})-(\ref{eq:AA}) is given by the substitution
\begin{equation} 
 \kket{ \bbra{ D^{(*)}| J^{\mu} |B^{(*)} }} \rightarrow   \kket{ \bbra{B^{(*)}D^{(*)}}| J^{\mu} |0} \, , 
\end{equation}
which can be simply realised by inverting the sign of $p_{D^{(*)}}$ and by analytic continuation of the FFs in $q^2$ from $m_\ell^2 \le q^2 \le (m_{B^{(*)}} - m_{D^{(*)}})^2$ to $(m_{B^{(*)}} + m_{D^{(*)}})^2 \le q^2  \le \infty$. 
Thus, the amplitudes entering in semileptonic decays are strongly correlated to the T-product in Eq.\,(\ref{eq:cf}), which is the reason why the latter is so important to constrain the form factors.

The Fourier transform of the T-product defines the following HVP tensors: 
\begin{eqnarray}
\Pi_V^{\mu\nu}(q)&=& i \,\int d^4x \, e^{iq\cdot x}\,  \bra{0} T\{V^{\mu\dagger}(x) V^{\nu}(0)\} \ket{0}\\
 &=& (q^{\mu}q^{\nu}-g^{\mu\nu}q^2) \, \Pi_{1^{-}}(q^2) + q^{\mu}q^{\nu}\, \Pi_{0^{+}}(q^2)\, , \nonumber \\
\Pi_A^{\mu\nu}(q)&=& i \, \int d^4x\,  e^{iq\cdot x} \, \bra{0} T\{A^{\mu\dagger}(x) A^{\nu}(0)\} \ket{0}\label{eq:Pis}\\
 &=& (q^{\mu}q^{\nu}-g^{\mu\nu}q^2)\,  \Pi_{1^{+}}(q^2) + q^{\mu}q^{\nu}\, \Pi_{0^{-}}(q^2)\, , \nonumber 
\end{eqnarray}
where $V^{\mu}$, $A^{\mu}$ are defined in Eq.\,(\ref{corr}) and the subscripts $0^{\pm}$,$1^{\pm}$ represent spin-parity quantum numbers of the intermediate states. Note that inserting a completeness sum between the vector or axial four-currents we are able to relate these expressions to Eq.\,(\ref{eq:cf}).

The quantities $\Pi_{0^{\pm}}$,$\Pi_{1^{\mp}}$ are called \emph{polarization functions}. In particular, the term proportional to $\Pi_{0^{+}}$ ($\Pi_{0^{-}}$) represents the \emph{longitudinal} part of the HVP tensor with vector (axial) four-currents, while the term proportional to $\Pi_{1^{-}}$ ($\Pi_{1^{+}}$) is the \emph{transverse} contribution to the HVP tensor with vector (axial) four-currents.

\subsection{Dispersion relations and analytic expressions for exclusive $B$ decays}
\label{subsec:disrelBD}

The imaginary parts of the longitudinal and  transverse polarization functions introduced in Eqs.\,(\ref{eq:Pis}) are related to 
their derivatives with respect to $q^2$ by the dispersion relations 
\begin{eqnarray}
\chi_{0^{+}} (q^2)&\equiv& \frac{\partial}{\partial q^2} [q^2  \Pi_{0^{+}} (q^2)]=\frac{1}{\pi} \int_0^{\infty} dz \frac{z \,{\rm Im }\Pi_{0^{+}}(z)}{(z-q^2)^2}\, , \nonumber \\[2mm]
\chi_{0^{-}} (q^2)&\equiv& \frac{\partial}{\partial q^2} [q^2  \Pi_{0^{-}} (q^2)]=\frac{1}{\pi} \int_0^{\infty} dz \frac{z\,{\rm Im }\Pi_{0^{-}}(z)}{(z-q^2)^2}\, ,
\label{veax21}\\[2mm]
\chi_{1^{-}} (q^2)&\equiv& \frac{1}{2} \left( \frac{\partial}{\partial q^2} \right)^2 [q^2  \Pi_{1^{-}} (q^2)]=\frac{1}{\pi} \int_0^{\infty} dz \frac{z \, {\rm Im }\Pi_{1^{-}}(z)}{(z-q^2)^3}\, , \nonumber\\[2mm]
\label{veax22}
\chi_{1^{+}} (q^2)&\equiv& \frac{1}{2} \left (\frac{\partial}{\partial q^2} \right)^2 [q^2  \Pi_{1^{+}} (q^2)]=\frac{1}{\pi} \int_0^{\infty} dz \frac{z \, {\rm Im}\Pi_{1^{+}}(z)}{(z-q^2)^3}\, .\nonumber
\end{eqnarray}

In what follows we will denote by $\chi$ a generic susceptibility. 
From a dimensional point of view note that the longitudinal (scalar/pseudoscalar)  susceptibilities $\chi_{0^{\pm}}$ are dimensionless, while the transverse (vector/axial) ones have dimension $[E]^{-2}$, where $E$ is an energy.  

The  two-particle contribution to the polarization functions can be expressed in terms of the FFs defined in the previous Section:
\begin{itemize}
\item scalar channel
\begin{eqnarray}\begin{aligned}
& q^2\, {\rm Im }\Pi_{0^{+},2p}\left[w(q^2)\right]=\frac{m_{B^{(*)}} m_{D^{(*)}}}{8  \pi } \sum_{i=1}^{3}(1 + \delta_{i2}) \frac{ (w^2-1)^{1/2}(w+1)}{4}^2  \frac{(\beta_i^2-1)\vert S_i \vert^2}{(\beta_i^2-\frac{w+1}{2})^2}\, ,   \label{eq:sch}
\end{aligned}
\end{eqnarray}
\item vector channel
\begin{eqnarray}
\begin{aligned}
&q^2\, {\rm Im }\Pi_{1^{-},2p}\left[w(q^2)\right]=\frac{m_{B^{(*)}} m_{D^{(*)}}} {96\pi} \left[ \sum_{i=1}^{3} (1 + \delta_{i2}) (w^2-1)^{3/2} \frac{ \beta_i^2\vert V_i \vert^2}{(\beta_i^2-\frac{w+1}{2})^2} \right. \\ 
& \hskip 3.6truecm+ \left. \sum_{i=4}^{7}  (w^2-1)^{3/2}  \frac{2\vert V_i \vert^2}{(\beta_i^2-\frac{w+1}{2})} \right] \, , \label{eq:vch}
\end{aligned}
\end{eqnarray}
\item pseudoscalar channel
\begin{eqnarray}\begin{aligned}
q^2\,{\rm Im }\Pi_{0^{-},2p}\left[w(q^2)\right]=\frac{m_{B^{(*)}} m_{D^{(*)}}}{32  \pi }\sum_{i=1}^{3} (1 + \delta_{i3})(w^2-1)^{3/2} \frac{ \beta_i^2\vert P_i \vert^2}{(\beta_i^2-\frac{w+1}{2})^2}\, , \label{eq:pch}
\end{aligned}\end{eqnarray}
\item axial channel
\begin{eqnarray}
\begin{aligned}
& q^2\,{\rm Im }\Pi_{1^{+},2p}\left[w(q^2)\right]= \frac{m_{B^{(*)}} m_{D^{(*)}}}{24\pi} \left[ \sum_{i=1}^{4} (w^2-1)^{1/2} \frac{(w+1)}{4}^2 \frac{2 \vert A_i \vert^2}{(\beta_i^2-\frac{w+1}{2})} \right. \\ 
&\hskip 3.6truecm+ \left. \sum_{i=5}^{7}  (w^2-1)^{1/2}  \frac{(w+1)}{4}^2 (1 + \delta_{i7})  \frac{(\beta_i^2-1)\vert A_i \vert^2}{(\beta_i^2-\frac{w+1}{2})^2} \right] \, , \label{eq:avch}
\end{aligned}
\end{eqnarray}
\end{itemize}
where $w(q^2)$ is given by Eq.\,(\ref{eq:w}) and the quantity $\beta_i$ is defined as 
\begin{equation}
\label{betai}
\beta_i=\frac{m_{B^{(*)}}+m_{D^{(*)}}}{2\sqrt{m_{B^{(*)}}m_{D^{(*)}}}}.
\end{equation}

\noindent Note that the above expressions hold for a single two-particle $B^{(*)}$-$D^{(*)}$ intermediate state. If there are several, essentially degenerate, such states, differing for example by the replacement of an up with a down quark (such as $B^-$ - $\bar D^0$ with $B^0$ - $D^-$),  then we have to sum up their contributions to ${\rm Im }\Pi_{J^P,2p}$.

We will also  need the expression for the one-particle contribution of mesons to the imaginary part of the polarization functions.
In the case of the vector $B_c^*$ and of the pseudoscalar $B_c$ mesons the one-particle current matrix elements are defined as 
\begin{eqnarray}
\bra{0} A^\mu \ket{B_c} &=& f_{B_c}\,  p^\mu_{B_c},\nonumber \\ 
\bra{0}V^\mu \ket{B_c^*,\lambda} &=& f_{B_c^*} M_{B_c^*} \epsilon^\mu_\lambda(p_{B^*_c}) \qquad (\lambda =1,2,3) 
\end{eqnarray}
with $f_{B_c^{(*)}}$ being the decay constants of the aforementioned mesons, $M_{B_c^{(*)}}$ their masses, $\lambda$  the $B_c^*$ polarization and $\epsilon_\lambda$ the corresponding polarization vector. 
The one-particle contributions to the polarization functions are then given by
\begin{eqnarray}
q^2\,{\rm Im }\Pi_{0^-,1p} (q^2) = \pi \delta(q^2-M_{B_c}^2) \, f_{B_c}^2 M_{B_c}^2\, , \nonumber \\[2mm]
q^2\,{\rm Im } \Pi_{1^-,1p} (q^2) = \pi \delta(q^2-M_{B_c^*}^2)\, f_{B_c^*}^2 M_{B_c^*}^2 \, .
\end{eqnarray}

These one-particle contributions can be opportunely subtracted from the corresponding susceptibilities through the dispersion relations. 
Thus, we can define
\begin{eqnarray}
\chi_{0^-}(q^2)\vert_{sub} &\equiv& \chi_{0^-}(q^2) - \sum^N_n \frac{f_{B_c,n}^2 M_{B_c,n}^2}{(M_{B_c,n}^2-q^2)^2}\, ,\nonumber \\[2mm]
\chi_{1^-}(q^2)\vert_{sub} &\equiv& \chi_{1^-}(q^2) - \sum^M_m \frac{f_{B_c^*,m}^2 M_{B_c^*,m}^2}{(M_{B_c^*,m}^2-q^2)^3}\, , \label{eq:subtra1p}
\end{eqnarray}
where in general we can consider $N$ pseudoscalar and $M$ vector poles (see, e.g., Ref.~\cite{Bigi:2017jbd} for reference values of the masses and the decay constants of such poles).
The generalization to possible one-particle states below the annihilation threshold in the scalar and axial-vector channels is straightforward.

\subsection{Dispersive bounds}
\label{subsec:dipdip}

First of all, we will describe the general ideas behind the dispersive method of Refs.~\cite{Okubo:1971jf}-\cite{Bourrely:1980gp} thanks to which one can obtain bounds for a generic form factor $f(q^2)$. 
We define 
\begin{equation}
t_{\pm} = \left( m_B \pm m_{D^{(*)}} \right)^2 
\end{equation}
and we use the analytic continuation of the amplitudes from the kinematical decay region, where $m_{\ell}^2 \leq q^2 \leq t_-$, to the single-meson or  pair production region, where $m_{B_c}^2 \leq q^2$ or $t_+ \leq q^2$. 
By means of the dispersion relations we can use the two-point correlation functions computed in QCD to obtain the constraints on $f(q^2)$ and then use analyticity to translate these constraints into the physical FFs relevant to semileptonic decays. 

The dispersion relations that we will consider in this work have already been introduced in Eqs.\,(\ref{veax21}). We also recall  that the derivatives of the various polarization functions can be determined in perturbative QCD only for values of  $q^2$ that are far from the region of production of  resonance states, namely 
\begin{equation}
(m_b + m_c) \Lambda_{QCD} \ll  (m_b + m_c)^2-q^2,
\end{equation}
where $m_b\sim 4.2$~GeV and $m_c\sim 1.3$~GeV are approximate values of the bottom and charm quark masses, respectively. A possible choice is thus also the value $q^2 = 0$, which has been widely used in the past, particularly in all the calculations that used the perturbative expression of the susceptibilities $\chi(q^2)$. 
On the contrary, with a non-perturbative determination of the two-point correlation functions we can use the most convenient value of $q^2$ at disposal, namely the value which will allow the most stringent bounds on the form factors. 

By inserting a complete set of states with the same quantum numbers of a generic current $J$ we have\footnote{For simplicity we omit Lorentz indices and other complications that are immaterial for the present discussion.}  
\begin{eqnarray}
{\rm Im} \Pi_{0^{\pm},1^{\mp}}=\frac{1}{2} \sum_n \int d\mu(n) (2\pi)^4 \delta^{(4)}(q-p_n) \vert \bra{0} J \ket{n}\vert^2 \, , 
\end{eqnarray}
where $d\mu(n)$ is the measure of the phase space for the set of states $n$. 
As the completeness sum is semi-positive definite, we can restrict our attention to a subset of hadronic states and thus produce a strict inequality.
This consideration allows us to rewrite the dispersion relations for $\chi(q^2)$ as
\begin{equation}
\label{eq:999}
\frac{1}{\pi\,\chi(q^2)} \int_{t_+}^{\infty} dt \frac{W(t)\vert f(t)\vert^2}{(t-q^2)^3} \leq 1,
\end{equation}
where $f(t)$ is the generic form factor and $W(t)$ is a computable function that depends on the particular form factor under consideration and is related to phase space factors. 
These have been given explicitly in Eqs.\,(\ref{eq:sch})-(\ref{eq:avch}) for all the possible bilinears. 

We can now use analyticity to turn the result~(\ref{eq:999}) into a constraint for the semileptonic region. To achieve this goal, it is necessary that the integrand is analytic below the pair-production threshold $t < t_+$. To this end we define
\begin{equation}
\label{zzz}
z(t,t_s) \equiv \frac{\sqrt{t_+ - t}-\sqrt{t_+-t_s}}{\sqrt{t_+ - t}+\sqrt{t_+-t_s}},
\end{equation}
which is real for $t_s < t_+$, zero for $t = t_s$ and a complex number on the unitary circle  for $t \geq t_+$. We can remove the poles of the integrand of Eq.(\ref{eq:999}) by multiplying it  by  appropriate powers of the  $z(t,t_s)$s, as determined by  the positions $t_s$ of the sub-threshold poles. Each pole has a distinct value of $t_s$, and the product $z(t , t_{s1})^{k_1} z(t, t_{s2})^{k_2} \, \dots$ removes all of them. Hence, we re-express Eq.\,(\ref{eq:999}) as
\begin{equation}
\label{eq:999888}
\frac{1}{\pi} \int_{t_+}^{\infty} dt \left| \frac{dz(t,t_0)}{dt} \right| \times  \left| \tilde{\Phi}(t,t_0)P(t)f(t) \right|^2 \leq 1,
\end{equation}
where $t_0$ is an arbitrary point that we will define below and  we have introduced the \emph{Blaschke factor} $P(t)$. The latter  is a product of many quantities of the form (\ref{zzz}) at the position of the sub-threshold poles (i.e., $t_s \leq t_+$), and the \emph{outer function} $\tilde{\Phi}(t,t_0)$, which is defined for the vector/axial channel as
\begin{equation}
\tilde{\Phi}(t,t_0) = \tilde{P}(t) \left[ \frac{W(t)}{\vert dz(t,t_0) /dt \vert \, \chi_{1^{\pm}}(q^2)(t-q^2)^3}    \right]^{1/2}.
\end{equation}
In the last expression $\tilde{P}(t)$ represents a product of the $z(t,t_s)$'s and $\sqrt{z(t,t_s)}$'s that remove the sub-threshold singularities and cuts in the kinematical part $W(t)$. 
Regarding the choice of $t_0$  in Eq.\,(\ref{eq:999888}), there are several possibilities which may be more or less convenient depending on the quantity at hand.
We have followed the common lore and, in the remaining  of the paper,  we used  $t_0 = t_-$,  so that the allowed  kinematic interval $[0, t_-]$  corresponds to the range $[z_{\rm max}, 0]$. 
The physical values of $z_{\rm max}$ for  different semileptonic  decays are given in Table~\ref{tab:zmax}. 
In a lattice calculation $z_{\rm max}$ will depend on the values of the initial and final meson masses at which the simulations are performed. 
\begin{table}[htb!]
\renewcommand{\arraystretch}{1.2}
\begin{center}
{\small
\begin{tabular}{|c|c|c|c|c|}
\hline
& $~D \to \pi~$ & $~D \to K~$ & $~B \to D~$ & $~B \to D^*~$\\
\hline
$~z_{\rm max}~$ & $~0.33~$& $~0.10~$& $~0.065~$ & $~0.056~$\\
\hline
\end{tabular}
}
\caption{\textit{Maximum values of the conformal variable $z = z(t, t_-)$ for various decay processes.}}
\label{tab:zmax}
\end{center}
\renewcommand{\arraystretch}{1.0}
\end{table}

A way to account for the bounds imposed by the susceptibilities on the form factors is provided by the two popular parameterisations BGL~\cite{ Boyd:1995cf,Boyd:1995sq,Boyd:1997kz} or CLN~\cite{Caprini:1995wq,Caprini:1997mu}.
Coming back to Eq.\,(\ref{eq:999888}), the quantity $\tilde{\Phi}(t,t_0) P(t) f(t)$ can be expanded in a set of orthonormal functions, proportional to powers of $z(t, t_0)$. The consequence of this strategy is that the form factor $f(t)$ in the semileptonic region can be expressed as
\begin{equation}
\label{222}
f(t)=\frac{1}{P(t)\tilde \Phi(t,t_0)} \sum_{n=0}^{\infty}\,  a_n\, z(t,t_0)^n,
\end{equation}
where, because of Eq.\,(\ref{eq:999888}), the coefficients $a_n$ have to satisfy the unitarity condition
\begin{equation}
\label{unitBGL111}
\sum_{n=0}^{\infty} \vert a_n\vert^2 \leq 1 ~ .
\end{equation}
In the case of $B\to D^{(*)}$ decays, since $z_{max} < 0.07$, the series present in Eq.\,(\ref{222}) are usually truncated after the first two or three terms, introducing only small uncertainties in the theoretical predictions.

In this work we do not adopt the BGL or CLN approach, since there is another alternative formulation~\cite{Lellouch:1995yv}, which is very convenient for translating the information given by the susceptibility $\chi(q^2)$ into a bound on the form factors. 
Indeed, we make the transformation
\beq 
\frac{1+z}{1-z}=\sqrt{\frac{t_+-t}{t_+-t_-}} ~ , ~
\eeq
or 
\beq 
\label{conf}
z = \frac{\sqrt{\frac{t_+-t}{t_+-t_-}}-1}{\sqrt{\frac{t_+-t}{t_+-t_-}}+1} ~ , ~
\eeq
which corresponds to $z = z(t, t_-)$. 
For values of $t$ in the range relevant for semileptonic decays (i.e.~$[0, t_ -]$) we map the complex $t$-plane into the unit disc in the variable $z$, whereas the integral around the cut in Eq.\,(\ref{eq:999888}) becomes an integral around the unit circle. 
Then, a generic integral of the form given in Eq.\,(\ref{eq:999}) can be written as an integral over $z$~\cite{Bourrely:1980gp}-\cite{Lellouch:1995yv}  
\begin{equation}
\label{eq:JQ2z}
\frac{1}{2\pi i } \int_{\vert z\vert =1} \frac{dz}{z}   \vert\phi(z,q^2) f(z)\vert^2 \leq \chi(q^2)\, , 
\end{equation}
where the kinematical functions $\phi(z,q^2)$ for the different form factors entering $B \to D^{(*)}$ decays will be specified in Eq.\,(\ref{eq:kinfun}) below.
At this point, by introducing an inner product defined as
\beq 
\langle g\vert h\rangle =\frac{1}{2\pi i } \int_{\vert z\vert=1 } \frac{dz}{z}   \bar {g}(z) h(z)\, ,  \label{eq:inpro}
\eeq
where $\bar{g}(z)$ is the complex conjugate of the function $g(z)$, the inequality~(\ref{eq:JQ2z}) can simply be written as
\begin{equation}
\label{eq:JQinpro}
  0\leq \langle\phi f\vert \phi f\rangle\leq \chi(q^2)\, , 
\end{equation}
where we have also used the positivity of the inner product.

\subsection{Extending the method of the dispersive bounds}
\label{subsec:exdip}

Following Refs.~\cite{Bourrely:1980gp}-\cite{Lellouch:1995yv}, we define the function $g_t(z)$ as
\begin{equation}
g_t(z) \equiv \frac{1}{1-\bar{z}(t) z}\, , 
\end{equation}
where $\bar{z}(t)$ is the complex conjugate of the variable $z(t)$ defined in Eq.\,(\ref{conf}) and $z$ is the integration variable of Eq.\,(\ref {eq:inpro}).
It is then straightforward to show that  
\begin{equation}
\label{CI}
\langle g_t|\phi f \rangle  = \phi(z(t),q^2)\, f\left(z(t)\right)\, , \qquad   \langle g_{t_m} | g_{t_l} \rangle  = \frac{1}{1- z(t_l) \bar{z}(t_m)}.
\end{equation}

Let us introduce  the matrix 
\begin{equation}
\label{Der}
\mathbf{M} = \left (
\begin{array}{ccccc}
\langle\phi f | \phi f \rangle  & \langle\phi f | g_t \rangle  & \langle\phi f | g_{t_1} \rangle  &\cdots & \langle\phi f | g_{t_n}\rangle  \\
\langle g_t | \phi f \rangle  & \langle g_t |  g_t \rangle  & \langle  g_t | g_{t_1} \rangle  &\cdots & \langle g_t | g_{t_n}\rangle  \\
\langle g_{t_1} | \phi f \rangle  & \langle g_{t_1} | g_t \rangle  & \langle g_{t_1} | g_{t_1} \rangle  &\cdots & \langle g_{t_1} | g_{t_n}\rangle  \\
\vdots & \vdots & \vdots & \vdots & \vdots \\ 
\langle g_{t_n} | \phi f \rangle  & \langle g_{t_n} | g_t \rangle  & \langle g_{t_n} | g_{t_1} \rangle  &\cdots & \langle g_{t_n} | g_{t_n} \rangle  \\
\end{array}\right )  \, .
\end{equation}
In a numerical simulations of lattice QCD the values ${t_1, \cdots, t_n}$ will correspond to  the squared 4-momenta at which the FFs have been computed non-perturbatively and that will be used as inputs for constraining the FF in regions non accessible to the calculation.

Note that the first matrix element in (\ref{Der}) is the quantity directly related to the susceptibility $\chi(q^2)$  through the dispersion relations, see Eq.\,(\ref{eq:JQinpro}).
To be more specific, in the case of $B \to D$ decays, in terms of  the longitudinal and  transverse susceptibilities $\chi_{0^+}(q^2)$ and $\chi_{1^-}(q^2)$ we have that:
\begin{eqnarray}
 && \langle \phi_0 f_0 | \phi_0 f_0 \rangle  \leq \,  \chi_{0^+}(q^2)\, ,  \nonumber  \\
&& \langle \phi_+ f_+ | \phi_+ f_+ \rangle \leq\, \chi_{1^-}(q^2) \, , 
\end{eqnarray}
where $\phi_{0,+}$ are kinematical functions 
\begin{eqnarray}
\phi_0(z,q^2) &=& \sqrt{\frac{2 n_I}{3}}\sqrt{\frac{3 t_+ t_-}{4 \pi}} \frac{1}{t_+ - t_-} \frac{1+z}{(1-z)^{5/2}} \left( \rho(0) + \frac{1+z}{1-z}\right)^{-2}\left( \rho(q^2) + \frac{1+z}{1-z}\right)^{-2}\, , \nonumber  \\
\phi_+(z,q^2) &=&  \sqrt{\frac{2 n_I}{3}}\sqrt{\frac{1}{\pi (t_+-t_-)}} \frac{(1+z)^2}{(1-z)^{9/2}} \left( \rho(0) + \frac{1+z}{1-z}\right)^{-2}\left( \rho(q^2) + \frac{1+z}{1-z}\right)^{-3}\, ,  \label{eq:kinfun}
\end{eqnarray}
where $n_I$  is an isospin Clebsch-Gordan factor and
\begin{equation}
\rho(q^2) \equiv \sqrt{\frac{t_+ - q^2}{t_+-t_-}}\, .
\end{equation}
When analyticity does not hold, i.e.~when a form factor has, for instance, $N$ poles at $t = t_{P1}, t_{P2}, \cdots...,t _{PN}$, it is sufficient to modify the kinematical function $\phi$ according to
\begin{equation}
\phi(z,q^2) \to \phi_P(z,q^2) \equiv \phi(z,q^2) \times \frac{z-z(t_{P1})}{1-\bar{z}(t_{P1})z}  \times \cdots  \times  \frac{z-z(t_{PN})}{1-\bar{z}(t_{PN})z}
\label{eq:modify} \end{equation}
and the previous definitions will continue to be valid.

The positivity of the inner products~(\ref{CI}) guarantees that the determinant of the matrix~(\ref{Der}) is positive semi-definite, namely
\begin{equation}
\label{detpos}
\det \mathbf{M} \geq 0.
\end{equation}
This condition can be rephrased in the second order inequality
\begin{equation}
\label{diseq2}
 \alpha \langle g_{t} | \phi f  \rangle^2 + 2 \beta \langle g_{t} | \phi f  \rangle \leq \, \gamma \, , 
\end{equation}
with 
\begin{eqnarray}
\alpha &\equiv& \det \mathbf{M}^{\{(1,1),(2,2)\}}\, , \nonumber \\
\label{beta1}
\beta &\equiv& \sum_{i=1}^n (-1)^{1+i} \det \mathbf{M}^{\{(1,1),(2,i+1)\}} \langle g_{t_i} | \phi f  \rangle \, , \\
\gamma &\equiv& \chi(q^2) \det \mathbf{M}^{\{(1,1)\}}-  \sum_{i,j=1}^n (-1)^{i+j} \det \mathbf{M}^{\{(1,1),(i+1,j+1)\}} \langle g_{t_i} | \phi f  \rangle\langle g_{t_j} | \phi f  \rangle \, , \nonumber 
\end{eqnarray}
where  $\mathbf{M}^{\{(i_1,j_1),(i_2,j_2),\cdots\}}$ is the minor obtained by deleting the rows $i_1,i_2,\cdots$ and the columns $j_1,j_2,\cdots$.
Calling $\Delta$ the discriminant of the inequality~(\ref{diseq2}), one can show that
\begin{equation}
\label{delta}
\Delta = \det \mathbf{M}^{\{(1,1)\}} \times \det \mathbf{M}^{\{(2,2)\}} \equiv \Delta_1 \times \Delta_2,
\end{equation}
so that at the end the relevant quantities will only be $\alpha,\beta, \Delta_1, \Delta_2$. 
Note that $\alpha$ and $ \Delta_2$ are $t$-independent, $i.e.$ they are given numbers once the susceptibility $\chi(q^2)$ and the lattice QCD inputs are chosen. On the contrary, $\beta$ and $ \Delta_1$ are $t$-dependent. Moreover, only the quantities $\beta$ and $\Delta_2$ depend on the chosen value of $q^2$.

At this point, since $\Delta_1 \geq 0$ by construction, the inequality (\ref{diseq2}) will have an acceptable solution only when $ \Delta_2 \geq 0$. If this condition is satisfied, by expressing the scalar product $ \langle g_{t} | \phi f  \rangle$ according to Eq.~(\ref{CI}) we obtain the following unitarity constraints on the form factor $f(t)$
\begin{equation}
\label{loup}
f_{lo}(t, q^2) \leq  f(t) \leq f_{up}(t, q^2)\, , 
\end{equation}
where 
\begin{equation}
\label{loup2}
f_{lo(up)}(t, q^2) \equiv \frac{-\beta(t, q^2) \mp \sqrt{\Delta_1(t) \, \Delta_2(q^2)}}{\alpha\,  \phi(z(t),q^2)\,}\, .
\end{equation}
Thus, by using a direct lattice measurement of the form factors at the points $t_1, t_2, \dots, t_n$ and the two-point functions of the suitable currents we can constrain the form factors in regions of momenta which for several reasons may not be accessible to lattice simulations.
The application to the case of the semileptonic $D \to K$ decays will be presented in Section~\ref{sec:proto}. 

The positivity of the inner product implies that $\alpha$ and $\Delta_1$ are strictly positive,  in such a way that, when also $\Delta_2(q^2) \geq 0$, the bounds in Eq.\,(\ref{loup2}) are well defined.
We stress that the unitarity filter $\Delta_2(q^2) \geq 0$ is $t$-independent, which implies that, when it is not satisfied, no prediction for $f(t)$ is possible at any value of $t$.

We point out an interesting feature of the dispersive approach based on the matrix~(\ref{Der}).
When the momentum transfer $t$ coincides with one of the data points, i.e.~when $t \to t_j$, the determinant $\Delta_1(q^2) \to 0$ and the quantity $\beta / \alpha \to \phi(z(t_j),q^2) f(z(t_j))$, so that $f_{lo(up)}(t, q^2) \to f(z(t_j))$.
In other words the form factor $f(t)$, obtained from the dispersive matrix method, reproduces exactly the given set of data points.
This is at variance with what may happen using the BGL or the CLN parameterisations, i.e.~when the number of powers of $z$ included in Eq.~(\ref{222}) is truncated below the number of data points. In this case there is no guarantee that the parameterization reproduces exactly the set of input data.

Explicit analytical expressions for $f_{lo(up)}(t, q^2) $, which are very useful for their direct numerical evaluation, are given in Appendix \ref{sec:detine}.

\section{Euclidean two-point functions}
\label{sec:lattice}


In this Section we discuss in details the approach that has been followed in order to constrain the values of the FFs from the two-point correlation functions computed non perturbatively by numerical QCD simulation on the lattice.  Many of the definitions and formulae introduced in this Section have been used on the lattice to compute the HVP function of two electromagnetic currents~\cite{Bernecker:2011gh} and its isospin-breaking corrections~\cite{Giusti:2017jof} contributing to the muon $g-2$.

\subsection{Basic definitions}
\label{subsec:basic}
 We compute the correlation functions at the  Euclidean four-momentum $Q\equiv(Q_0,\vec{Q})$, given in terms of   the Minkowskian momentum  $q\equiv(q_0,\vec{q})$  by  the relations $Q_0=iq_0$ and $\vec{Q}=\vec{q}$. With this choice  $Q^2=-q^2$. Furthermore, we perform a Wick rotation on the coordinates, so that we  pass from  the Minkoskian coordinates  $x_M=(\tau,\vec{x})$ to the Euclidean ones $x=(t,\vec{x})$, with $t = i\,\tau$.
The  vector and axial HVP tensors  take the form 
\begin{eqnarray}
\Pi_V^{\mu\nu}(Q)&=&  \int d^4x e^{-iQ\cdot x} \bra{0} T\{V_E^{\mu\dagger}(x) V_E^{\nu}(0)\} \ket{0}\nonumber \\
 &=& (-Q^{\mu}Q^{\nu}+\delta^{\mu\nu}Q^2) \Pi_{1^{-}}(Q^2) - Q^{\mu}Q^{\nu}\Pi_{0^{+}}(Q^2)\, , \\
\Pi_A^{\mu\nu}(Q)&=&  \int d^4x e^{-iQ\cdot x} \bra{0} T\{A_E^{\mu\dagger}(x) A_E^{\nu}(0)\} \ket{0}\nonumber\\
 &=& (-Q^{\mu}Q^{\nu}+\delta^{\mu\nu}Q^2) \Pi_{1^{+}}(Q^2) - Q^{\mu}Q^{\nu}\Pi_{0^{-}}(Q^2) \nonumber \, , 
\end{eqnarray}
where we have introduced the  currents 
\begin{equation}
V_E^{\mu}= \bar{c} \gamma_E^{\mu} b, \,\,\,\,\, A^{\mu}= \bar{c} \gamma_E^{\mu} \gamma_E^5 b \, .
\end{equation}
defined in terms of Hermitian,  Euclidean  Dirac matrices   satisfying  the anticommutation relations
\begin{equation}
\{\gamma^{\mu}_E,\gamma^{\nu}_E\}=2\delta^{\mu\nu}\, . 
\end{equation}
 In the following we will omit the explicit subscript $E$ in the definition of the Euclidean currents and $\gamma$-matrices.

\subsection{Correlators and derivatives of  the polarization functions on the lattice}
\label{subsec:correlator}

A convenient choice is  to work with  the momentum $Q=(Q^0,\vec{0})$ so that 
\begin{eqnarray}
Q^2\Pi_{0^+}(Q^2)&=&-\int_{-\infty}^{\infty} dt^\prime\, e^{-iQt^\prime} C_{0^+}(t^\prime)\,  , \nonumber\\
Q^2\Pi_{1^-}(Q^2)&=&-\int_{-\infty}^{\infty} dt^\prime \,e^{-iQt^\prime} C_{1^-}(t^\prime)\, ,\label{eq:tC}\\
Q^2\Pi_{0^-}(Q^2)&=&-\int_{-\infty}^{\infty} dt^\prime \, e^{-iQt^\prime} C_{0^-}(t^\prime)\, , \nonumber \\
Q^2\Pi_{1^+}(Q^2)&=&-\int_{-\infty}^{\infty} dt^\prime \, e^{-iQt^\prime} C_{1^+}(t^\prime)\, , \nonumber
\end{eqnarray}
where the explicit expressions of the various correlators  computed at the time distance $t$ are 
\begin{eqnarray}
C_{0^+}(t)&=&\int d^3\vec x\,  \bra{0} T\{ \bar{b}(t,\vec x) \gamma_0 c(t,\vec x) \bar{c}(0) \gamma_0 b(0)\} \ket{0}\, , \nonumber\\
C_{1^-}(t)&=&\frac{1}{3} \sum_{i=1}^3 \int d^3\vec x\, \bra{0} T\{ \bar{b}(t,\vec x) \gamma_i c(t,\vec x) \bar{c}(0) \gamma_i b(0)\} \ket{0}\, , \label{dpf2}\\
C_{0^-}(t)&=&\int d^3\vec x\, \bra{0} T\{ \bar{b}(t,\vec x) \gamma_0\gamma_5 c(t,\vec x) \bar{c}(0) \gamma_0\gamma_5 b(0)\} \ket{0}\, ,  \nonumber\\
C_{1^+}(t)&=&\frac{1}{3} \sum_{i=1}^3 \int d^3\vec x\, \bra{0} T\{ \bar{b}(t,\vec x) \gamma_i\gamma_5 c(t,\vec x) \bar{c}(0) \gamma_i\gamma_5 b(0)\} \ket{0}\, . \nonumber
\end{eqnarray}

By recalling the definition of the spherical Bessel functions
\begin{equation}
j_0(z) = \frac{\sin(z)}{z} ~ , ~ \qquad j_1(z)=\frac{\sin(z)}{z^2}-\frac{\cos(z)}{z} ~   
\end{equation}
and given that
\begin{equation}
\frac{\partial}{\partial Q^2}\cos(Qt)=-\frac{t^2}{2}j_0(Qt),
\end{equation}
we get 
\begin{eqnarray}
\label{dpf1}
\chi_{0^{+}} (Q^2)&=&\int_0^{\infty} dt^\prime\, t^{\prime \, 2} j_0(Qt^\prime) C_{0^+}(t^\prime)\, ,  \nonumber\\
\chi_{1^{-}} (Q^2)&=&\frac{1}{4} \int_0^{\infty} dt^\prime\, t^{\prime\, 4} \frac{j_1(Qt^\prime)}{Qt} C_{1^-}(t^\prime)\, , \label{dpf2b}\\
\chi_{0^{-}} (Q^2)&=&\int_0^{\infty} dt^\prime\, t^{\prime\, 2} j_0(Qt^\prime) C_{0^-}(t^\prime)\, ,  \nonumber\\
\chi_{1^{+}} (Q^2)&=&\frac{1}{4} \int_0^{\infty} dt^\prime\, t^{\prime\,  4} \frac{j_1(Qt^\prime)}{Qt^\prime} C_{1^+}(t^\prime)\, .  \nonumber
\end{eqnarray}
In this work, in view of a comparison with the results obtained by using the perturbative calculation of the susceptibilities, we will take $Q^2=0$.  
In this case $j_0(0)=1$ and $\lim_{x\rightarrow0}j_1(x)/x=1/3$, so that the derivatives of the longitudinal and transverse polarization functions are equal to the second and the fourth moments of the longitudinal and transverse Euclidean correlators, respectively.
We point out again that, by using the two-point correlation functions determined non-perturbatively, we may constrain the form factors also at $Q^2 \neq 0$.

\subsection{Ward Identities}
\label{subsec:WI}

Some relations,  which  will be particularly useful in the analysis of the numerical results, can be derived using the Ward Identities (WIs) that the vector and axial  vector quark currents satisfy
 \begin{eqnarray}
\partial_{\mu} \bar{b}(x)\gamma_{\mu}c(x)&=&(m_b-m_c)\bar{b}(x)c(x)\, , \nonumber  \\
\partial_{\mu} \bar{b}(x)\gamma_{\mu}\gamma_5c(x)&=&(m_b+m_c)\bar{b}(x)\gamma_5c(x)\, , 
\end{eqnarray}
where $m_b$ and $m_c$ are the (bare)  masses  of  the bottom and charm quarks respectively. 
Hence, by defining two further (Euclidean) polarization functions connected to the scalar and the pseudoscalar currents, namely
 \begin{eqnarray}
\Pi_S(Q^2)&\equiv&  \int d^4x\,  e^{-iQ\cdot x} \bra{0} T\{\bar{b}( x)c(x) \bar{c}(0)b(0)\} \ket{0}\, , \nonumber \\
\Pi_P(Q^2)&\equiv&  \int d^4x\, e^{-iQ\cdot x} \bra{0} T\{\bar{b}(x)\gamma_5c(x) \bar{c}(0)\gamma_5b(0)\} \ket{0}\, , 
\end{eqnarray}
the WIs imply that 
\begin{eqnarray}
Q_{\mu}Q_{\nu}{\Pi}^V_{\mu\nu}(Q)&=&(m_b-m_c)^2 ~ \Pi_S(Q^2)\, , \nonumber \\
Q_{\mu}Q_{\nu}{\Pi}^A_{\mu\nu}(Q)&=&(m_b+m_c)^2 ~ \Pi_P(Q^2)\, , 
\end{eqnarray}
from which we obtain
\begin{eqnarray}
-Q^4\Pi_{0^+}(Q^2)&=&(m_b-m_c)^2\Pi_S(Q^2)\, ,  \nonumber \\
-Q^4\Pi_{0^-}(Q^2)&=&(m_b+m_c)^2\Pi_P(Q^2)\, . 
\end{eqnarray}
Moreover, by performing a double derivative with respect to $Q^2$ we get
\begin{eqnarray}
\left( -2\frac{\partial}{\partial Q^2}-Q^2 \frac{\partial^2}{\partial^2 Q^2} \right) [Q^2 \Pi_{0^+}(Q^2)] &=& (m_b-m_c)^2\frac{\partial^2}{\partial^2 Q^2} \Pi_S(Q^2) \, , \nonumber \\
\left( -2\frac{\partial}{\partial Q^2}-Q^2 \frac{\partial^2}{\partial^2 Q^2}  \right) [Q^2 \Pi_{0^-}(Q^2)] &=& (m_b+m_c)^2\frac{\partial^2}{\partial^2 Q^2} \Pi_P(Q^2) \, . 
\end{eqnarray}

At this point, we can define the scalar and pseudoscalar analogues of Eqs.\,(\ref{dpf2}):
\begin{eqnarray}
\chi_S (Q^2)&=&-\frac{1}{2}\left(\frac{\partial^2}{\partial^2 Q^2}\right) \Pi_{S} (Q^2) = \frac{1}{4} \int_0^{\infty} dt^\prime\, t^{\prime  \, 4} ~ \frac{j_1(Qt^\prime)}{Qt^\prime} C_S(t^\prime), \nonumber \\
\chi_P (Q^2)&=&-\frac{1}{2}\left(\frac{\partial^2}{\partial^2 Q^2}\right) \Pi_{P} (Q^2) = \frac{1}{4} \int_0^{\infty} dt^\prime\, t^{\prime \, 4} ~ \frac{j_1(Qt^\prime)}{Qt^\prime} C_P(t^\prime),
\end{eqnarray}
where the scalar and pseudoscalar Euclidean correlators are defined as
\begin{eqnarray}
C_S(t)&=&\int d^3\vec x \, \bra{0} T\{ \bar{b}(t,\vec x) c(t,\vec x) \bar{c}(0)  b(0)\} \ket{0},  \nonumber \\
C_P(t)&=&\int d^3\vec x \, \bra{0} T\{ \bar{b}(t,\vec x)\gamma_5 c(t,\vec x) \bar{c}(0)\gamma_5   b(0)\} \ket{0}\, . 
\end{eqnarray}

The WIs offer thus the possibility to express the derivatives of vector longitudinal and axial longitudinal polarization functions in a different way, namely
\begin{eqnarray}
\label{eq:longix} 
\chi_{0^{+}} (Q^2)&=&(m_b-m_c)^2\chi_S (Q^2)-\frac{1}{2} Q^2\frac{\partial}{\partial Q^2} \chi_{0^{+}} (Q^2)\nonumber \\
&=& \frac{1}{4} \int_0^{\infty} dt^\prime\, t^{\prime \, 4} ~ \frac{j_1(Qt^\prime)}{Qt^\prime} \left[ (m_b-m_c)^2 C_S(t^\prime) + Q^2 C_{0^+}(t^\prime) \right] \\ 
\chi_{0^{-}} (Q^2)&=&(m_b+m_c)^2\chi_P (Q^2)-\frac{1}{2} Q^2\frac{\partial}{\partial Q^2} \chi_{0^{-}} (Q^2) \nonumber \\
&=& \frac{1}{4} \int_0^{\infty} dt^\prime\, t^{\prime \, 4}~  \frac{j_1(Qt^\prime)}{Qt^\prime} \left[ (m_b+m_c)^2 C_P(t^\prime) + Q^2 C_{0^-}(t^\prime) \right] \, .  \nonumber 
\end{eqnarray}
Thus, when  setting $Q^2 = 0$, we can compute the derivatives of the longitudinal vector and axial-vector polarization functions directly through the fourth moments of the scalar and pseudoscalar correlators, respectively.
The advantage of the above procedure based on the WIs will be clarified in Section~\ref{sec:susceptibilities}.

Eqs.\,(\ref{dpf2b}) and (\ref{eq:longix}) represent the basic formulae for the evaluation of the dispersive bounds in terms of Euclidean correlators (calculable on the lattice) for space-like values of $q^2 = -Q^2$, i.e.~$q^2 \leq 0$.
Time-like values of $q^2$, which are relevant for the semileptonic form factors, corresponds to $Q^2 < 0$.
This requires the use of imaginary values for the Euclidean four-momentum $Q$, i.e.~$Q = (iq, \vec{0})$.  
Thus, the formulae from which we can extract the derivatives of the polarization functions on the lattice for $0 \leq q^2 \lesssim (m_b - m_c)^2$ are
\bea
     \chi_{0^+}(q^2) & = & \frac{1}{12} \int_0^\infty dt^\prime ~ t^{\prime\, 4} H_1(qt^\prime) ~ \left\{ (m_b - m_c)^2 C_S(t^\prime) - q^2 C_{0^+}(t^\prime) \right\} ~ ,
      \nonumber \\
        \chi_{1^-}(q^2) & = & \frac{1}{12} \int_0^\infty dt^\prime ~ t^{\prime\, 4} H_1(qt^\prime) ~ C_{1^-}(t^\prime) ~ , \\   \label{eq:finalEucl}
      \chi_{0^-}(q^2) & = & \frac{1}{12} \int_0^\infty dt^\prime ~ t^{\prime\, 4} H_1(qt^\prime) ~ \left\{ (m_b + m_c)^2 C_P(t^\prime) - q^2 C_{0^-}(t^\prime) \right\} ~ ,
      \nonumber \\
      \chi_{1^+}(q^2) & = & \frac{1}{12} \int_0^\infty dt ~ t^{\prime\, 4} H_1(qt^\prime) ~ C_{1^+}(t^\prime) ~ , ~ \nonumber
\eea
where 
\be
   H_1(x) \equiv \frac{3}{x^2} \left[ \mbox{cosh}(x) - \frac{\mbox{sinh}(x)}{x}\right ] 
    \label{eq:H1x}
\ee
with $H_1(0) = 1$.
For $q \lesssim m_b - m_c$ all the  above integrals over the Euclidean time $t^\prime$ are finite, because at large time distances the Euclidean correlators behave as $\mbox{exp}[-(m_b + m_c) t^\prime]$, while $H_1(qt^\prime) \propto \mbox{exp}(qt^\prime)$. 

In general, on the lattice, the WIs are modified by discretisation terms~\cite{Bochicchio:1985xa} that vanish in the continuum limit, namely when the lattice spacing $a \to 0$. 
For example, with improved fermion actions like Wilson improved~\cite{Sheikholeslami:1985ij,Luscher:1996sc}, Maximal Twisted~\cite{Frezzotti:2000nk,Frezzotti:2003xj,Frezzotti:2003ni} and Domain Wall Fermions~\cite{Kaplan:1992bt},  we have 
\begin{eqnarray}
Z_V \partial_{\mu} \bar{b}(x)\gamma_{\mu}c(x)&=&(m_b-m_c)\bar{b}(x)c(x)+O(a^2)\, , \nonumber \\
Z_A \partial_{\mu} \bar{b}(x)\gamma_{\mu}\gamma_5c(x)&=&(m_b+m_c)\bar{b}(x)\gamma_5c(x)+O(a^2)\, , 
\end{eqnarray}
where $Z_V$ and $Z_A$ are appropriate renormalization constants for the lattice vector and axial-vector currents~\cite{Bochicchio:1985xa}.

Further discretisation effects may enter in the T-products computed on the lattice, that we use in our analysis. We propose to reduce discretisation errors by using  a combination of non-perturbative and perturbative subtractions which were found very effective in the past (see  later Section~\ref{sec:susceptibilities}).

\section{Statistical and systematic errors with dispersive bounds in the presence of kinematical constraints}
\label{sec:errors}

In this Section we discuss the treatment of the statistical errors and of the systematic effects when the bounds on the values of the form factor $f(t)$ are derived using the values $f(t_i)$ computed at the points $t_i$ with $i = 1, 2, \dots, n$, and the appropriate susceptibility $\chi$ discussed in the previous Sections.  Besides the statistical errors which affect all the quantities computed in numerical simulations of lattice QCD, we have also systematic effects. In particular discretisation errors, which may become rather severe when studying heavy quark transitions, can modify in an important way the continuum unitarity relations and introduce inconsistencies in the equations introduced in Section~\ref{sec:dispbound}.

Without any loss of generality we consider in this Section the case of the semileptonic scalar $f_0(t)$ and vector $f_+(t)$ form factors, which have to fulfill the kinematical constraint $f_0(0) = f_+(0)$ at zero four-momentum transfer.

Although the treatment of the statistical errors in the presence of kinematical constraints was already discussed in Ref.~\cite{Lellouch:1995yv}, in this work we introduce a different method which looks to us simpler to implement. This method, together with the skeptical Bayesian approach of Refs.~\cite{DAgostini:2020vsk,DAgostini:2020pim}, allows us also to treat the effects induced by discretisation and other systematic effects.

\subsection{Generation of the bootstrap events}
\label{subsec:boot}

The machinery  described in Section~\ref{sec:dispbound} allows us to compute the lower/upper bounds of $f_{0(+)}(t)$, once we have chosen our set of input data, i.e.~$\{\chi_{0^+(1^-)}, f_{0(+)}(t_1),\cdots,  f_{0(+)}(t_n)\}$. 
Thus, the input data set is made of $2n+2$ quantities: the $n$ values of the scalar form factor $f_0$, the $n$ values of the vector form factor $f_+$ and the two susceptibilities $\chi_{0^+}$ and  $\chi_{1^-}$. For sake of simplicity we are considering the same number of data points for both the scalar and the vector form factors evaluated at the same series of values $t_i$ ($i = 1, \cdots, n$).

The crucial question is, however, how to propagate the uncertainties related to these quantities into the evaluation of the  FFs $f_{0(+)}(t)$ at a generic value of $t$. 
To answer this question, we propose a method different from the one described in Ref.~\cite{Lellouch:1995yv}. 
We start by building up a multivariate Gaussian distribution with mean values and covariance matrix given respectively by $\{ f_{0}(t_1),\cdots,f_{0}(t_n), f_{+}(t_1),\cdots,f_{+}(t_n)\}$ and $\Sigma_{ij} = \rho_{ij} \sigma_i \sigma_j$, where $f_{0(+)}(t_i)$ are the form factors extracted from the three-point functions in our numerical simulation on a given set of gauge field configurations, $\sigma_i$ are the corresponding uncertainties, and $\rho_{ij}$ is their correlation matrix (including also correlations between the two form factors).

There are two possible analyses that can be applied to the lattice results:
\begin{enumerate}
\item[A)] When we have direct access to the data of the simulations, as it is the case of the $D\to K $ transitions discussed in Section\,\ref{sec:proto}, we can generate by ourself jackknife or bootstrap sets of all the quantities defined above. At the same time we use the non-perturbative susceptibilities evaluated on the same jackknife/bootstrap sets used to compute the values of the form factors $f_{0(+)}(t_i)$;
 \item[B)] When we use data produced by other groups that  provide  their values of $f_{0(+)}(t_i)$, $ \sigma_i$ and $\rho_{ij}$, we generate $N_{boot}$ bootstrap events according to the expected probability distributions. In this case we generate $N_{boot}$ values of the non-perturbative susceptibilities $\chi_{0^+(1^-)}$ through normal distributions defined by their mean values and standard deviations. 
\end{enumerate}

The first option is certainly to be preferred to reduce the statistical noise by taking properly into account all the correlations of the data.

For each jackknife/bootstrap event we consider the $(n+1) \times (n+1)$ matrices $\mathbf{M}^0 $ and $\mathbf{M}^+$ (see Eq.~(\ref{Der})) corresponding to the scalar and vector form factors, respectively.
Since $\Delta_1^0 = \Delta_1^+$ is non-negative by definition, the positivity condition\,(\ref{delta}) implies that both $\Delta_2^0$ and  $\Delta_2^+$ should be positive. Thus, we compute $\Delta_2^{0(+)}$ and verify their signs. If either $\Delta_2^0$ or  $\Delta_2^+$ results to be negative, then the event is eliminated from the sample. From the physical point of view, this step can be read as a consistency check between all the input data, namely the susceptibilities and the FFs for that particular bootstrap. At the end of the procedure, we will be left with $\widetilde{N}_{boot} \le  {N}_{boot}$ events.

\subsection{Implementation of the constraints}
\label{subsec:imp}

At $t=0$ the FFs $f_0$ and $f_+$ are subject to the constraint 
\begin{equation}
\label{KC}
f_0(0) = f_+(0).
\end{equation}
In order to satisfy this condition, in the subset of the $\widetilde{N}_{boot}$ events satisfying the unitarity filters $\Delta_2^{0(+)} \geq 0$, we select only the $N_{boot}^* \leq \widetilde{N}_{boot}$ events for which the dispersive bands for $f_0$ and $f_+$ overlap each other at $t=0$.  This corresponds to impose the conditions
\begin{eqnarray}
f_{0,up}(0, q^2) & > & f_{+,lo}(0, q^2) ~ , ~ \nonumber \\
f_{+,up}(0, q^2) & > & f_{0,lo}(0, q^2) ~ ,  ~ 
\label{eq:disegua}
\end{eqnarray}
where $f_{lo,up}(t, q^2)$ were defined in Eq.\,(\ref{loup2}) for a generic form factor $f$.
Omitting for simplicity the argument $q^2$ at which the susceptibilities $\chi_{0(+)}$ are calculated, the conditions (\ref{eq:disegua})  can then be rephrased as 
\begin{equation}\label{eq: KC_MM}
\big| \phi_+(z(0)) \, \beta_+(0) - \phi_0(z(0)) \beta_0(0) \big| \leq \sqrt{\Delta_1(0)}\bigg[ \phi_+(z(0)) \, \sqrt{\Delta_2^+} + 
\phi_0(z(0)) \, \sqrt{\Delta_2^0} \bigg] \,  .
\end{equation}

As already said, the above condition select $N_{boot}^* \leq \widetilde{N}_{boot}$ events.
Following Ref.~\cite{Lellouch:1995yv} for each of the $N_{boot}^*$ events we define
\begin{eqnarray}
\label{philo}
f_{lo}^*(0) &=& \max[f_{+,lo}(0),f_{0,lo}(0)]\, ,\nonumber \\
\label{phiup}
f_{up}^*(0)&=& \min[f_{+,up}(0),f_{0,up}(0)]\, , 
\end{eqnarray}
so that, putting $f(0) \equiv f_0(0)=f_+(0)$, one has
\begin{equation}
\label{eq:KCstar}
f_{lo}^*(0) \leq f(0) ~ \leq f_{up}^*(0) ~ . ~
\end{equation}
We now consider the form factor $f(0)$ to be uniformly distributed in the range given by Eq.~(\ref{eq:KCstar}) and we add it to the input data set as a new point at $t_{n+1} = 0$.
To be more precise, for each of $N_{boot}^*$ events we generate $N_0$ values of $f(0)$ with uniform distribution in the range $[f_{lo}^*(0), f_{up}^*(0)]$, obtaining a new sample having $\overline{N}_{boot} = N_{boot}^* \times N_0$ events, each of them satisfying by construction both the unitarity filters $\Delta_2^{0(+)} \geq 0$ and the kinematical constraint~(\ref{KC}).

We then consider two modified $(n+2) \times (n+2)$ matrices, $\mathbf{M}_C^0$ and $\mathbf{M}_C^+$, that have one more row and one more column with respect to matrices $\mathbf{M}^0$ and $\mathbf{M}^+$ and contain the common form factor $f(t_{n+1} = 0)$, namely matrices of the form
 \begin{equation} 
\label{DerKC}
\mathbf{M}_C = \left (
\begin{array}{cccccc}
\langle \phi f | \phi f \rangle  & \langle \phi f | g_t \rangle  & \langle \phi f | g_{t_1} \rangle  &\cdots & \langle \phi f | g_{t_n}\rangle  & \langle \phi f | g_{t_{n+1}}\rangle  \\
\langle g_t | \phi f \rangle  & \langle g_t |  g_t \rangle  & \langle g_t | g_{t_1} \rangle  &\cdots & \langle g_t | g_{t_n}\rangle  & \langle g_t  | g_{t_{n+1}}\rangle  \\
\langle g_{t_1} | \phi f \rangle  & \langle g_{t_1} | g_t \rangle  & \langle g_{t_1} | g_{t_1} \rangle  &\cdots & \langle g_{t_1} | g_{t_n}\rangle  & \langle g_{t_1}  | g_{t_{n+1}}\rangle   \\
\vdots & \vdots & \vdots & \vdots & \vdots & \vdots \\
\langle g_{t_{n}} | \phi f \rangle  & \langle g_{t_{n}} | g_t \rangle  & \langle g_{t_{n}} | g_{t_1} \rangle  &\cdots & \langle g_{t_{n}} | g_{t_n} \rangle  & \langle g_{t_{n}}  | g_{t_{n+1}}\rangle   \\
\langle g_{t_{n+1}} | \phi f \rangle  & \langle g_{t_{n+1}} | g_t \rangle  & \langle g_{t_{n+1}} | g_{t_1} \rangle  &\cdots & \langle g_{t_{n+1}} | g_{t_n} \rangle  & \langle g_{t_{n+1}}  | g_{t_{n+1}}\rangle \\
\end{array}
\right ) ~ . ~
\end{equation}

For any point $t$ at which  we want to predict the allowed dispersive band of the form factor $f(t)$ (which can be either $f_0(t)$ or $f_+(t)$) without  directly computing it in our simulation, we compute the matrix $\mathbf{M}_C$ and using Eq.\,(\ref{loup2}) we get $f_{lo}(t)$ and  $f_{up}(t)$. 
This can be done for each of the $N_0$ events. 
Let us indicate the result of the $k$-th extraction by $f_{lo}^k(t)$ and $f_{up}^k(t)$, respectively.
Then, for each of the $N^*_{boot}$ events the lower and upper bounds $\overline{f}_{lo}(t)$ and $\overline{f}_{up}(t)$ can be defined as  
\begin{eqnarray}
\overline{f}_{lo}(t) & = & \min[f^1_{lo}(t),f^2_{lo}(t),\dots,f^{N_{0}}_{lo}(t)]\, , \nonumber \\
\overline{f}_{up}(t) & = & \max[f^1_{up}(t),f^2_{up}(t),\dots,f^{N_{0}}_{up}(t)] ~ . ~
\end{eqnarray}

\begin{figure}[htb!]
\centering
 \includegraphics[scale=0.50]{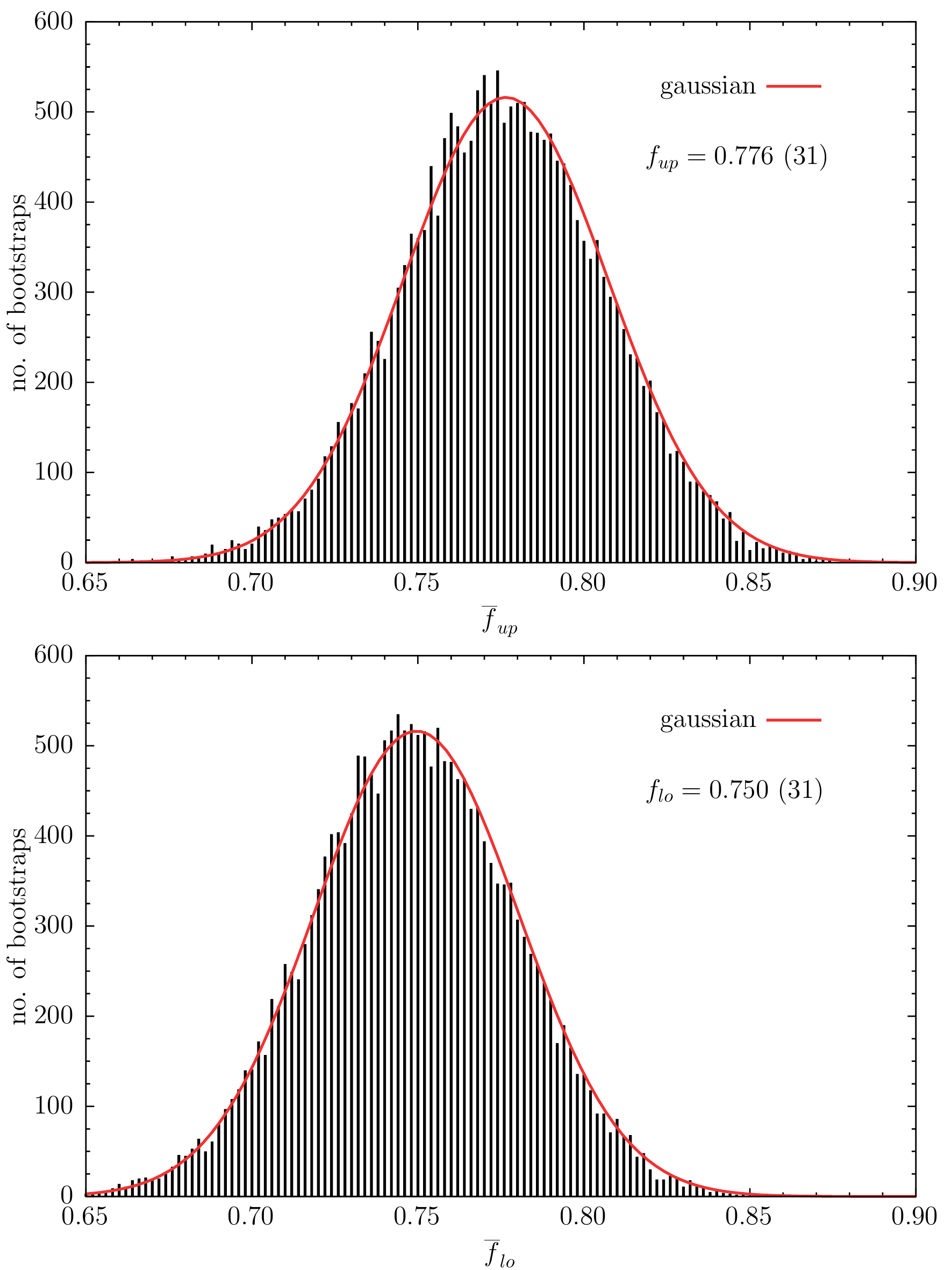}
 \centering
\caption{\textit{Histograms of the values of $\overline{f}_{up}$ (upper panel) and $\overline{f}_{lo}$ (lower panel) for the bootstrap events that pass the unitarity filter in the case of the vector form factor $f_+(t = 0~{\rm GeV}^2)$ of the $D \to K$ transition.}\hspace*{\fill} \small}
\label{Histo_Fit_B}
\end{figure}

At this point we can generate the bounds of the form factor $f(t)$. 
To achieve this goal, we combine all the $N_{boot}^*$ results $\overline{f}_{lo,up}^i(t)$ ($i = 1, \cdots, N_{boot}^*$) to generate the corresponding histograms and fit them with a Gaussian Ansatz, as it is shown in Fig.\,\ref{Histo_Fit_B} in an illustrative case.
From these fits we extract the average values $f_{lo(up)}(t)$, the standard deviations $\sigma_{lo(up)}(t)$ and the corresponding correlation factor $\rho_{lo,up}(t) = \rho_{up,lo}(t)$, namely
\begin{eqnarray}
    \label{eq:avgs}
    f_{lo(up)}(t) & = & \frac{1}{N_{boot}^*} \sum_{i=1}^{N_{boot}^*} \overline{f}_{lo(up)}^i ~ , ~ \\[2mm]
    \label{eq:sigmas}
    \sigma_{lo(up)}^2(t) & = & \frac{1}{N_{boot}^* - 1}\sum_{i=1}^{N_{boot}^*} ~ \left[ \overline{f}_{lo(up)}^i(t) - f_{lo(up)}(t) \right]^2 ~ , ~ \\[2mm]
    \label{eq:rhos}
    \rho_{lo,up}(t) = \rho_{up,lo}(t) & = & \frac{1}{N_{boot}^* - 1}\sum_{i,j=1}^{N_{boot}^*} ~ \left[ \overline{f}_{lo}^i(t)  - f_{lo}(t) \right] ~ 
                                                              \left[ \overline{f}_{up}^j(t)- f_{up}(t) \right] ~ . ~
\end{eqnarray}

It is understood that the above procedure is applied for both the scalar $f_0(t)$ and the vector $f_+(t)$ form factors.

An important aspect of the full procedure is a good determination of the correlations among the matrix elements in Eq.~(\ref{DerKC}), namely among the corresponding form factors. This is automatically achieved by generating the jackknife/bootstrap events with the procedure A) described in subsection \ref{subsec:boot}, whereas an accurate determination of the correlation matrix of the form factors  must be provided when one wants to use data produced by other groups.

\subsection{Combination of the lower and upper bounds for each FF}
\label{subsec:comb}

After the steps described before, for any choice for $t$ we obtain from the bootstrap events (pseudogaussian) distributions for $f_{0,lo}(t)$,  $f_{0,up}(t)$,  $f_{+,lo}(t)$ and $f_{+,up}(t)$  as well as the corresponding mean values, standard deviations and correlations. 
We combine them according to the following procedure.

Let us consider a single bootstrap event in which $f_L$ is  the lower bound and $f_U$ is the upper one for a generic FF at the given value of $t$ (for sake of simplicity we omit the $t$-dependence for a while). 
We associate to the FF $f$ a flat distribution between $f_L$ and $f_U$
\begin{equation}
\label{PDinitial}
P(f)=\frac{1}{f_{U}-f_{L}} \Theta(f-f_L)  \Theta(f_{U}-f) ~ , ~ 
\end{equation} 
where $f = f_{0(+)}$,  $f_U = f_{ 0(+), up}$,  $f_L = f_{ 0(+), lo}$ and $\Theta$ is the Heaviside step function. 
The mean value and the variance associated to the distribution (\ref{PDinitial}) are respectively given by
\begin{eqnarray}
\label{meaninit}
&& \frac{1}{f_{U}-f_{L}} \int_{f_{L}}^{f_{U}}df ~ f \, \Theta(f-f_L) \, \Theta(f_U-f) = \frac{f_U+f_L}{2} ~ , ~ \\[2mm]
\label{sigmainit}
&& \frac{1}{f_{U}-f_{L}} \int_{f_{L}}^{f_{U}} df ~ \left(f - \frac{f_U+f_L}{2} \right)^2 \, \Theta(f-f_L) \, \Theta(f_U-f) = \frac{(f_U - f_L)^2}{12} ~ . ~
\end{eqnarray}

It is however necessary to average over the whole set of bootstrap events. Since the lower and the upper bounds of a generic FF are strongly correlated, we  adopt a multivariate Gaussian distribution to describe them, $i.e.$
\begin{equation}
\label{PDloeup}
P_{LU}(f_L,f_U) = \frac{\sqrt{\det C^{-1}}}{2 \pi} ~ e^{-\frac{1}{2} \left[ C_{LL}^{-1} (f_L - f_{lo} )^2 + 2C_{LU} ^{-1}(f_U - f_{up}) (f_L - f_{lo}) + C_{UU}^{-1} (f_U - f_{up})^2 \right]} ~ , ~ 
\end{equation}
where $f_{lo(up)}$ represents the mean of the lower(upper) bound over all the bootstrap events, given by Eq.~(\ref{eq:avgs}), and $C$ is the covariance matrix
\be
C = \left( 
\begin{tabular}{cc}
$\sigma_{lo}^2$ & $\rho_{lo,up} ~ \sigma_{lo} \sigma_{up}$ \\[2mm] 
$\rho_{up,lo} ~ \sigma_{lo} \sigma_{up}$ & $\sigma_{up}^2$ \\[2mm]
\end{tabular}
\right) ~
\ee
with $\sigma_{lo(up)}$ and $\rho_{lo,up} = \rho_{up,lo}$ being given by Eqs.~(\ref{eq:sigmas}) and~(\ref{eq:rhos}), respectively.
In Eq.~(\ref{PDloeup}) the normalization has been chosen so that  
\begin{equation}
 \int_{-\infty}^{+\infty}\int_{-\infty}^{+\infty}df_U\,df_L\,  P_{LU}(f_L,f_U)  = 1 ~ .
\end{equation}
Using the product of the distributions (\ref{PDinitial}) and (\ref{PDloeup}) we can compute the final values of the form factor $f(t)$ and its variance $\sigma_f^2(t)$ as
\begin{eqnarray}
f(t) & = & \frac{f_{lo}(t) + f_{up}(t)}{2} ~ , ~ \\
\sigma_f^2(t) & = & \frac{1}{12} \left[ f_{up}(t) - f_{lo}(t) \right]^2 + \frac{1}{3} \left[ \sigma_{lo}^2(t) + \sigma_{up}^2(t) + \rho_{lo,up}(t) \, \sigma_{lo}(t) \, \sigma_{up}(t) \right] ~ . ~
\end{eqnarray}

\subsection{Systematic effects and the skeptical approach}
\label{subsec:skept}

In various analyses of the lattice data that we performed to test our method we have encountered the following phenomenon. For some of the bootstrap events no solution could be found, either because $\Delta_2<0$ or because there is no overlap between the allowed regions for $f_0(0)$ and $f_+(0)$.  This may obviously happen for a statistical fluctuation of the sample at hand. When, however, the fraction of rejected bootstrap events is large, say much larger than $50\%$, one may argue that systematic effects come into play (e.g.~lattice artefacts), that have not been corrected for, which may jeopardise the unitary relations. In cases like these, however, it is still possible to extract the relevant information, which in our case is the allowed interval for the form factors, by using different statistical  approaches like the {\it skeptical} one discussed in Refs.~\cite{DAgostini:2020vsk,DAgostini:2020pim}.  

We stress that the above issue is not important in the case of the $D \to K$ semileptonic decays for the analyses of the lattice data either extrapolated to the physical point and to the continuum limit (see Section~\ref{sec:continuum}) or obtained at finite lattice spacing and unphysical pion masses (see Section~\ref{sec:ensembles}).  In these  cases  we see that practically all  the bootstraps generated for the FFs survive after both the unitarity and the kinematical constraints. We anticipate that the skeptical approach, instead, turns out to be useful in the study of the $b \to c$ transitions, where we expect much larger discretisation effects which may manifest as apparent violation of the unitarity constraints. Focusing our attention for instance onto the $B \to D$ case, the skeptical approach is useful when the matrix procedure explained in this work is applied to the FFs  presented by MILC~\cite{Lattice:2015rga}.
Indeed only the $\sim 15\%$ of the initially generated events survive to the simultaneous application of the unitarity and  kinematical constraints. On the contrary, the introduction of the skeptical method described below allows the largest part of the generated bootstraps events to pass both filters. This fact will be properly illustrated in a forthcoming paper about the application of the matrix method to the $b \to c$ transitions.

In the reminder of this Section we explain the general idea of the skeptical approach.  

Probability theory helps us in building up a model in which the values and uncertainties of the physical quantities about which we are in doubt are allowed to vary from the nominal ones. Obviously, the model is not unique, as not unique are the probability distributions that can be used. The simplest choice consists in enlarging the reported standard deviations $\sigma_{i}$ of the measured points, by assuming the {\it true} standard deviations, $\sigma_{i}^t$ are related to the $\sigma_{i}$ by a factor $r_i$, one for each of the measured points, $\sigma_{i}^t = r_i \, \sigma_{i}$, whereas the average values $\bar f_i$ is the same.  All the points $i$  are treated democratically and fairly, i.e. our prior belief of each  $r_i$ has expected value equal to one; its prior distribution does not depend on  the point; we are skeptical, and hence each $r_i$ has {\it a priori}  a wide range of possibilities described by a probability distribution, with a prior $100\%$ standard
uncertainty on $r_i$, i.e.~$\sigma(r_i)/E[r_i] = 1$.

The simplest model is to introduce a Gamma probability distribution for the variable $r \geq 0$ 
\begin{equation}
\label{eq:skeptical}
P_{skept}(r) = \frac{1}{B ~ \Gamma(A)} \left( \frac{r}{B} \right)^{A-1}  e^{-r/B} ~ . ~ 
\end{equation}
The parameters $A$ and $B$ are fixed by imposing that this distribution has both mean value and variance equal to $1$. A simple calculation shows that this request corresponds to the choice $A = B = 1$, i.e.~$P_{skept}(r) = e^{-r}$. At this point, we slightly modify the procedure described in Sections~\ref{subsec:imp}--\ref{subsec:comb}, namely we build up a multivariate Gaussian distribution whose covariance matrix now is
\begin{equation}
\label{eq:cov_r}
 \Sigma_{ij} = \rho_{ij} \sigma_{i} \sigma_{j} \times r^2,
\end{equation}
where the $\sigma$'s are the uncertainties of the measured points in the numerical simulation and $\rho_{ij}$ the corresponding  correlation matrix. 

Thus, to summarize, according to Eq.~(\ref{eq:skeptical}) we extract $N_r$ values of $r$ and we produce $N_{boot}$ bootstrap events for the FFs using the covariance matrix~(\ref{eq:cov_r}). Thus, in the skeptical case we have a sample of $N_r \times N_{boot}$ events. At this point, we proceed with the unitarity filter related to the sign of $\Delta_2$ and with the implementation of the constrained matrix $\mathbf{M_{C}}$ as explained in the previous Sections.

\section{Vacuum polarization functions in perturbation theory and lattice artefacts}
\label{sec:pertlatt}

In this Section we discuss the perturbative calculation of the susceptibilities in the continuum and on the lattice, and the subtraction/reduction of lattice artefacts that can be obtained by using the perturbative calculation of the polarization functions. Although in what follows we will consider lattice QCD in the Twisted Mass Fermion (TMF) regularisation, the main arguments of our discussion are general and can be applied to any lattice regularisation of the theory. One peculiarity, which is however common to other regularisations, is the on-shell $O(a)$ improvement of the physical particle spectrum and of the matrix  elements of local bilinear operators. Thus the lattice artefacts for physical quantities related to these matrix elements are of $O(a^2)$. 
 
In lattice simulations performed at finite lattice spacing one can attempt to obtain the physical results either by extrapolating the lattice quantities to the continuum or by reducing the discretisation effects by a subtraction procedure based on perturbation theory.  A combination of the two strategies is indeed the most effective one. Some of the lattice artefacts can also be eliminated non perturbatively using, for example, the WIs of the theory~\cite{Bochicchio:1985xa}.
 In this Section we shall deal with the perturbative approach which can be implemented in one-loop (or higher-loops) order by computing for  a given quantity, say the polarization function or its derivatives, the corresponding Feynman diagrams, at finite lattice spacing. In the analysis of lattice data we will also discuss the extrapolation of the results to the continuum theory which can be combined with the perturbative subtraction of lattice artefacts discussed in this Section.

\begin{figure}[htb!]
\begin{center}
\includegraphics[width=0.50\textwidth]{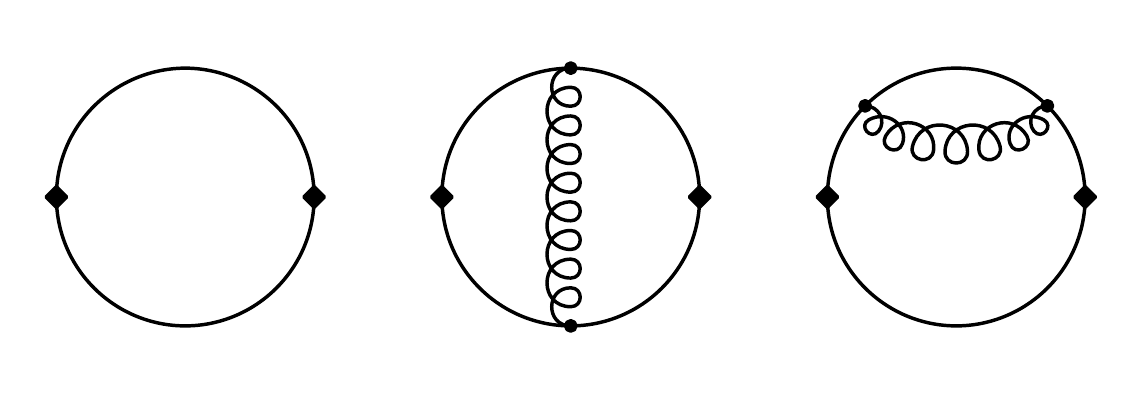}
\caption{{\it One- and two-loop Feynman diagrams for the polarization function. The crosses represent the bilinear operators and the curly line the gluon propagator.} \hspace*{\fill}}
\label{fig:2loops}
\end{center}
\end{figure}

Recalling the definitions of $\Pi_{V,A}^{\mu\nu}$ given in Section~\ref{sec:lattice}, we analyse the structure of a twisted fermions loop, the graphical representation of which, at lowest order, corresponds to the first Feynman diagram on the left in  Fig.\,\ref{fig:2loops}. 

Given the on-shell $O(a)$ improvement of the vector current correlators at physical distances we focus on the impact of contributions to the Fourier  sum from small and zero distance. Formally, we are interested in the  expansion of the generic polarization function $\Pi(Q^2)\equiv \Pi_{0^\pm,1^\mp}(Q^2)$
\beq \Pi(Q^2) =\sum_{k\ge -6} C_k a^k \, \eeq
where one can show that $C_1 = 0$ with  maximally TMF. 

In the following we discuss as an example the case of the vector current. In this case  the coefficients of the expansion can be derived from the generic form of the lattice polarization tensor using the symmetries of the lattice action~\cite{Burger:2014ada}
\bea \Pi_V^{\mu\nu}(Q^2) &=& \Pi_{V\,{\rm con}}^{\mu\nu}(Q^2) +\left(\frac{Z_1}{a^2} +\tilde Z_1 \langle -S_6+\frac{1}{2}S_5^2\rangle_0 \right)\, \delta^{\mu\nu}+
Z_{\mu^2}^+ \, \left(\mu_1^2+\mu_2^2\right)\, \delta^{\mu\nu} + Z_{\mu^2}^- \, \left(\mu_1^2-\mu_2^2\right)\, \delta^{\mu\nu}\nonumber \\ &&+ \left( Z_{Q^2}Q^2 \, \delta^{\mu\nu}+Z_{QQ}\, Q^\mu Q^\nu \right) + O(a^2,\,\,\, {\rm operators} \,\,\, {\rm of}\,\,\, {\rm  higher }\,\,\, {\rm dimension}) \, , \label{eq:expPI}\eea
where $\Pi_{V\,{\rm con}}(Q^2)$ is the continuum polarization tensor, $S_5$ and $S_6$ are defined by the expansion of the lattice action close to the continuum limit 
\beq S_{eff }= S_4 + a\, S_5 + a^2\, S_6 + a^2 \, S_7 +\dots \, , \eeq 
where $S_k=\int d^4x\, {\cal L}_k$ with the terms ${\cal L}_k$ containing linear combinations of  fields with mass dimension $k$, and $\langle \dots \rangle_0$ stands for vacuum expectation value of some combination of operators, either local, e.g.~$\langle S_6 \rangle_0$, or non local, $\langle S_5^2 \rangle_0$. We will indicate explicitly the dependence of  $\Pi^{\mu\nu}(Q^2)$ or of the $\chi$'s on the quark masses only when it will be necessary for the discussion of the results.

Luckily enough all the divergent or  mass dependent  lattice artefacts in the first line of Eq.~(\ref{eq:expPI}) disappear when we apply the derivative with respect to $Q^2$ to obtain the susceptibilities $\chi$'s, see Eqs.\,(\ref{veax21}). There are however terms of $O(a^0)$  which remain and that, in some cases,  can  even make the longitudinal polarization function different form zero even with degenerate quark masses $\mu_1=\mu_2$. Besides these terms there are discretisation errors of $O(a^2)$ or higher which remain.  The strategy to reduce their effect is that  widely used  in literature, see for example Ref.~\cite{Constantinou:2013ada}.  In our case it is  even simpler since the quantities that we consider, namely the $\chi$'s, are finite in perturbation theory. Let us call  $\chi_{\rm LAT}(Q^2,a)$ the  generic susceptibility computed non perturbatively on the lattice, $\chi(Q^2,a)$ the corresponding susceptibility computed in lattice perturbation theory, $\tilde \chi(Q^2)$ the expression resulting from $\chi(Q^2,a)$ by neglecting all contributions which vanish for $a \to 0$,  $\lim_{a \to 0}\chi(Q^2,a) \to \tilde \chi(Q^2)$, $\chi_{\rm con}(Q^2)$ the susceptibility computed  perturbatively in the continuum theory. We introduce the following quantities
\beq   \Delta_{1\chi}(Q^2, a) = \chi(Q^2, a)-\tilde \chi(Q^2)\, ,  \qquad  \Delta_{2\chi}(Q^2) = \tilde \chi(Q^2) - \chi_{\rm con}(Q^2)\, , \eeq
where $\Delta_{1\chi}(Q^2,a) $ represents the discretisation errors that we want to subtract, and $\Delta_{2\chi}(Q^2,a)$ the finite terms which are different in the continuum with respect to the lattice case. For local operators a typical example of these kind of corrections is represented by the current renormalisation constants, $J^\mu(x) = Z_J\,J_{L}^\mu(x) $.  
In order to extract the subtracted  susceptibility we construct then the combination
\beq \chi_s(Q^2, a) = \chi_{\rm LAT}(Q^2, a) - \Delta_{1\chi}(Q^2, a) - \Delta_{2\chi}(Q^2) =  \chi_{\rm LAT}(Q^2, a) - \chi(Q^2,a) + \chi_{\rm con}(Q^2,\mu)\, .  \label{eq:schemesub}\eeq
Thus, up to a certain order in perturbation theory and up to non perturbative effects,  $\chi_s(Q^2,a)$ reduces to the continuum result without discretisation errors.   
If one uses  the one-loop  perturbative calculations, the discretisation error then reduces to $O(\alpha_s a^2)$.

This procedure can easily  extended to higher orders in $\alpha_s$. In higher orders in perturbation theory, for the transverse susceptibilities, it will be also  necessary to renormalize the quark masses. For two-point correlation functions involving scalar or pseudoscalar densities, it will be also necessary to renormalise these bilinear operators in some chosen renormalisation scheme. We assume that we use the same renormalisation scheme, e.g.~$\overline{{\rm MS}}$ or the on-shell scheme for the quark masses. Strictly speaking then, we should write $\Delta_{2\chi}(Q^2,\mu) = \tilde \chi(Q^2,\mu) - \chi_{\rm con}(Q^2,\mu)$ to account for the scale(s) at which the strong coupling constant and, eventually, the quark masses are renormalised. In this respect we should also write $\chi_{\rm LAT}(Q^2,a,\mu)$  and $\tilde \chi(Q^2,\mu)$  since these quantities will contain the contribution of the counter terms necessary to renormalise the strong coupling constant or the quark masses.

After the subtraction procedure we expect then  to have discretisation errors of $O(\alpha_s^n a^2)$, where $n$ depends on the order at which we have computed perturbatively the current-current correlation functions, or non-perturbative discretisation errors of $O(a^2\Lambda_{QCD}^2)$.  Note that $\chi_s(Q^2, a)$ is the quantity to be used at finite lattice spacing for extracting the bounds on the form factors $f(t)$ according to the procedure explained in Section~\ref{sec:dispbound}. A detailed discussion of the counter terms up to two-loop perturbation theory will be given in a future publication where the subtraction procedure will be extended to $O(\alpha_s a^2)$.

\subsection{An instructive example: the vector current polarization tensor at one loop}
\label{subsec:instructive}

In order to  illustrate the procedure followed to reduce the discretisation errors, as an instructive example we discuss in details the one-loop perturbative calculation of the vector current polarization tensor and of the corresponding susceptibilities. 

At lowest order in perturbation theory, by calling $k$ the internal momentum, we may easily compute the correlator of two local vector currents on the lattice
\begin{equation}
\label{eq:VP}
\Pi_V^{\mu\nu}(Q,a)=\int_{-\pi/a}^{+\pi/a} \frac{d^4 k}{(2\pi)^4}\, {\rm Tr}\left[\gamma^{\mu} G_1(k + \frac{Q}{2})  \gamma^{\nu}G_2(k - \frac{Q}{2} )\right],
\end{equation}
where the integration interval represents the first Brillouin zone. Here $G_{i=1,2}$ indicates the tree-level Wilson twisted-mass propagator, namely
\begin{equation}
\label{propag}
G_i(p)=\frac{-i\gamma_{\mu}\mathring{p}_{\mu} + \mathcal{M}_i (p) -i \mu_{q,i} \gamma_5 \tau^3 }{\mathring{p}^2 + \mathcal{M}_i^2 (p)+\mu_{q,i}^2}, \,\,\,\,\,\,\,\, i=1,2
\end{equation}
where we have defined on the lattice
\begin{equation}
\label{latt}
\mathring{p}_{\mu} \equiv \frac{1}{a} \sin(ap_{\mu}), \,\,\,\,\,\,\,\,  \mathcal{M}_i (p) \equiv m_{i} + \frac{r_i}{2} a \hat{p}^2_{\mu}, \,\,\,\,\,\,\,\, \hat{p} \equiv \frac{2}{a} \sin(\frac{ap_{\mu}}{2}).
\end{equation}
In order to make the calculation it is convenient to define the dimensionless quantities 
\begin{equation}
\rho_\mu  \equiv p_\mu a,\,\,\,\,\, \tilde{m} \equiv m a,\,\,\,\,\,\tilde{\mu} \equiv \mu a\, , 
\end{equation}
and express Eq.(\ref{propag}) as
\begin{equation}
G_i(\rho) = a \,  \frac{-i\gamma_{\mu}\mathring{\rho}_{\mu} + \mathcal{M}_i (\rho) -i \tilde{\mu}_{q,i} \gamma_5 \tau^3 }{\mathring{\rho}^2 + \mathcal{M}_i^2 (\rho)+\tilde{\mu}_{q,i}^2}, \,\,\,\,\,\,\,\, i=1,2 \, . \label{eq:admG}
\end{equation}
Taking into account the change of the integration variables, we have that 
\be
\label{dimens0}
\Pi_V^{\mu\nu}(Q,a) = \frac{1}{a^2}\,  \mathcal{P}_V^{\mu\nu}(Q  a) = \frac{1}{a^2}\, \int_{-\pi}^{+\pi} \frac{d^4 \rho}{(2\pi)^4}\, {\rm Tr}\left[\gamma^{\mu} G_1(\rho + \frac{Q a}{2})  \gamma^{\nu}G_2(\rho- \frac{Q  a }{2} )\right],
\ee
where $\mathcal{P}_V^{\mu\nu}(Q a)$ is a dimensionless quantity which can only depend on dimensionless quantities ($Q a$, $m_1 a$, $m_2 a$, $\dots$). At this point we may obtain the $\chi$'s by applying the appropriate derivatives with respect to $Q_\mu$  to the expression given in Eq.\,(\ref{dimens0}).  Note that any  derivative with respect to $Q_\mu$ we  make to obtain the $\chi$'s implies the appearance of a factor $a$ in front of the r.h.s.~of Eq.\,(\ref{dimens0}), since the integral only depends on the product $Q a$.  A particularly convenient choice of $Q$ in the evaluation of the lattice integral (\ref{dimens0}), and the corresponding expression at two loops, is $Q=(Q_0,\vec 0)$, $Q^2=Q^2_0$, $\partial /\partial Q^2 = 1/(2Q_0)\partial /\partial Q_0$, see Eqs.\,(\ref{eq:tC})--(\ref{dpf2b}).   

When we want to obtain the continuum expression (at this order we do not need to define the renormalisation scheme since everything is finite) it is enough to take the limit $a\to 0$ in the integrand (\ref{eq:VP}) and apply to $\mathcal{P}_V^{\mu\nu}(Q a)$ the derivatives with respect to $Q_0$.

It is useful to start by discussing the case of the susceptibilities at $Q=0$. In this case, in the continuum, one obtains
\bea  
\left(m_2^2\,  \chi_{1^-}(Q^2=0)\right)_{\rm con}&=& \frac{N_c}{96
 \pi ^2 \left(1-u^2\right)^5}  \, \left\{\left(1-u^2\right) \left(3+4 u-21 u^2+40 u^3-21 u^4+4 u^5+3 u^6\right)\right. \nonumber \\ ~ && \left. +12 u^3 \left(2-3 u+2 u^2\right) \log\left[u^2\right]\right\} \\
\left(\chi_{0^+}(Q^2=0)\right)_{\rm con}&= &\frac{N_c}{24 \pi ^2 \left(1-u^2\right)^3}\, 
\left\{\left(1-u^2\right) \left(1-4 u+u^2\right) \left(1+u+u^2\right)-6 u^3 \log \left[u^2\right]\right\} \nonumber
\, , \eea
where $N_c$ is the number of colours and  the quantities on the l.h.s.~being dimensionless can only depend on the ratio $u \equiv m_1/m_2$.  In what follows $m_2$ will always denote the heavier of the two valence quarks in the decaying meson, namely the $b$ quark for $B \to D^{(*)}$ decays, the charm for $D\to K^{(*)}$ decays and the strange for kaon semileptonic decays. Note that in the limit $m_1 \to m_2$ (i.e.~$u \to 1$) the longitudinal susceptibility $\chi_{0^+}(Q^2)$ vanishes because the currents are conserved in this limit. Also on the lattice, as $a \to 0$, the $\chi$'s can only depend on $u$ and thus, in perturbation theory we expect 
\bea \left(m_2^2\,  \chi_{1^-}(Q^2=0, a )\right)_{\rm LAT}&=& \left(m_2^2\,  \chi_{1^-}(Q^2=0)\right)_{\rm con} \,
+ a^2 m_2^2 \,  \delta \chi^\prime_{1^-}(u, a^2 m_2^2)\, , \nonumber \\
\left(  \chi_{0^+}(Q^2=0, a )\right)_{\rm LAT}&=& \left( \chi_{0^+}(Q^2=0)\right)_{\rm con} + \delta \chi_{0^+}(u)+ a^2 m_2^2 \,  \delta  \chi^\prime_{0^+}(u,  a^2 m_2^2) \, , \eea
where the quantities $\delta \chi^{(\prime)}_i$ can be eliminated in perturbation theory following the scheme described in Eq.\,(\ref{eq:schemesub}).
The lattice susceptibilities $\left(m_2^2\,  \chi_{1^-}(Q^2=0, a )\right)_{\rm LAT}$ and $\left(  \chi_{0^+}(Q^2=0, a )\right)_{\rm LAT}$ are obtained by applying the appropriate derivatives with respect to $Q_0$  to the expression in Eq.\,(\ref{dimens0}) and putting $Q_0=0$. The four dimensional integral can be performed numerically without difficulties. 

Since we are able to compute  the polarization tensor  non perturbatively,  in principle we are also able to enforce the unitarity constraints on the FFs at $Q^2 \neq 0$. Thus in order to reduce the lattice artefacts we also need the lattice and continuum perturbative calculation in this case.

At one loop, in the continuum, for $Q^2 \neq 0$ we may construct two dimensionless quantities, namely $u$ as before and $z = \sqrt{Q^2/m_2^2}$.  In terms of these variables we obtain~\cite{Caprini:1997mu}
\bea 
\chi_{1^-}(Q^2=z^2 m_2^2) & = & \frac{N_c}{16 m_2^2 \pi ^2 (z^2)^3} \Bigl\{ -\frac{\left(1-u^2\right) \log\left[u^2\right] \left(2 z^4+2 \Lambda _- \Lambda _++z^2 \left(3 \Lambda _-+\Lambda _+\right)\right)}{2 z^2} \nonumber \\ 
&-& \frac{12 \Lambda _- \Lambda _+^2+3 z^2 \Lambda _+ \left(7 \Lambda _-+3 \Lambda _+\right)+z^4 \left(3 \Lambda _-+19 \Lambda _+\right)}{6  \Lambda _+} \\ 
&-& \log\left[\frac{\Lambda _-+\Lambda _++2 \sqrt{\Lambda _- \Lambda _+}}{\Lambda _-+\Lambda _+-2\sqrt{\Lambda _- \Lambda _+}}\right] \frac{\sqrt{\Lambda _- \Lambda _+}}{8z^2 \Lambda _- \Lambda _+^2} \left[ 8 \Lambda _-^2 \Lambda _+^3+8 z^2 \Lambda _- \Lambda _+^2 \left(2 \Lambda _-+\Lambda_+\right) \right. \nonumber \\
&+&\left. z^4 \Lambda _+ \left(5 \Lambda _-^2+18 \Lambda _- \Lambda _++\Lambda _+^2\right)+z^6 \left(-\Lambda _-^2+6 \Lambda _- \Lambda _++3 \Lambda _+^2\right) \right] \Bigl\} \nonumber \, , 
\eea
\bea
\chi_{0^+}(Q^2=z^2 m_2^2) & = & \frac{N_c}{16 \pi ^2 (z^2)^2} \Bigl\{ 2 \left(z^2+\Lambda _-\right) \left(z^2+2 \Lambda _+\right) \nonumber \\ 
&+& \frac{\log\left[\frac{\Lambda _-+\Lambda_++2 \sqrt{\Lambda _- \Lambda _+}}{\Lambda _-+\Lambda _+-2 \sqrt{\Lambda _- \Lambda _+}}\right] \left(z^2+\Lambda _-\right) \sqrt{\Lambda _- \Lambda _+} \left(4\Lambda _- \Lambda _++z^2 \left(3 \Lambda _-+\Lambda _+\right)\right)}{2 z^2 \Lambda _-} \nonumber \\ 
&+& \frac{\left(1-u^2\right) \log\left[u^2\right] \left(4 \Lambda_- \Lambda _++z^2 \left(\Lambda _-+3 \Lambda _+\right)\right)}{2 z^2} \Bigl\} \, , 
\eea
where $\Lambda _\pm = (1\pm u)^2 - z^2$.  Also in this case $\chi_{0^+}(Q^2=z^2 m_2^2) $ vanishes for $m_1 \to m_2$ ($u \to 1$). 
The corresponding lattice quantities are easily obtained as before by applying the appropriate derivatives with respect to $Q_0$ to the expression in Eq.\,(\ref{dimens0}) and putting $Q_0=\sqrt{Q^2}\neq 0$. 

On the lattice there is another equivalent way to compute the correlation function in perturbation theory, which is more suited for perturbative calculations at higher orders (more loops). Let us continue the discussion with the example of the vector current. 
We define the free propagator of the quark (and in higher orders also of the gluon propagator) by numerical Fourier transform of the momentum-space propagator given in Eq.\,(\ref{eq:admG})
\beq G_i(x,y) = \frac{1}{a^4} \, \int_{-\pi}^{+\pi} \,  \frac{d^4\rho}{(2\pi)^4}\, e^{-i \rho \cdot (x-y) }\, G_i(\rho)  \label{eq:Gx}\, , \eeq
where $x\equiv(t, \vec x  )$, $y\equiv(t_0, \vec y)$ and we have elected one of the lattice directions to our Euclidean time.
At one loop we define 
\bea 
\label{eq:VPx}
\Pi_V^{\mu\nu}(t, \vec x  ) &=&{\rm Tr}[\gamma^{\mu} G_1(x,0 )  \gamma^{\nu}G_2(0,x )] \, , \label{eq:pmunux}\nonumber  \\
G_V^{\mu\nu}( t) &=& a^3\,  \sum_{\vec x }\, \Pi_V^{\mu\nu}( t,\vec x ) \, . \label{eq:pmunut}
\eea
It is straightforward to show the relations  
\beq G_V^{00}(t) = C_{0^+}(t) \, , \qquad \frac{1}{3} \sum_{i=1,2,3}\, G_V^{ii}(t)  =C_{1^-}(t)\, , \eeq
where the quantities $C_{0^+}(t) $ and $C_{1^-}(t)$ are those defined in Eq.\,(\ref{dpf2}). We can then obtain the susceptibilities using the expressions in Eq.\,(\ref{dpf2b}).  

The difference between this latter way of computing the $\chi$'s, that we will call the x-space method, and the calculation of the $\chi$'s from the derivatives applied to Eq.\,(\ref{dimens0}), that we denote as the Q-space method, is the following. Eq.\,(\ref{dimens0}) refers to a discretised, infinite volume lattice. The Fourier transform in Eq.\,(\ref{eq:Gx}), however, must be done, in practice, on a finite lattice of volume $L^3$ and time extent $T$
\beq G_i(x,0) = \frac{1}{a^4 T L^3 } \sum_{\rho_i = 2\pi n_i /L, \rho_0 = 2\pi n_0/T} \, e^{-i \rho \cdot x }\, G_i(\rho)  \label{eq:Gxd}\, . \eeq
Thus, in order to compare the results with the two different methods we must extrapolate the results obtained  from $G_V^{\mu\nu}( t)$ on a finite lattice to the infinite volume limit. We have done the calculation in the two ways  for different values of $Q$  and found very good agreement between them. 
A further advantage of the x-space  method  is that we can perform the calculation on the same finite volume used for the non perturbative calculation of the two-point function.

In order to further reduce lattice artefacts we may extend the subtraction procedure of Eq.\,(\ref{eq:schemesub}) by computing the lattice and continuum quantities at $O(\alpha_s)$ for arbitrary quark masses and values of $Q^2$. The relevant Feynman diagrams are shown in Fig.\,(\ref{fig:2loops}) where the crosses represent  bilinear operators and  the curly line  the gluon propagator. 
 This will be implemented in the future study of $B \to D^{(*)}$ decays for which discretisation effects are much larger than in the $D\to K$ decays considered in the present study.

\section{Longitudinal and transverse susceptibilities}
\label{sec:susceptibilities}

In this Section we present a detailed description of the non-perturbative, lattice QCD calculation of the longitudinal and transverse susceptibilities.
The information about the gauge field configurations, quark masses, extrapolations to the physical point and to the continuum, as well as about all the relevant renormalization constants (RCs) used in this work, are given in Appendix~\ref{sec:simulations}.
Using the ETMC gauge ensembles of Table~\ref{tab:simudetails},  we have evaluated the following two-point correlation functions 
\bea
    \widetilde{C}_{0^+}(t) & = & \widetilde{Z}_V^2 ~ \int d^3x  \langle 0 | T\left[ \bar{q}_1(x) \gamma_0 q_2(x) ~ \bar{q}_2(0) \gamma_0 q_1(0) \right] | 0 \rangle ~ , \nonumber \\
     \widetilde{C}_{1^-}(t) & = & \widetilde{Z}_V^2 ~ \frac{1}{3} \sum_{j=1}^3 \int d^3x  \langle 0 | T\left[ \bar{q}_1(x) \gamma_j q_2(x) ~ \bar{q}_2(0) \gamma_j q_1(0) \right] | 0 \rangle ~ , \nonumber \\
     \widetilde{C}_{0^-}(t)& = & \widetilde{Z}_A^2 ~ \int d^3x  \langle 0 | T\left[ \bar{q}_1(x) \gamma_0 \gamma_5 q_2(x) ~ \bar{q}_2(0) \gamma_0 \gamma_5 q_1(0) \right] | 0 \rangle ~ , \label{eq:rencor}  \\ 
     \widetilde{C}_{1^+}(t) & = & \widetilde{Z}_A^2 ~ \frac{1}{3} \sum_{j=1}^3 \int d^3x \langle 0 | T\left[ \bar{q}_1(x) \gamma_j \gamma_5 q_2(x) ~ \bar{q}_2(0) \gamma_j \gamma_5 q_1(0) \right] | 0 \rangle ~ ,\nonumber  \\
   \widetilde{C}_S(t) & = & \widetilde{Z}_S^2 ~ \int d^3x \langle 0 | T\left[ \bar{q}_1(x) q_2(x) ~  \bar{q}_2(0) q_1(0) \right] | 0 \rangle ~ , 
   \nonumber \\
    \widetilde{C}_P(t) & = & \widetilde{Z}_P^2 ~ \int d^3x  \langle 0 | T\left[ \bar{q}_1(x) \gamma_5 q_2(x) ~ \bar{q}_2(0) \gamma_5 q_1(0) \right] | 0 \rangle ~ , \nonumber 
\eea
where $q_1$ and $q_2$ are the two valence quarks with bare masses $a \mu_1$ and $a \mu_2$ given in Table~\ref{tab:simudetails}, while the multiplicative factors $\widetilde{Z}_\Gamma$ ($\Gamma = \{ V, A, S, P \}$) are the appropriate RC of the bilinear currents, which will be specified in a while.
We consider either opposite or equal values for the Wilson parameters $r_1$ and $r_2$ of the two valence quarks, namely either the case $r_1 = - r_2$ or the case $r_1 = r_2$. 
Since our twisted-mass setup is at its maximal twist, in the case $r_1 = - r_2$ we have $\widetilde{Z}_\Gamma = \{ Z_A, Z_V, Z_P, Z_S \}$, while in the case $r_1 = r_2$ we have $\widetilde{Z}_\Gamma = \{ Z_V, Z_A, Z_S, Z_P \}$.
Upon  renormalisation of the bilinear currents,  the correlation functions $\widetilde{C}_j(t)$  with $j = \{0^+, 1^-, 0^-, 1^+, S, P \}$ corresponding to either opposite or equal values of the Wilson parameters $r_1$ and $r_2$ differ only by effects of order ${\cal{O}}(a^2)$.

The statistical accuracy of the meson correlators (\ref{eq:rencor}) can be significantly improved by adopting the one-end trick stochastic method~\cite{Foster:1998vw, McNeile:2006bz}, which employs spatial stochastic sources at a single time slice chosen randomly. 

We start by considering the longitudinal and transverse susceptibilities of both the vector and the axial-vector currents evaluated at $Q^2 = 0$, namely $\chi_j(0)$ with $j = \{0^+, 1^-, 0^-, 1^+\}$, defined in Eqs.~(\ref{eq:rencor}) as either the second or the fourth moments of the corresponding longitudinal and transverse Euclidean correlators $C_j(t)$.
For each gauge ensemble the values of $\chi_j(0)$ have been evaluated for the various combinations of the two valence quark masses $m_1 = a \mu_1 / (Z_P a)$ and $m_2 = a \mu_2 / (Z_P a)$, chosen in the light, strange and charm regions of Table~\ref{tab:simudetails}.

\subsection{The $c \to s$ transition}
\label{sec:ctos}

In this work we limit ourselves to the quark mass combinations $a \mu_1 = a \mu_c$ and $a \mu_2 = a \mu_s$, which correspond to the $c \to s$ transition. 
The simulated susceptibilities $\chi_j(0)$ ($j = \{0^+, 1^-, 0^-, 1^+\}$) are smoothly interpolated at $m_1 = a \mu_c / (Z_P a) = m_c^{phys}$ and $m_2 = a \mu_s / (Z_P a) = m_s^{phys}$, i.e.~at the physical charm and strange quark masses given in Table~\ref{tab:8branches}.

\begin{figure}[htb!]
\begin{center}
\includegraphics[scale=0.70]{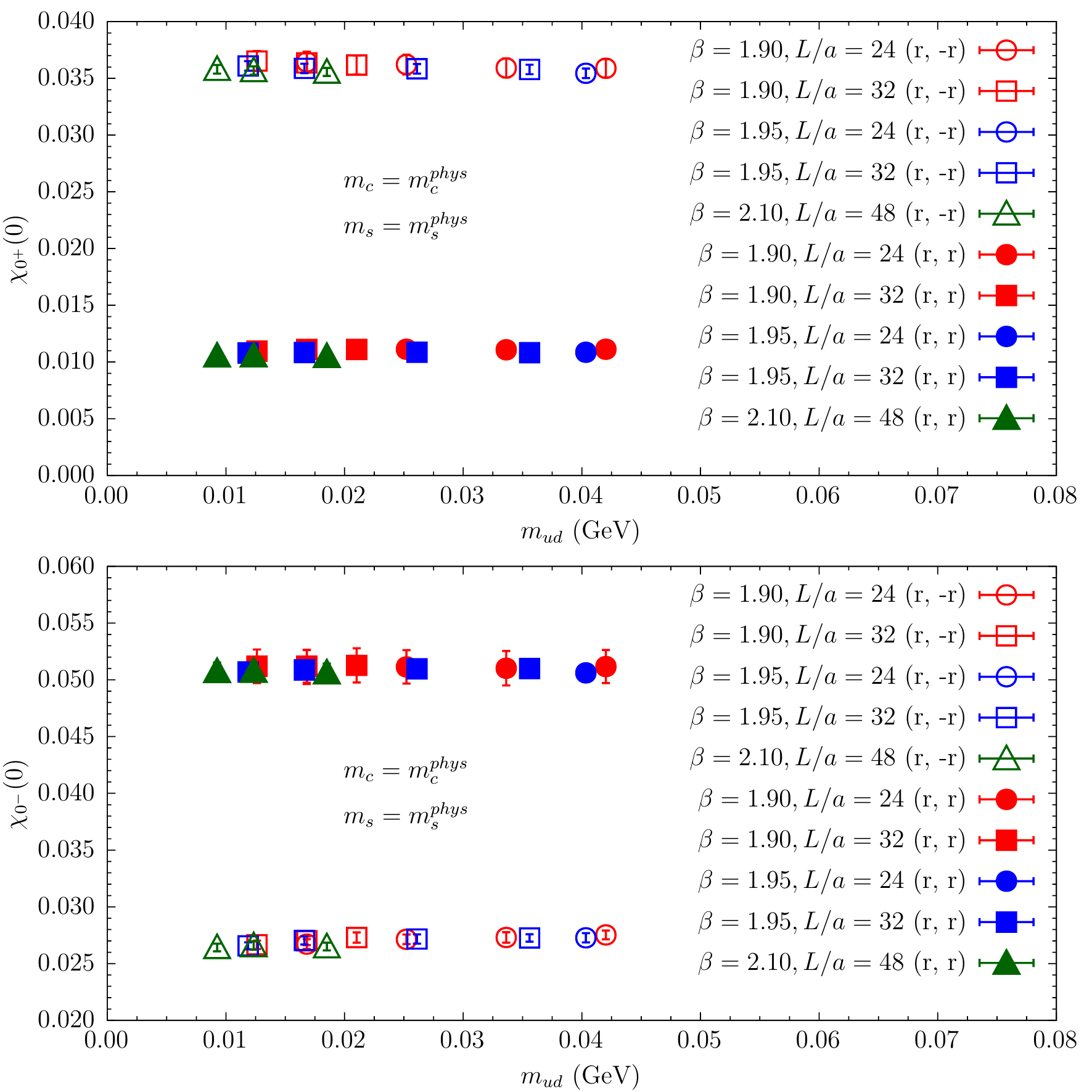}
\end{center}
\vspace{-0.80cm}
\caption{\it \small Vector and axial longitudinal susceptibilities $\chi_{0^+}(0)$ (upper panel) and $\chi_{0^-}(0)$ (lower panel) corresponding to the ETMC gauge ensembles for the $c \to s$ transition. The susceptibilities are obtained after a smooth interpolation at $m_1 = m_c^{phys}$ and $m_2 = m_s^{phys}$  given in Table~\ref{tab:8branches}. The empty markers correspond to the choice of opposite values $(r, -r)$ of the valence-quark Wilson parameters, while the full ones refer to the case of equal values $(r, r)$.\hspace*{\fill} }
\label{fig:VLAL_0}
\end{figure}

\begin{figure}[htb!]
\begin{center}
\includegraphics[scale=0.70]{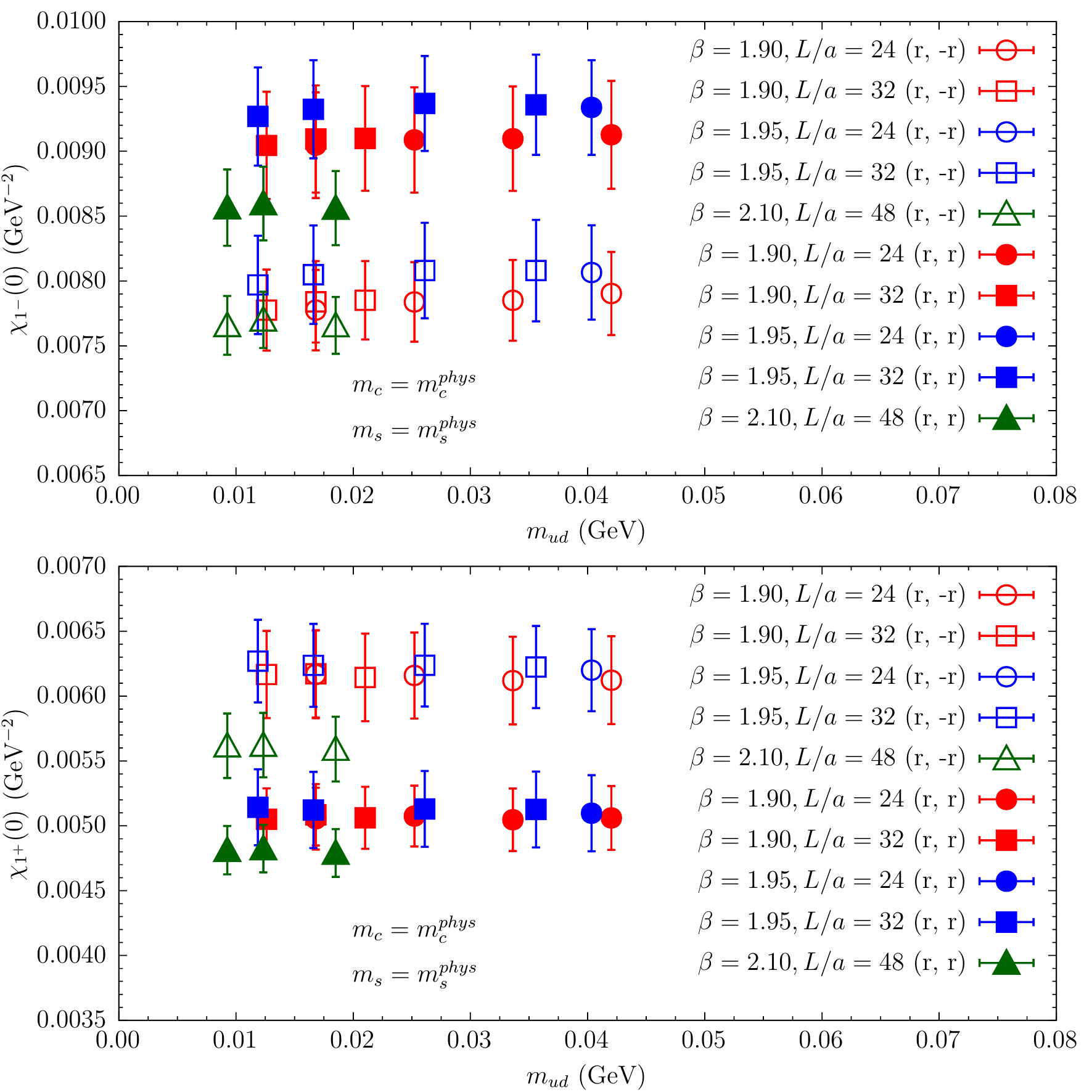}
\end{center}
\vspace{-0.80cm}
\caption{\it \small The same as in Fig.\,\ref{fig:VLAL_0}, but for the vector and axial transverse susceptibilities $\chi_{1^-}(0)$ (upper panel) and $\chi_{1^+}(0)$ (lower panel). \hspace*{\fill}} 
\label{fig:VTAT_0}
\end{figure}

The results for the vector and axial longitudinal susceptibilities are shown in Fig.\,\ref{fig:VLAL_0} and  those for the transverse ones in Fig.\,\ref{fig:VTAT_0} at either opposite or equal values of the valence-quark Wilson parameters, which will be denoted hereafter by $(r, -r)$ and $(r, r)$.
Differences among the results corresponding to the two $r$-combinations are expected to occur because of (twisted-mass) discretization effects, but it can clearly be seen that such differences are much larger in the longitudinal channels with respect to the transverse cases.

Moreover, because of charge conservation, the susceptibility $\chi_{0^+}(0)$ evaluated for $m_1 = m_2$ should vanish in the continuum limit. 
This is strongly violated for the $(r, -r)$ combination, as shown in Table~\ref{tab:CT} for the degenerate case $m_1 = m_2 = m_c^{phys}$.
\begin{table}[hbt!]
\begin{center}
\renewcommand{\arraystretch}{1.20}
\begin{tabular}{||c||c|c|c||}
\hline
 $m_1 = m_2 = m_c^{phys}$ & A30.32 & B25.32 & D20.48 \\ \hline
$\chi_{0^+}$(r, -r) & 2.55~~(9) $\cdot 10^{-2}$ & 2.56~(4) $\cdot 10^{-2}$ & 2.58~(4) $\cdot 10^{-2}$ \\ \hline
$\chi_{0^+}$(r, r)  & 4.30~(10) $\cdot 10^{-4}$ & 4.30~(8) $\cdot 10^{-4}$ & 4.06~(4) $\cdot 10^{-4}$ \\ \hline
\hline   
\end{tabular}
\renewcommand{\arraystretch}{1.0}
\end{center}
\vspace{-0.250cm}
\caption{\it \small Values of the vector longitudinal susceptibility $\chi_{0^+}(0)$ corresponding to either opposite (r, -r) or equal (r, r) values of the valence-quark Wilson parameters in the degenerate case $m_1 = m_2 = m_c^{phys}$. The three ETMC gauge ensembles A30.32, B25.32 and D20.48 (see Table~\ref{tab:simudetails}) correspond to nearly the same pion mass and differ in the values of the lattice spacing. \hspace*{\fill} }
\label{tab:CT}
\end{table}

The above observations point toward the presence of extra contributions coming from possible contact terms related to the product of two currents, which appear in all the correlators (\ref{eq:rencor}).  The issue of contact terms, which may affect the evaluation of the correlators for any lattice formulation of QCD, has been throughoutly investigated for our ETMC setup in Refs.~\cite{Burger:2014ada,Giusti:2017jof} in the case of the HVP contribution to the muon ($g-2$), which as known involves the product of two electromagnetic currents (i.e.~the degenerate case $m_1 = m_2$).  The presence of contact terms is also evident from the explicit calculation in lattice  perturbation theory of Section~\ref{sec:pertlatt}. A strategy to subtract non perturbatively the largest contamination due to  these contact terms will be discussed in the following.

The main outcome is that contact terms may not vanish in the continuum limit due to the mixing of the product of two currents with terms proportional to second derivatives of the Dirac delta function.
Therefore, a quick inspection of Eqs.\,(\ref{eq:rencor}) reveals that the longitudinal susceptibilities $\chi_{0^+}(0)$ and $\chi_{0^-}(0)$ are affected by contact terms (being second moments), while the transverse ones $\chi_{1^-}(0)$ and $\chi_{1^+}(0)$ are not (being fourth moments).
A way to avoid contact terms is to replace  the longitudinal susceptibilities in Eqs.\,(\ref{eq:rencor})  with the corresponding expressions (\ref{eq:finalEucl})  derived using WI identities at $q^2 = 0$, namely
\bea
     \chi_{0^+}(0) & = & \frac{1}{12} (m_1 - m_2)^2 \int_0^\infty dt^\prime ~ t^{\prime \, 4} ~ C_S(t^\prime) ~ , ~ \nonumber \\
     \label{eq:chiAL_WTI}
     \chi_{0^-}(0) & =  & \frac{1}{12} (m_1 + m_2)^2 \int_0^\infty dt^\prime ~ t^{\prime \, 4} ~ C_P(t^\prime) ~ , ~ 
\eea 
where $C_S(t)$ and $C_P(t)$ are given by Eqs.\,(\ref{eq:rencor}).
In this way the longitudinal susceptibilities are evaluated using the fourth moments of the scalar and pseudoscalar correlators and they become free from contact terms.
Note that in the degenerate case $m_1 = m_2$ the susceptibility $\chi_{0^+}(0)$ given by Eq.\,(\ref{eq:chiAL_WTI}) vanishes at any finite value of the lattice spacing.

\begin{figure}[htb!]
\begin{center}
\includegraphics[scale=0.70]{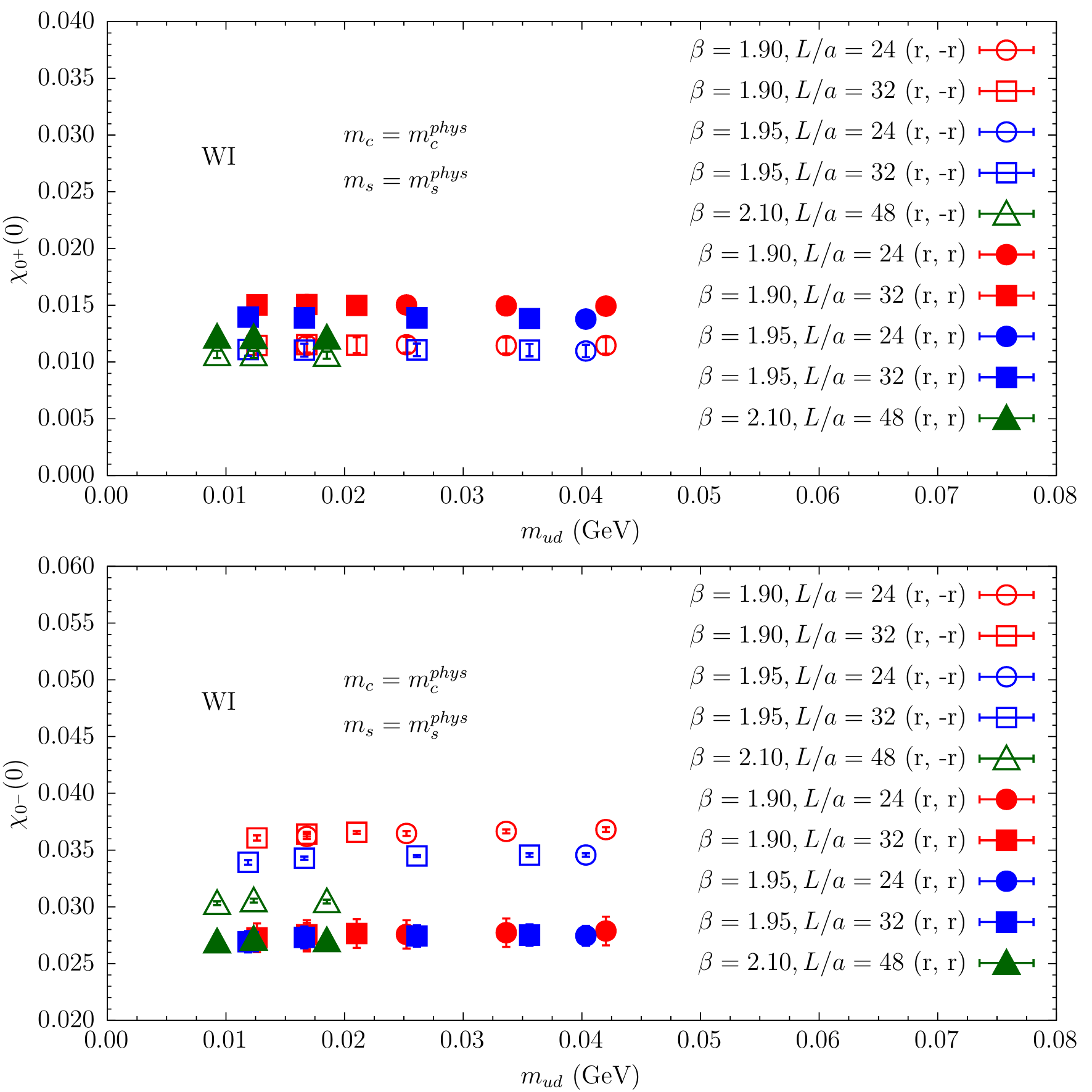}
\end{center}
\vspace{-0.80cm}
\caption{\it \small The same as in Fig.\,\ref{fig:VLAL_0}, but using the definitions (\ref{eq:chiAL_WTI}) based on the use of the WI. The vertical scales are kept the same as in Fig.\,\ref{fig:VLAL_0} to show the impressive improvement obtained by the use of correlations based on the fourth moments. \hspace*{\fill}} \label{fig:VLAL_WI}
\end{figure}

The results for the {\it contact-free} longitudinal susceptibilities $\chi_{0^+}(0)$ and $\chi_{0^-}(0)$ are shown in Fig.\,\ref{fig:VLAL_WI}.
The comparison with the results of Fig.\,\ref{fig:VLAL_0} indicates that the differences between the two $r$-combinations are significantly reduced when  the effects of the contact terms are eliminated. 
Moreover, for both $r$-combinations the  relative  impact of the discretization effects on the {\it contact-free} longitudinal susceptibilities appear to be similar to the one corresponding to the case of the transverse susceptibilities shown in Fig.\,\ref{fig:VTAT_0}. For these reasons  in this work we rely only on the calculation of the longitudinal susceptibilities based on Eqs.\,(\ref{eq:chiAL_WTI}).  

Even if Eqs.\,(\ref{eq:chiAL_WTI}) are essentially free from contact terms, it is, however,  very interesting to investigate  the impact of the contact terms on the longitudinal susceptibilities in  Eqs.\,(\ref{eq:rencor}).  In particular,  from the observation that $\chi_{0^+}(0)$ is much smaller  for the combination $(r, r)$, see Table~\ref{tab:CT}, we deduce that in this case  the contact terms  are much smaller than for the other combination $(r, -r)$.  

In general a sizeable reduction of the contact terms  can be achieved by using the subtraction procedure developed in  Section~\ref{sec:pertlatt}.  The perturbative contribution, evaluated at order ${\cal{O}}(\alpha_s^0)$,  can already explain the large impact of the contact terms for the combination $(r, -r)$.
An even more effective cancellation of the contact terms is obtained by using the non perturbative subtraction proposed below.  
A detailed, general presentation of the numerical implementation of the subtraction procedure will not be given here. We will discuss it in details in a forthcoming study of the $b \to c$ transition, for which also discretisation effects are much larger than in the case of the $c \to s(\ell)$ decays considered in the present work.   We find that the subtraction turns out to be beneficial also for reducing the discretization effects, that for $b \to c $ are much larger,  for both $r$-combinations.  Here we anticipate that the perturbative contribution to the susceptibility $\chi_{0^+}(0)$, evaluated at order ${\cal{O}}(\alpha_s^0)$ in the degenerate case $m_1 = m_2 = m_c^{phys}$, turns out to vanish for the combination $(r, r)$, while in the case $(r, -r)$ it is equal to $\approx 0.016$ for the three ensembles A30.32, B25.32 and D20.48, see Appendix~\ref{sec:simulations}, which represents $\approx 60 \%$ of the corresponding non-perturbative value ($\approx 0.026$) shown in Table~\ref{tab:CT}. 

An alternative, rather effective, way to get rid of the contact terms in the susceptibility $\chi_{0^+}(0)$ is to subtract the contact terms evaluated non perturbatively at $m_1 = m_2$, more precisely by using the formula 
\be
    \label{eq:CTsub}
    \overline{\chi}_{0^+}(0; m_1, m_2) \equiv \chi_{0^+}(0; m_1, m_2) - \frac{\chi_{0^+}(0; m_1, m_1) + \chi_{0^+}(0; m_2, m_2)}{2} ~ , ~
\ee
obtaining in this way that $\overline{\chi}_{0^+}(0; m_1 = m_2) = 0$ as in the case of the WI-based formula (\ref{eq:chiAL_WTI}).

\begin{figure}[htb!]
\begin{center}
\includegraphics[scale=0.70]{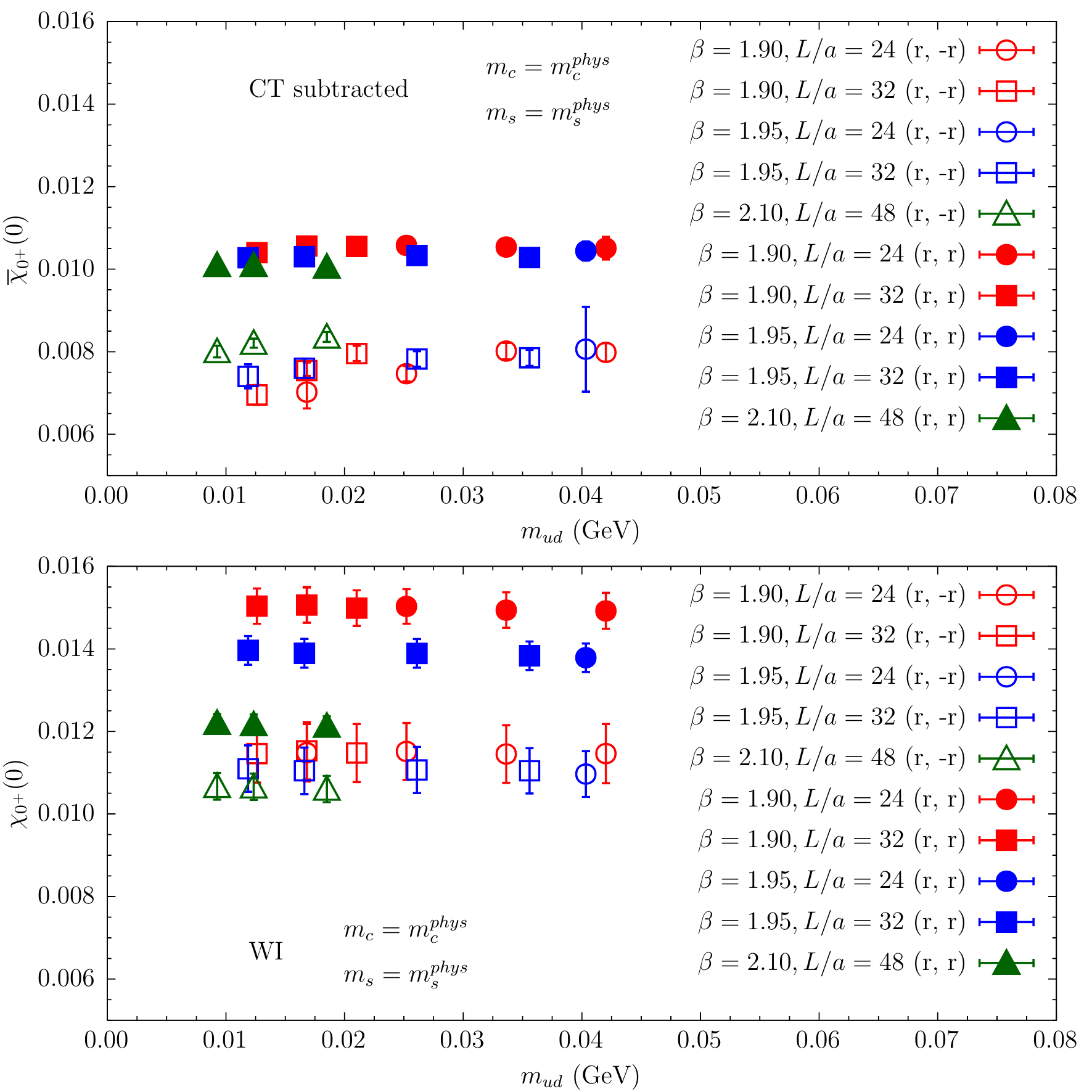}
\end{center}
\vspace{-0.80cm}
\caption{\it \small Vector longitudinal susceptibility $\overline{\chi}_{0^+}(0)$ obtained using the subtraction procedure given by Eq.\,(\ref{eq:CTsub}) (upper panel) and $\chi_{0^+}(0)$ based on the Ward identity (\ref{eq:chiAL_WTI}) (lower panel). \hspace*{\fill}} 
\label{fig:VL_comp}
\end{figure}

In Fig.\,\ref{fig:VL_comp} the results obtained using the non-perturbative subtraction in Eq.\,(\ref{eq:CTsub}), upper panel, are compared with the ones based on the WI identity (\ref{eq:chiAL_WTI}), lower panel, already given in  the upper panel of Fig.\,\ref{fig:VLAL_WI}.  
We give the two plots in the same figure, using the same scale,  in order to make the comparison between the two determinations easier.
The discretization effects appear to be different within the two procedures.
In particular, the subtraction procedure leads to quite small cutoff effects in the case of the $(r, r)$ combination.
After Eq.\,(\ref{eq:chiAT_DK}) below it will be shown   that the continuum limits of the data for $\chi_{0^+}(0)$ based on the two possible different determinations  agree very nicely.

\begin{figure}[htb!]
\begin{center}
\includegraphics[scale=0.70]{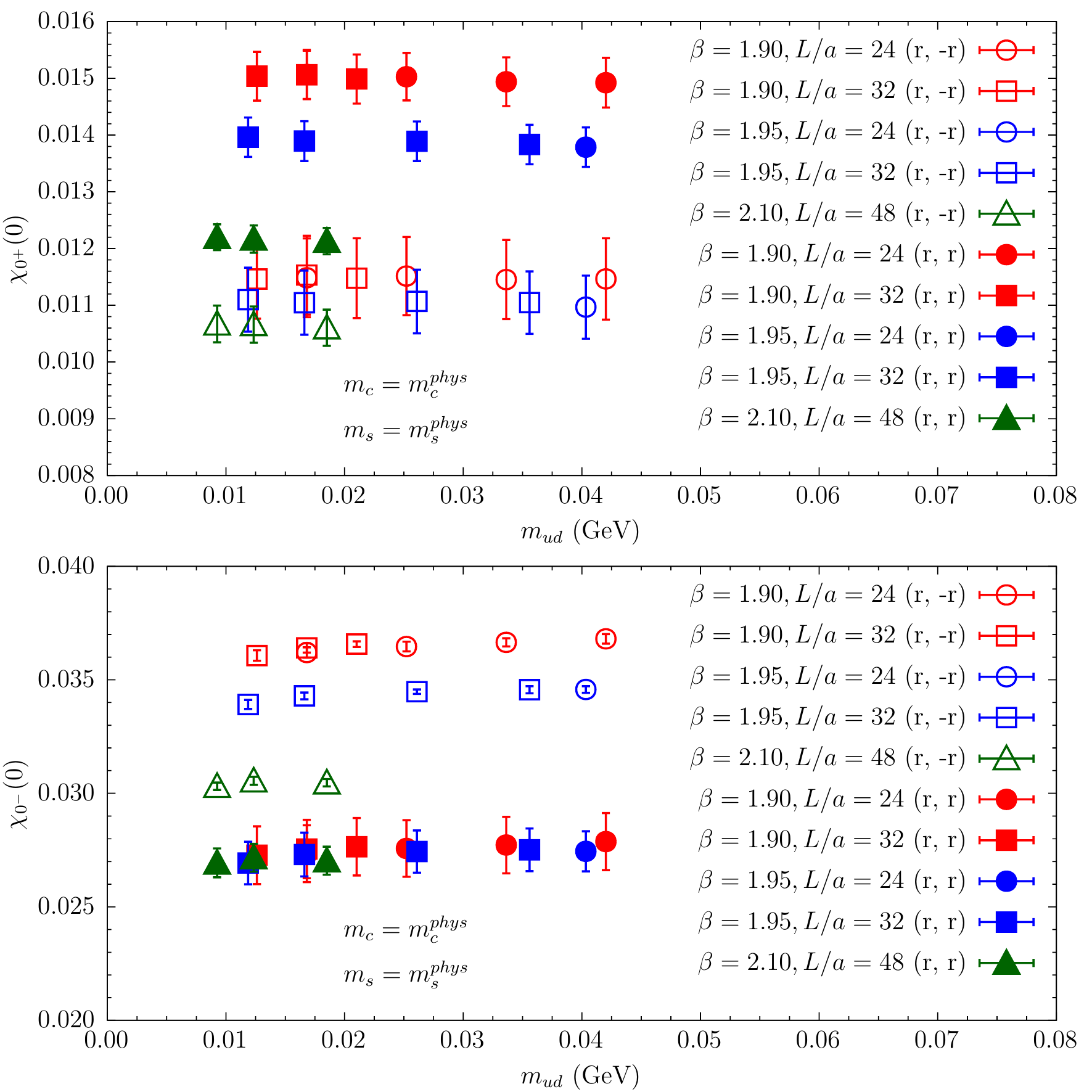}
\end{center}
\vspace{-0.80cm}
\caption{\it \small Vector and axial longitudinal susceptibilities $\chi_{0^+}(0)$ (upper panel) and $\chi_{0^-}(0)$ (lower panel) given respectively by Eqs.\,(\ref{eq:chiAL_WTI}), based on the use of the Ward identities.\hspace*{\fill}} 
\label{fig:VLAL}
\end{figure}

Since the subtraction procedure given in Eq.\,(\ref{eq:CTsub}) is not applicable to the axial longitudinal susceptibility $\chi_{0^-}(0)$, in what follows we make use of the longitudinal susceptibilities based on the Ward identities and shown in Fig.\,\ref{fig:VLAL}.
Note that for both $r$-combinations the impact of the discretization effects on the WI-based longitudinal susceptibilities appear to be similar to the one corresponding to the case of the transverse susceptibilities shown in Fig.\,\ref{fig:VTAT_0}.

The transverse and  longitudinal susceptibilities shown in Fig.\,\ref{fig:VTAT_0} and\,\ref{fig:VLAL_WI} exhibit a quite mild dependence on the light-quark mass $m_{ud}$, since the latter comes entirely from the light sea quarks. Therefore, we fit the lattice data separately for each of the two $r$-combinations by adopting a simple linear ansatz in the light-quark mass $m_{ud}$  as well as in the values of the squared lattice spacing $a^2$, since in our lattice setup the susceptibilities are ${\cal{O}}(a)$-improved, namely
\be
    \label{eq:chi_fit}
    \chi_j(0; m_{ud}, a^2) = \chi_j(0) \left[ 1 + A_1 \left( m_{ud} - m_{ud}^{phys} \right) + D_1 ~ a^2 \right] ~ , ~
\ee
where, for sake of simplicity, $\chi_j(0)$ stands for $\chi_j(0; m_{ud}^{phys}, 0)$ and we have not written explicitly  dependence   of the coefficients $A_1$ and $D_1$ on the specific channel $j$ ($j = \{ 0^+, 1^-, 0^-, 1^+ \}$). 
The quality of the various fits is always very good ($\chi^2/\mbox{(d.o.f.)} \lesssim 0.6$) and our findings for the extrapolated quantities $\chi_j(0)$ are collected in Table~\ref{tab:chi_DK}.

\begin{table}[hbt!]
\renewcommand{\arraystretch}{1.20}
\begin{center}
\begin{tabular}{||c||c|c||c|c||}
\hline
\multicolumn{1}{||c||}{$\chi_j(0)$} & \multicolumn{2}{|c||}{$1^{st}-4^{th}$ branches} & \multicolumn{2}{|c||}{$5^{th}-8^{th}$ branches} \\ 
\hline
                                                                     & $(r, -r)$       & $(r, r)$ & $(r, -r)$ & $(r, r)$ \\ \hline \hline
$\chi_{0^+}(0) ~ \cdot 10^3$                        & 9.89~(65) & 9.50~(17) & 8.47~(30) & 9.31~(15) \\ \hline
$\chi_{1^-} (0) ~ \cdot 10^3$~(GeV$^{-2}$) & 7.52~(40) & 8.10~(29) & 7.72~(27) & 8.17~(25) \\ \hline
$\chi_{0^-} (0) ~ \cdot 10^2$                        & 2.46~~(4) & 2.66~(19) & 2.47~~(3) & 2.35~~(9) \\ \hline
$\chi_{1^+}(0) ~ \cdot 10^3$~(GeV$^{-2}$) & 5.11~(13) & 4.57~(22) & 5.15~(11)  & 4.71~(13)  \\ \hline \hline
\end{tabular}
\end{center}
\renewcommand{\arraystretch}{1.0}
\caption{\it \small Values of the longitudinal and transverse susceptibilities $\chi_j(0)$ with $j = \{ 0^+, 1^-, 0^-, 1^+ \}$ relevant for the $c \to s$ transition averaged over the $1^{st}-4^{th}$ and $5^{th}-8^{th}$ branches of our bootstrap analysis after extrapolation to the physical pion point and to the continuum limit of the lattice data corresponding separately to the two $r$-combinations.\hspace*{\fill}} 
\label{tab:chi_DK}
\end{table}

Averaging over all the eight branches~\cite{Carrasco:2014cwa} of our bootstrap analysis, see also Appendix~\ref{sec:simulations}, and over the results corresponding separately to the two $r$-combinations our final results for the longitudinal and transverse susceptibilities relevant for the $c \to s$ transition are
\bea
     \chi_{0^+}(0) & = & 9.29 ~ (64) \cdot 10^{-3} ~ , ~ \nonumber \\[2mm]
     \label{eq:chiVT_DK}
     \chi_{1^-}(0)  & = & 7.88 ~ (41) \cdot 10^{-3} ~ \mbox{GeV}^{-2} ~ , ~ \\[2mm]
     \chi_{0^-}(0)  & = & 2.48 ~ (15) \cdot 10^{-2} ~ , ~ \nonumber \\[2mm]
     \label{eq:chiAT_DK}
     \chi_{1^+}(0) & = & 4.89 ~ (29) \cdot 10^{-3} ~ \mbox{GeV}^{-2} ~ , ~\nonumber 
\eea
where the errors include both the statistical uncertainties related to the Monte Carlo simulations and the systematic errors, which are mainly dominated by the uncertainties due to discretization effects.

\subsection{Subtraction of the ground-state contribution}
\label{sec:grs}

The susceptibilities $\chi_{\{0^+, 1^-, 0^-, 1^+ \}}(0)$ obtained in the previous Section represent upper limits to the dispersive bounds on the form factors relevant in the semileptonic $D \to K(K^*) \ell \nu_\ell$ decays.
Such limits can be improved by removing the contributions of the bound states from the Euclidean correlators $C_j(t)$ for $j = \{0^+, 1^-, 0^-, 1^+ \}$, Eqs.\,(\ref{dpf2}).
In particular, according to the Particle Data Group (PDG)~\cite{Zyla:2020zbs}  the meson states $D_{s0}^*$, $D_s^*$, $D_s$ and $D_{s1}$, which are relevant for the channels $j = 0^+, 1^-,  0^-, 1^+$, have masses below the threshold of the production of a pair of $D$ and $K$($K^*$) mesons.
Thus, their contribution to the susceptibilities $\chi_{\{0^+, 1^-, 0^-, 1^+ \}}(0)$ can be removed in order to improve the dispersive bounds on the semileptonic $D \to K(K^*) \ell \nu_\ell$ decays.
In this Section we describe such a subtraction.

As well known, at large time distances one has
\be
    \label{eq:larget} 
     C_j(t)_{ ~ \overrightarrow{t  \gg a, ~ (T - t) \gg a} ~ } \frac{\mathcal{Z}_j}{2M_j} \left[ e^{ - M_j  t} + e^{ - M_j (T - t)} \right] ~ ,  
\ee
where $M_j$ is the mass of ground-state meson $H_{12}^j$ and $\mathcal{Z}_j$ is the matrix element $ \mathcal{Z}_j \equiv | \langle H_{12}^j | \overline{q}_1 \Gamma_j q_2 | 0 \rangle|^2$ with $\Gamma_j = \{ \gamma_0, \vec{\gamma}, \gamma_0 \gamma_5, \vec{ \gamma} \gamma_5, \mathbb{1} , \gamma_5 \}$ for $j = \{0^+, 1^-, 0^-, 1^+, S, P \}$.  

\begin{figure}[htb!]
\begin{center}
\includegraphics[scale=0.70]{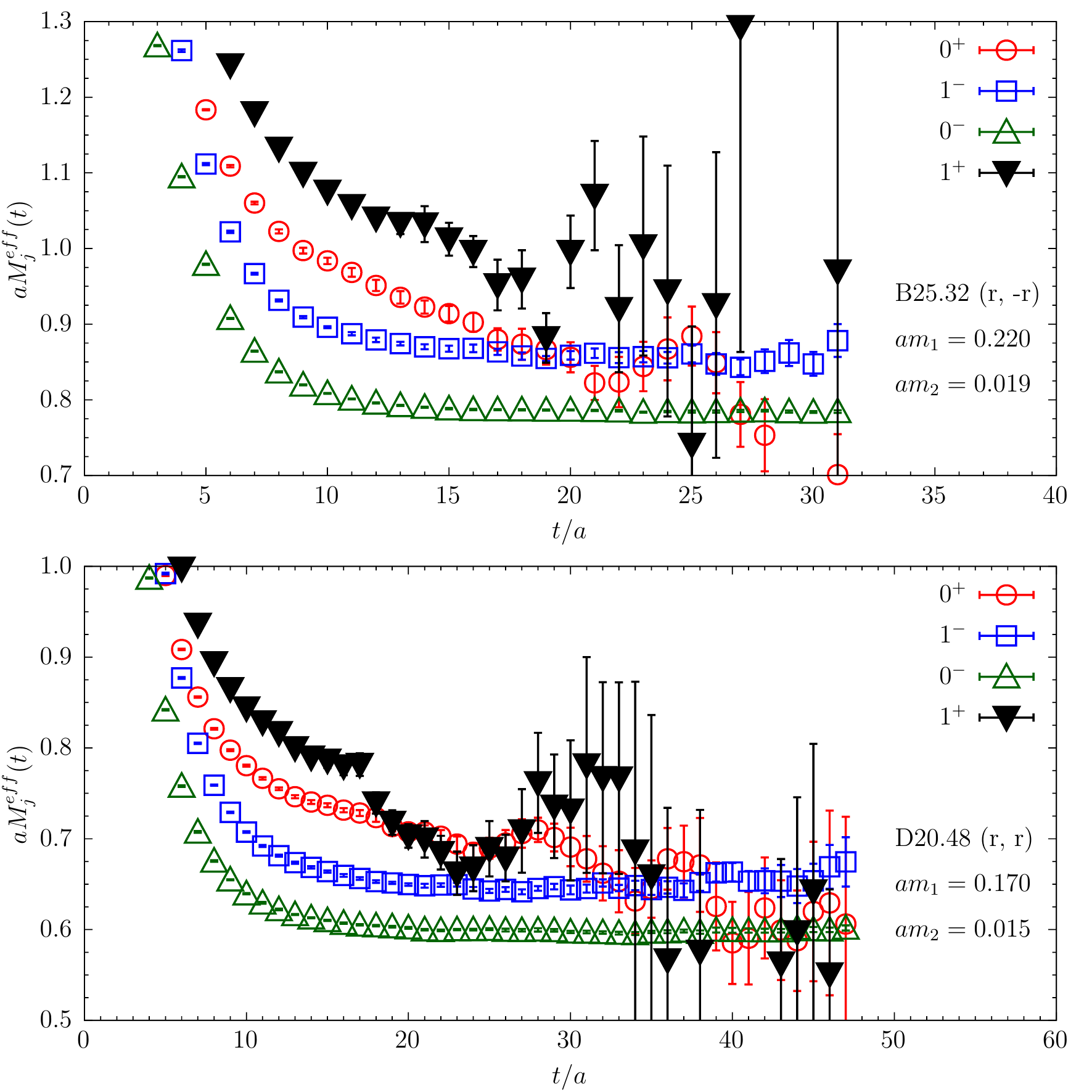}
\end{center}
\vspace{-0.80cm}
\caption{\it \small The temporal behavior of the effective mass (\ref{eq:Meff}) for the correlators $C_j(t)$ with $j = \{0^+, 1^-, 0^-, 1^+ \}$ in the case of the ensembles B35.32 (upper panel) and D20.48 (lower panel). Due to the use of the WT identities the effective mass $M_j^{eff}(t)$ for $j = 0^+$ and $j = 0^-$ correspond to the scalar and pseudoscalar correlators  given by Eqs.\,(\ref{eq:rencor}). The (bare) quark masses and the combinations of the Wilson $r$-parameters are specified in the insets and they roughly correspond to the case $m_1 \approx m_c^{phys}$ and $m_2 \approx m_s^{phys}$.\hspace*{\fill} }
\label{fig:plateaux}
\end{figure}

Thus, the ground-state mass $M_j$ and the matrix element $\mathcal{Z}_j$ can be extracted from the exponential fit given in the r.h.s.~of Eq.\,(\ref{eq:larget}) performed in the temporal region $ t = [t_{min}, t_{max}]$, where the effective mass $M_j^{eff}(t)$
\be
    \label{eq:Meff}
    M_j^{eff}(t) \equiv \mbox{log}\left( \frac{C_j(t-1)}{C_j(t)} \right)
\ee
may exhibit a plateau.
The temporal behavior of the effective masses $M_j^{eff}(t)$ is shown in Fig.\,\ref{fig:plateaux} in two illustrative cases.
Since we use the WI  in Eqs.\, (\ref{eq:finalEucl}) at $q^2=0$  for evaluating the longitudinal susceptibilities, the effective mass $M_j^{eff}(t)$ for $j = 0^+$ and $j = 0^-$ correspond to the scalar and pseudoscalar correlators given by Eqs.\,(\ref{eq:chiAL_WTI}).

At large time distances the quality of the plateaux is good for $j = 1^-$ and $j =  0^-$, while it is definitely poor in the cases $j = 0^+$ and $j = 1^+$.
The latter ones are likely to be plagued by the effects of parity breaking (mixing with $j = 0^-$ and $j = 1^-$) present in our lattice formulation.
We stress that we do not use $M_j^{eff}(t)$ to extract the ground-state masses $M_j$, but we perform the exponential fit given in the r.h.s.~of Eq.\,(\ref{eq:larget}) in the temporal regions shown in Table~\ref{tab:plateaux} for the various ETMC ensembles.
The quality of the exponential fits turns out to be always acceptable ($\chi^2 / \mbox{d.o.f.} \lesssim 0.5$).
\begin{table}[!hbt]
\begin{center}	
\begin{tabular}{||c|c||c||}
\hline
$\beta$ & $V / a^4$ & $[t_{\rm min} / a, \, t_{\rm max} / a]$ \\
\hline \hline
1.90 & $24^3 \times 48$ & [16, 22] \\ \cline{2-3}
        & $32^3 \times 64$ & [16, 22] \\ \hline
1.95 & $24^3 \times 48$ & [17, 22] \\ \cline{2-3}
        & $32^3 \times 64$ & [17, 22] \\ \hline
2.10 & $48^3 \times 96$ & [22, 30] \\ \hline
\end{tabular}
\end{center}
\caption{\it Temporal regions chosen for performing the exponential fit of the r.h.s.~of Eq.\,(\ref{eq:larget}) for the various ETMC ensembles. The values of $t_{min} / a$ for the different values of the gauge coupling $\beta$ correspond to the same value of $t_{min}$ in physical units, namely $t_{min} \simeq 1.4$ fm.\hspace*{\fill}} 
\label{tab:plateaux}
\end{table}

\begin{figure}[htb!]
\begin{center}
\includegraphics[scale=0.55]{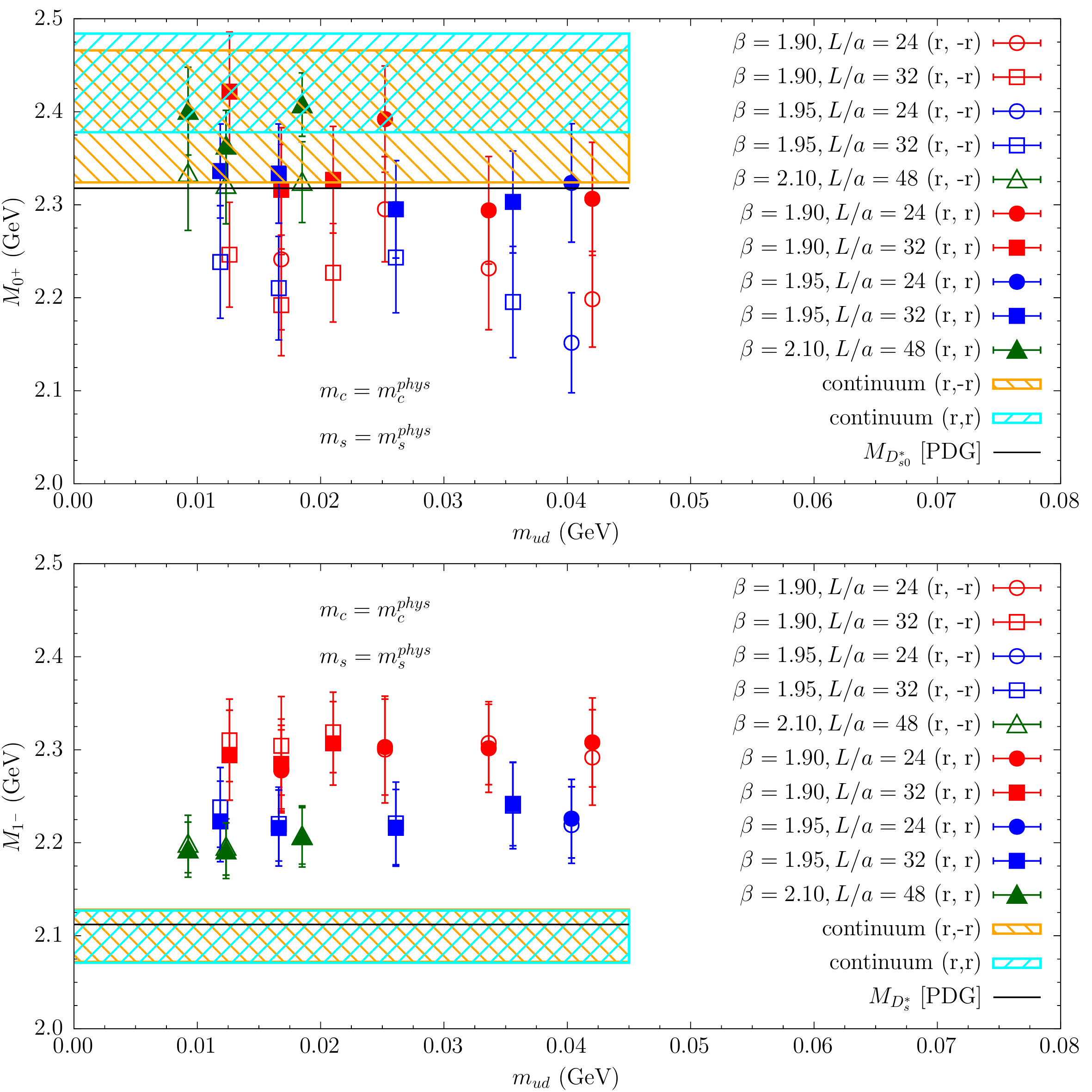}
\end{center}
\vspace{-0.80cm}
\caption{\it \small Ground-state masses of the channels $0^+$ (upper panel) and $1^-$ (lower panel) extracted from the exponential fit~(\ref{eq:larget}) and interpolated at $m_1 = m_c^{phys}$ and $m_2 = m_s^{phys}$. The black solid  lines correspond to the experimental masses of the mesons $D_{s0}^*$ and $D_s^*$ from the PDG~\cite{Zyla:2020zbs}, namely: $M_{D_{s0}^*} ^{exp.}= 2.3178~(5)$ GeV and $M_{D_s^*}^{exp.} = 2.1122~(4)$ GeV , respectively. The shaded areas correspond to the extrapolated values of the  $D^*$ masses to the physical pion point and to the continuum limit, averaged over all the eight branches of the analysis (see Table~\ref{tab:masses_DK}). \hspace*{\fill} }
\label{fig:MVLVT}
\end{figure}

The values of the ground-state masses, interpolated at $m_1 = m_c^{phys}$ and $m_2 = m_s^{phys}$, are shown in Fig.\,\ref{fig:MVLVT} in the case of the channels $0^+$ and $1^-$. 
Then, we extrapolate the lattice data to the physical pion point and to the continuum limit using the fitting function
\be
    \label{eq:mass_fit}
    M_j(m_{ud}, a^2) = M_j \left[ 1 + A_1^M \left( m_{ud} - m_{ud}^{phys} \right) + D_1^M ~ a^2 \right] ~ , ~
\ee
where, for sake of simplicity, $M_j$ stands for $M_j(m_{ud}^{phys}, 0)$.
Our results for $M_j$ are collected in Table~\ref{tab:masses_DK}.
\begin{table}[hbt!]
\renewcommand{\arraystretch}{1.20}
\begin{center}
\begin{tabular}{||c||c|c||c|c||}
\hline
\multicolumn{1}{||c||}{$M_j$} & \multicolumn{2}{|c||}{$1^{st}-4^{th}$ branches} & \multicolumn{2}{|c||}{$5^{th}-8^{th}$ branches} \\ 
\hline
                             & $(r, -r)$       & $(r, r)$       & $(r, -r)$       & $(r, r)$ \\ \hline \hline
$M_{0^+}$ (GeV) & 2.410~(62) & 2.445~(48) & 2.380~(76) & 2.416~(53) \\ \hline
$M_{1^-}$ (GeV)  & 2.104~(27) & 2.106~(27) & 2.095~(28) & 2.092~(26) \\ \hline
$M_{0^-}$ (GeV)  & 1.959~(29) & 1.952~(29) & 1.946~(25) & 1.938~(26) \\ \hline
$M_{1^+}$ (GeV) & 2.43~~(15) & 2.38~~(11) & 2.39~~(15) & 2.40~~(11) \\ \hline \hline
\end{tabular}
\end{center}
\renewcommand{\arraystretch}{1.0}
\caption{\it \small Values of the ground-state masses $M_j(0)$ with $j = \{ 0^+, 1^-, 0^-, 1^+ \}$ relevant for the $c \to s$ transition averaged over the $1^{st}-4^{th}$ and $5^{th}-8^{th}$ branches of our bootstrap analysis after extrapolation to the physical pion point and to the continuum limit of the lattice data corresponding separately to the two $r$-combinations.\hspace*{\fill} }
\label{tab:masses_DK}
\end{table}

Averaging over all the eight branches of our bootstrap analysis and over the results corresponding separately to the two $r$-combinations our results for the ground-state masses relevant for the $c \to s$ transition are
\bea
     M_{0^+}(0) & = & 2.413 ~ (65) ~ \mbox{GeV} ~ , \qquad [M_{D_{s0}^*}^{exp.} = 2.3178~(5) ~ \mbox{GeV} ]~ , ~ \nonumber \\
     M_{1^-}(0)  & = & 2.099 ~ (28) ~ \mbox{GeV} ~ , \qquad [M_{D_s^*}^{exp.} = 2.1122~(4) ~ \mbox{GeV} ]~ , ~\nonumber \\
     \label{eq:mds} \\
     M_{0^-}(0)  & = & 1.949 ~ (29) ~ \mbox{GeV} ~ , \qquad [M_{D_s}^{exp.} = 1.96834~(7) ~ \mbox{GeV} ]~  , ~ \nonumber \\
     M_{1^+}(0) & = & 2.40 ~~ (13) ~ \mbox{GeV} ~ , \qquad [M_{D_{s1}}^{exp.} = 2.4595~(6) ~ \mbox{GeV}] ~ . ~\nonumber
\eea
and within the uncertainties they compare nicely  with the experimental results given by the PDG~\cite{Zyla:2020zbs}.

We can now proceed to the evaluation of the ground-state contributions to the susceptibilities, $\chi_j^{(gs)}(0)$, using the results of the exponential fit~(\ref{eq:larget}), and to their subtraction from the global susceptibilities $\chi_j(0)$, namely
\be
    \chi_j^{(sub)}(0) = \chi_j(0) - \chi_j^{(gs)}(0) \qquad \mbox{for}~j = \{ 0^+, 1^-, 0^-, 1^+ \} ~ . ~
\label{eq:subtra} \ee
By repeating the analysis made in Section~\ref{sec:ctos} for the global susceptibilities $\chi_j(0)$ in the case of  the subtracted ones defined in Eq.\,(\ref{eq:subtra}), we obtain for the $\chi_j^{(sub)}(0)$ (after extrapolation to the physical pion point and to the continuum limit) the following values
\bea
     \chi_{0^+}^{(sub)}(0) & = &   4.33 (1.33 ) \times 10^{-3} ~ , ~ \nonumber \\
     \chi_{1^-}^{(sub)}(0)  & = & 4.19 (36 ) \times 10^{-3} ~ \mbox{GeV}^{-2} ~ , ~\nonumber \\ \label{eq:xsub} \\
     \chi_{0^-}^{(sub)}(0)  & = & 9.42 (91 ) \times 10^{-3} ~ , ~ \nonumber\\[2mm]
     \chi_{1^+}^{(sub)}(0) & = & 3.74 (56 ) \times 10^{-3} ~ \mbox{GeV}^{-2} ~ . ~\nonumber
\eea

As a check of our subtraction procedure, we consider  the ground-state contribution  to $\chi_{1^-}^{(gs)}(0)$ and to $\chi_{0^-}^{(gs)}(0)$ 
given by 
\be
      \label{eq:chiALVT_grs}
      \chi_{1^-}^{(gs)}(0) = \frac{f_{1^-}^2}{M_{1^-}^4} ~ , ~ \qquad \chi_{0^-}^{(gs)}(0) = \frac{f_{0^-}^2}{M_{0^-}^2} 
\ee
where $f_{1^-}$ and $f_{0^-}$ are the (leptonic) decay constants of the $D_s^*$ and $D_s$ mesons, respectively.
Using the ETMC results $f_{D_s^*} = 268.8~(6.6)$ MeV from Ref.~\cite{Lubicz:2017asp} and $f_{D_s} = 247.2~(4.1)$ MeV from Ref.~\cite{Carrasco:2014poa} as well as the experimental values of the $D_s^*$ and $D_s$ meson masses (see Eqs.\,(\ref{eq:mds})), we get $\chi_{1^-}^{(gs)} = 3.63~(19) \times 10^{-2}$\,GeV$^{-2}$ and $\chi_{0^-}^{(gs)} = 1.58~(5) \times 10^{-2}$.  By subtracting the above values, which we stress  have been evaluated using results for the decay constants from other lattice calculations and experimental values for the meson masses, from the corresponding results~(\ref{eq:chiVT_DK}) one obtains $\chi_{1^-}^{(sub)} = 4.25 (45)\times 10^{-3}$\,GeV$^{-2}$ and $\chi_{0^-}^{(sub)} = 9.0 ( 1.5) \times  10^{-3}$ in very good agreement with the results (\ref{eq:xsub}), which were obtained, instead, by using only data extracted from our simulations. 

Equations~(\ref{eq:xsub})  represent our non-perturbative results for the vector and axial susceptibilities relevant for the semileptonic $D \to K(K^*) \ell \nu_\ell$ decays.  We are planning to study the case $q^2 \neq 0 $ in a future study of the FFs in $B \to D^{(*)}$ semileptonic decays where there is a hope that 
the strength of the bounds may increase as $q^2$   approaches this region as suggested by Eqs.\,(16)-(19) of Ref.~\cite{Lellouch:1995yv}.

\section{The prototype: Extraction of the form factors for  $D\to K $ decays}
\label{sec:proto}

Together with Section \ref{sec:susceptibilities}, this is the central Section of our work where we show that from the knowledge of the form factors in the large $q^2$ region and of the susceptibilities it is possible to determine the form factors with good precision, without making any assumption on their functional dependence on the squared momentum transfer $q^2$.  
As an illustration of the method we have used the recent results calculation of the form factors in $D \to K$ decays from Ref.\,\cite{Lubicz:2017syv}.  
In the next Sections we consider two analyses.
In the first one we adopt the results of Ref.\,\cite{Lubicz:2017syv} already extrapolated to the physical point and to the continuum limit. 
In the second analysis we make use directly of the results obtained at finite values of the lattice spacing and for unphysical pion masses.

\subsection{Extraction of the form factors in the continuum}
\label{sec:continuum}

The values of $f_0(q^2_i)$ and $f_+(q^2_i)$ with the corresponding uncertainties, extrapolated at several values of $q^2_i$ to the physical point and to the continuum limit in Ref.\,\cite{Lubicz:2017syv}, are given in the second and third columns of Table~\ref{DKFFs}.

\begin{table}[htb!]
\begin{center}
\begin{tabular}{|c||c|c||c|c|}
\hline
$q^2$\,(GeV$^2$) & $f_+(q^2)\vert_{LQCD}$ &$f_0(q^2)\vert_{LQCD}$ & $f_+(q^2)\vert_{unit}$ & $f_0(q^2)\vert_{unit}$\\
\hline
\hline
0.0 & 0.765(31) & 0.765(31) & 0.772(30) & 0.772(30)\\
0.2692 & 0.815(31) & 0.792(28) & 0.822(29) & 0.800(26)\\
0.5385 & 0.872(31) & 0.820(25) & 0.878(30) & 0.826(25)\\
0.8077 & 0.937(32) & 0.849(23) & 0.942(31) & 0.853(21)\\
1.0769 & 1.013(34) & 0.879(21) & 1.015(34) & 0.882(20)\\
$\mathbf{1.3461}$ & $\mathbf{1.102(38)}$  & $\mathbf{0.911(19)}$ &$1.102(38)$ & $0.911(19)$ \\
$\mathbf{1.6154}$ & $\mathbf{1.208(44)}$  & $\mathbf{0.944(19)}$ &$1.208(44)$ & $0.944(19)$ \\
$\mathbf{1.8846}$ & $\mathbf{1.336(54)}$  & $\mathbf{0.979(19)}$ &$1.336(54)$ & $0.979(19)$ \\
\hline
\end{tabular}
\end{center}
\caption{{\it  Lattice determinations of the FFs entering the $D \to K \ell \nu$ decay extrapolated to the physical pion point and to the continuum limit in Ref.\,\cite{Lubicz:2017syv}. The bold values are those adopted as inputs for our study. The fourth and fifth columns contain the results obtained in this work by using the dispersive matrix method. For this transition the kinematical range is $0 \leq q^2 \leq t_- = (M_D-M_K)^2 \simeq 1.88$\,GeV$^2$. \hspace*{\fill}}}\hspace*{\fill} 
\label{DKFFs}
\end{table}

On the same set of configurations we have computed the two-point functions of the relevant vector and scalar operators and extracted the longitudinal and transverse susceptibilities, extrapolated to the physical point and to the continuum  limit.  The longitudinal and transverse susceptibilities, $\chi_{0^+}$  and $\chi_{1^-}$,  have been  evaluated in Section\,\ref{sec:susceptibilities}, according to the strategy described in Sections~\ref{sec:dispbound} and~\ref{sec:lattice}. Their numerical values and uncertainties are
\bea \chi_{0^+}(0) = 0.0043(13)   \, ,   \qquad     \chi_{1^-}(0) = 0.00419(36)~{\rm GeV}^{-2} \label{chiLT}\, .  \eea 

As already  stated  in the introduction one of the main reasons to implement the dispersive bounds is that in the case of $B$ decays ($B \to D^{(*)}$ and $B\to \pi$), in most of present lattice simulations, one is able to access only the kinematical region close to $q^2_{max}$ where the final meson has a small momentum.  The dispersive bounds instead, following the method discussed in Sections~\ref{sec:dispbound} and~\ref{sec:lattice}, allow us to determine with good accuracy also unaccessible points, close to the minimal $q^2$, i.e.~$q^2  \sim m_\ell^2 \sim 0$, without assuming any specific functional form for the $q^2$-dependence of the FFs.  
Bearing in mind possible differences and further difficulties, in order to check the validity of this strategy, we have used only the lattice QCD points for $D \to K$ 
decays at high-$q^2$ (in boldface in the Table\,\ref{DKFFs}). We have implemented our unitarity method and then compared our results with the direct lattice computations in the region at smaller $q^2$  (see the first five rows of Table \ref{DKFFs}). 

\begin{figure}[htb!]
\begin{center}
\includegraphics[scale=0.70]{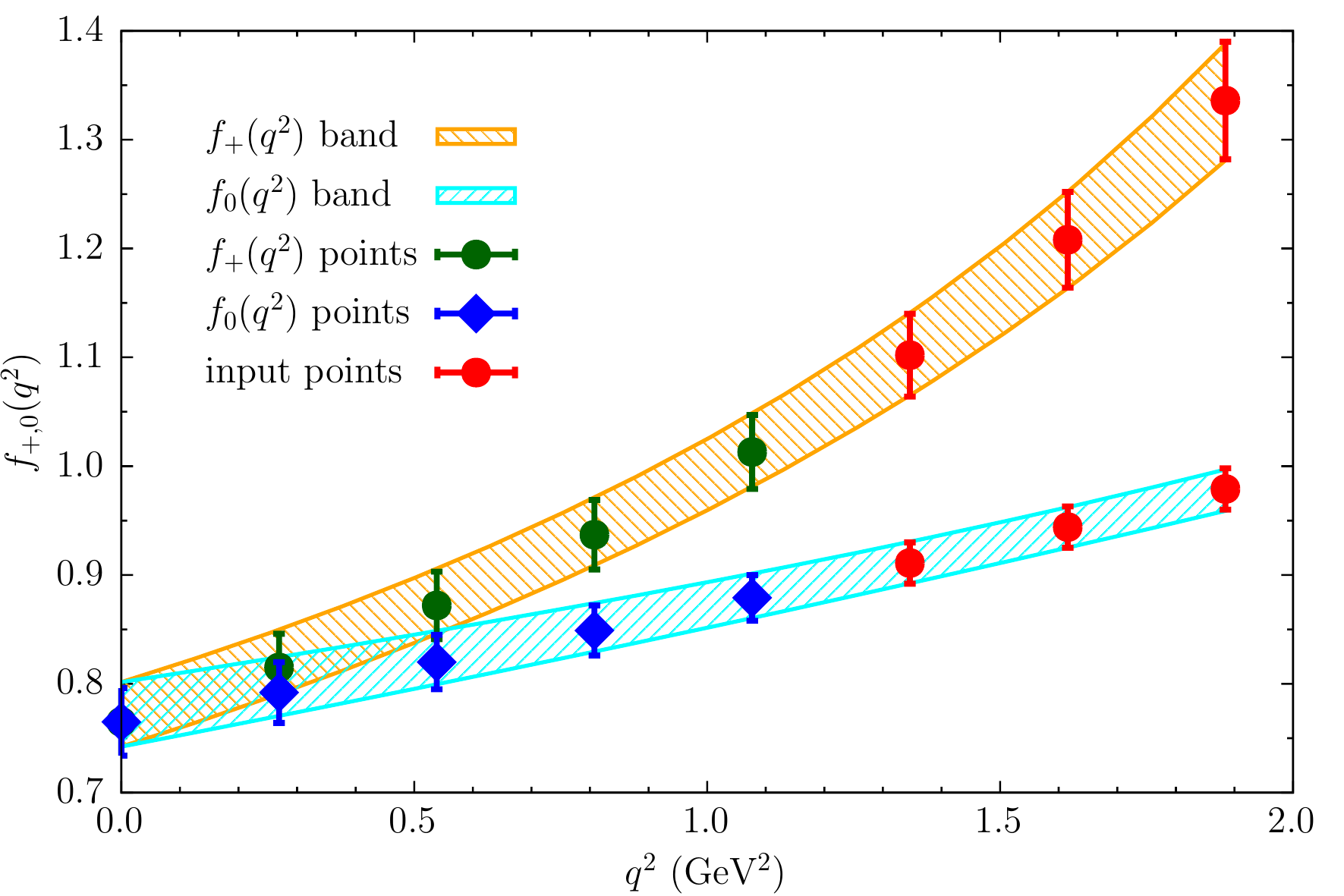}
\caption{{\it The $D \to K$ form factors $f_+(q^2)$ (orange band) and $f_0(q^2)$ (cyan band) obtained in this work and in Ref.\,\cite{Lubicz:2017syv} (dots and diamonds). For each of the form factors, the three red points at the largest values of $q^2$ have been used as inputs for our study, while the five points at lower $q^2$ for each band are not. The latter ones are plotted in order to show the agreement between the lattice QCD data of Ref.\,\cite{Lubicz:2017syv} and the results for the FFs computed with our method.} \hspace*{\fill}}
\label{DKFFfinallast}
\end{center}
\end{figure}

Using the average values and errors given in Ref.~\cite{Lubicz:2017syv} we have first applied the procedure B) given in the Subsection~\ref{subsec:boot}.   Using $20000$ bootstrap events we observe that only a fraction of the events equal to $\sim 15\%$ is rejected by the $\Delta_2^{0(+)} \geq 0$ condition (see Eq.\,(\ref{delta})), while almost all the survived ones respect the kinematical constraint $f_+(0) = f_0(0)$ at $q^2=0$.
In Fig.~\ref{DKFFfinallast} we show our results. The orange band represents the extrapolated vector FF $f_+(q^2)$, while the cyan one corresponds to the scalar FF $f_0(q^2)$. The three red points at the largest values of $q^2$, for each form factor, are the bolded values in Table~\ref{DKFFs} used as inputs for our method, while the green and blue ones are the unbolded values shown for  comparison.

Our unitarity results shown in Table~\ref{DKFFs} and Fig.\,\ref{DKFFfinallast} exhibit an excellent agreement with the lattice QCD values from Ref.~\cite{Lubicz:2017syv} and also a quite similar precision in the whole kinematical range.
In particular, our value of the form factors at $q^2=0$ is
\begin{equation}
f_+(0) = f_0(0) = 0.772 (30)\, ,
\end{equation}
which agrees very nicely with the corresponding result $0.765 (31)$ from Ref.~\cite{Lubicz:2017syv} , having, we stress, a comparable error even if only three points at large $q^2$ have been used as input.

\subsection{Extraction of the form factors from each ensemble}
\label{sec:ensembles}

Since we have access to the original data of Ref.~\cite{Lubicz:2017syv}, we can redo the analysis of that paper having computed in Section~\ref{sec:ctos} on the same ensembles also the susceptibilities. The goal is now to implement the matrix method directly on the lattice data points bootstrap by bootstrap and make a totally model-independent extraction of the form factors following  the procedure A) of Subsection~\ref{subsec:boot}.  

\begin{figure}[htb!]
\centering
\includegraphics[scale=0.45]{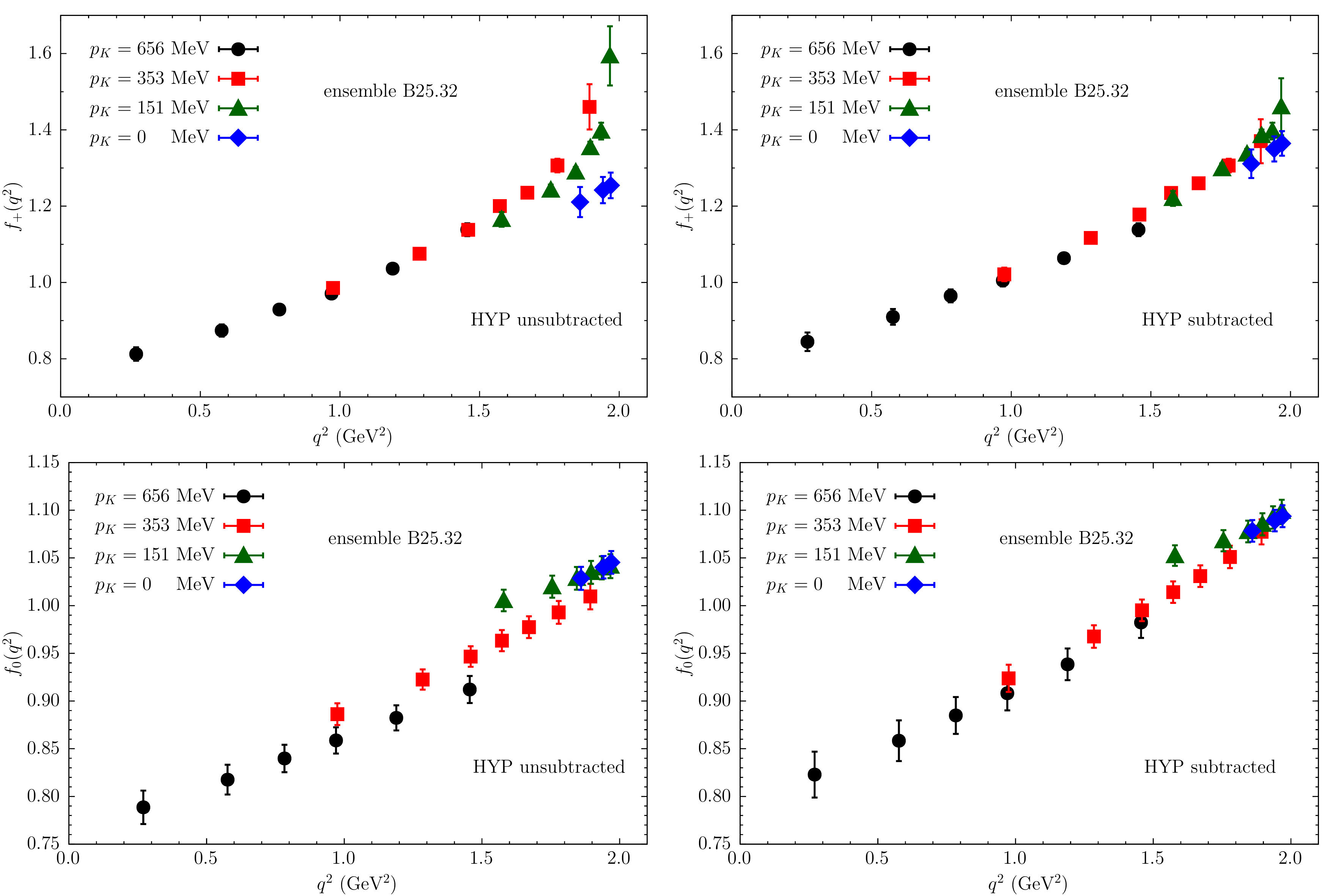}
\caption{\textit{The form factors $f_+(q^2)$ (upper panels) and $f_0(q^2)$ (lower panels) as a function of $q^2$ before (left panels) and after  (right panels) the subtraction of the hypercubic terms for the ensemble B25.32. Different markers represent different values of the final state  meson momentum. By Lorentz symmetry the extracted form factors should only depend on $q^2$. This is not the case and an extra dependence on the value of the  meson momentum is clearly visible beyond the statistical uncertainties. After the subtraction of the hypercubic terms this extra dependence is sensibly reduced. The form factors are already interpolated to the physical charm and strange quark masses.\hspace*{\fill}}}
\label{Hyp_f+}
\end{figure}

As in Ref.~\cite{Lubicz:2017syv}, our analysis is divided in eight branches which differ by the choice of the scaling variable, the fitting procedures and the choice of the method used to determine non perturbatively the values of the mass renormalization constant (see Appendix~\ref{sec:simulations}). For every branch we generate 100 bootstrap events in order to take into account the statistical uncertainties. We keep separate all the branches until the continuum and physical pion point is reached. There we combine the bootstrap events and also the branches using Eq.~(28) of Ref.~\cite{Carrasco:2014cwa}.
As extensively described in Ref.~\cite{Lubicz:2017syv}, the lattice data obtained using the three-point correlation functions with the insertion of vector and scalar densities are affected by non-negligible hypercubic effects, which break Lorentz symmetry. 
The latter ones can be sensibly reduced following the strategy of Ref.~\cite{Lubicz:2017syv}. An example of this procedure is shown in Fig.\,\ref{Hyp_f+}. To implement this procedure, however, some functional form for subtracting hypercubic effects, and consequently some model dependence, was introduced.

\begin{figure}[htb!]
\centering
\includegraphics[scale=0.70]{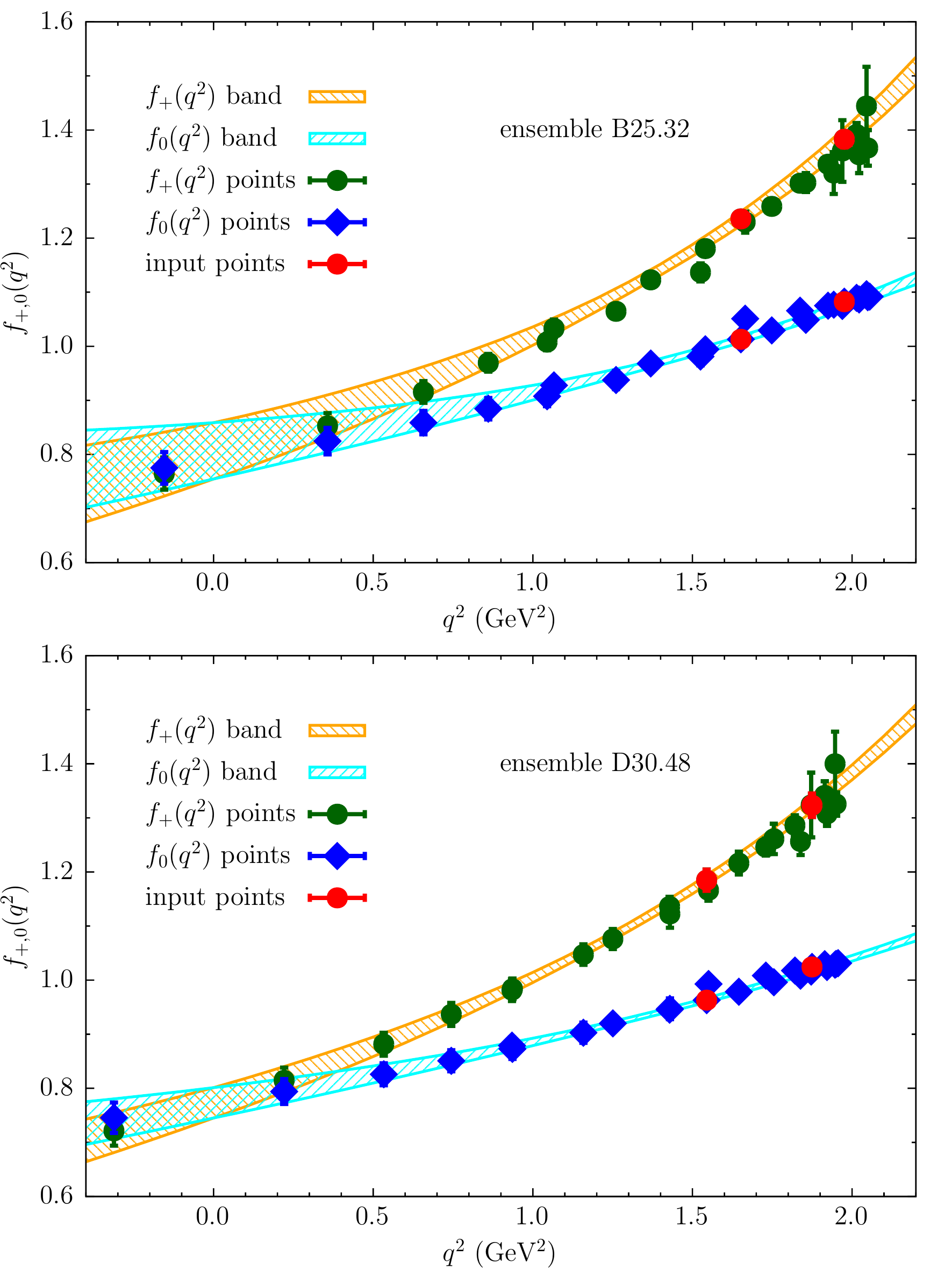}
\caption{\textit{The $D \rightarrow K$ form factors $f_+(q^2)$ (orange band) and $f_0(q^2)$ (cyan band) obtained in this work and in Ref.~\cite{Lubicz:2017syv} (dots and diamonds) in the case of the ETMC ensembles B25.32 (upper panel) and D30.48 (lower panel). The red markers (two points at large values of $q^2$ for each form factor) have been used as inputs for our study, while the other ones are not. The lattice data of Ref.~\cite{Lubicz:2017syv} are interpolated to the physical values of the charm and strange quark masses determined in Ref.\,\cite{Carrasco:2014cwa}.\hspace*{\fill}}}
\label{Sovrapposizione}
\end{figure}

For the $D \rightarrow K$ decay the lattice data of Ref.~\cite{Lubicz:2017syv} (already interpolated to the physical charm and strange quark masses) cover indeed all the kinematical region in $q^2$. 
The idea is to mimic what happens in lattice calculations of $B$ decays where all the lattice data are concentrated at $q^2 \sim q^2_{max}$.
Thus, we have chosen to use, for each form factor, only two points at large values of $q^2$ corresponding to the $D$-meson at rest, shown as red markers in Fig.\,\ref{Sovrapposizione}. The great advantage of studying the $D \rightarrow K$ decay is that we can compare our results obtained with the unitarity procedure to the ones obtained from a direct calculation of the form factors.
We applied the matrix method described in the previous Sections to the determination of the FFS using 31 bins in $q^2$  in the range $[-0.5 {\rm GeV}^2, q^2_{max}]$. The susceptibilities $\chi_{0^+,1^-}$ are those computed non perturbatively for each ensemble in Section~\ref{sec:susceptibilities}. They have been obtained by eliminating the one particle state both for $\chi_{0^+}$ and $\chi_{1^-}$. Thus, the kinematical functions $\phi_{0(+)}$ have been modified accordingly to Eq.\,(\ref{eq:modify}) by including respectively the $D^{*}_s$ and the $D^*_{0}$(2400) poles. Their masses have been calculated on the same ensembles in   Section\,\ref{sec:grs} (see Fig.\,\ref{fig:MVLVT}).

To illustrate the procedure we show in Fig.\,\ref{Sovrapposizione} the comparison between our predictions for the allowed bands of the form factors, obtained by using as inputs only the  points denoted as red markers at large $q^2$, and the rest of the lattice points that are not used as input in our analysis in the case of the ETMC ensembles B25.32 and D30.48 (see Appendix~\ref{sec:simulations}). 
The agreement is excellent. 
These results suggest that it will be possible to obtain quite precise determinations of the form factors for $B$ decays by combining form factors at large $q^2$ with the non perturbative calculation of the susceptibilities.

We now combine the results for all the ensembles and perform the extrapolation to the continuum limit and to the physical pion point adopting the following ansatz  
\begin{equation}
\label{eq:fit_f}
f(q^2, m_\ell, a^2) = c_0 \left[ 1+A^K \xi_{\ell}\log \xi_{\ell} + c_1\xi_\ell + c_2 \xi_\ell^2 + c_3 a^2 + c_4 \xi_\ell a^2 \right] ~ , ~
\end{equation}
where 
\begin{equation}
\xi_\ell = \frac{2 {\cal{B}} m_\ell}{16 \pi^2 {\cal{F}}^2} ~ , ~ 
\end{equation}
being $m_\ell$ the renormalized light-quark mass, and ${\cal{B}}$ and ${\cal{F}}$ the $SU(2)$ low-energy constants entering the chiral Lagrangian at leading order, whose values were determined in Ref.~\cite{Carrasco:2014cwa}. In the fitting procedure the parameters $c_0, c_1, c_2, c_3$  and $c_4$, which depend on $q^2$ and on the form factor, are treated as free independent parameters for each bin in $q^2$, and the correlations among the different bins are automatically taken into account by generating events within the jackknife/bootstrap procedure. Differently, the parameter $A^K$ is the coefficient of the chiral-log. We have checked that it can be safely fixed at the value $A^K = 1/2$ predicted by the hard-pion $SU(2)$ chiral perturbation theory at $q^2 = 0$ (see Ref.~\cite{Lubicz:2017syv}). For each value of $q^2$ the quality of the fit~(\ref{eq:fit_f}) with a total of 5 free parameters turns out to be quite good being $\chi^2/{\rm d.o.f.} \sim 1$. We stress that no assumption has been made concerning the $q^2$-dependence of the parameters appearing in Eq.\,(\ref{eq:fit_f}).

\begin{figure}[htb!]
\centering
\includegraphics[scale=0.70]{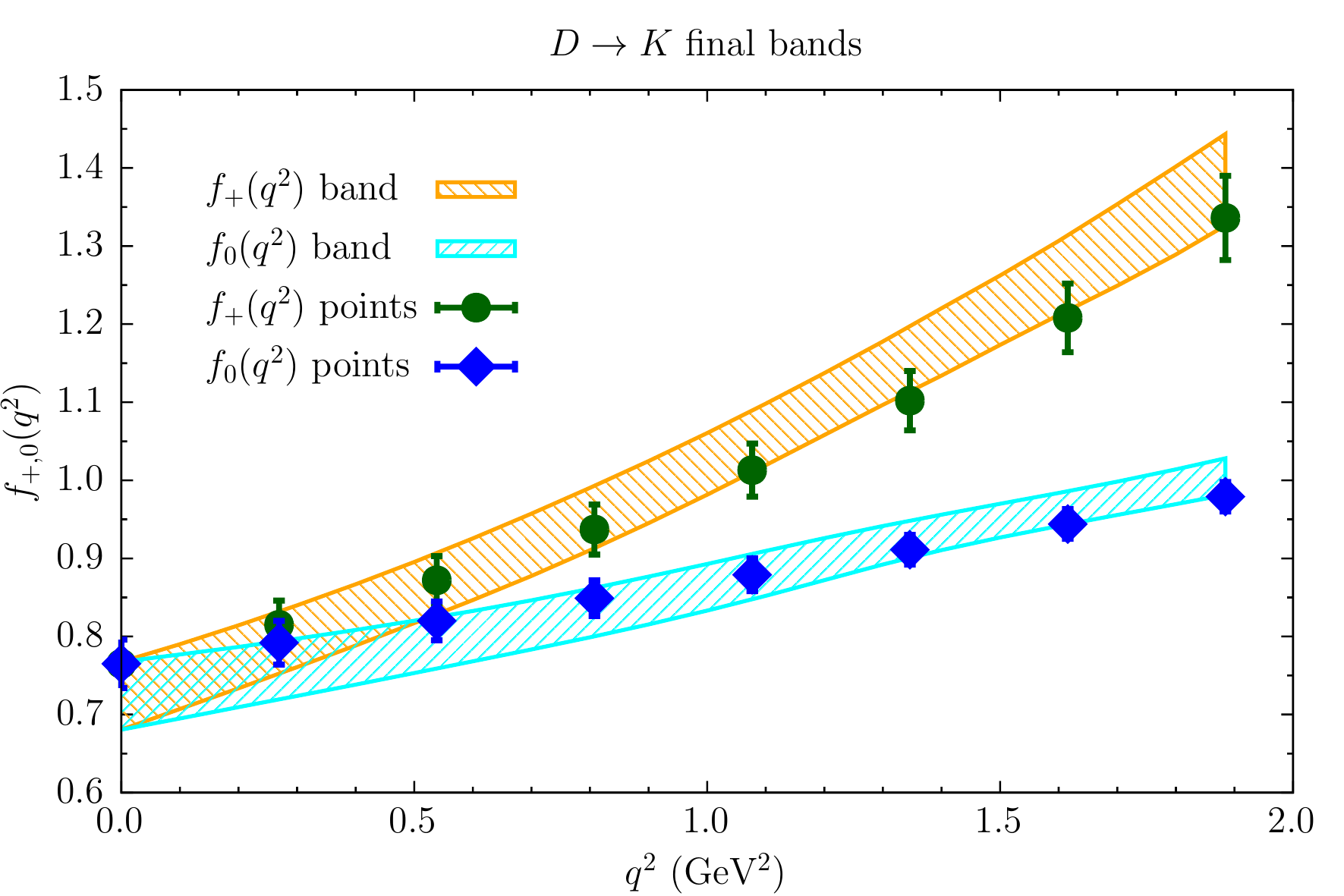}
\centering
\caption{\textit{Momentum dependence of the form factors $f_+(q^2)$ (orange band) and $f_0(q^2)$ (cyan band), extrapolated to the physical point and to the continuum limit, obtained using the dispersive matrix method of this work. The markers represent the lattice results computed in Ref.~\cite{Lubicz:2017syv}. \hspace*{\fill}}\small}
\label{Continuo}
\end{figure}

At this point, we recombine the bootstrap events and the branches of the analysis to obtain the final results.
In Fig.\,\ref{Continuo} we present the final bands for the vector and scalar form factors, extrapolated to the physical value of the pion  mass and to the continuum  limit. The bands agree with the results of Ref.~\cite{Lubicz:2017syv} and exhibit a good precision. This demonstrates that the dispersive matrix method allows to determine the semileptonic form factors in their whole kinematical range with a quality comparable to the one obtained by the direct calculations, even if only a quite limited number of input lattice data for each FF (and the non-perturbative susceptibilities) are used\footnote{We have explicitly checked that the results at $q^2 = 0$ shown in Fig.~\ref{Sovrapposizione} are stable against the addition of (red) points provided they are taken in the large $q^2$ region.}.

In Table \ref{TABLE} we provide explicitly our final results for the vector and scalar form factors, computed at the eight values of $q^2$ adopted in Ref.~\cite{Lubicz:2017syv}, including their total uncertainties. The latter ones take into account: i) statistical Monte  Carlo errors of the simulations and their propagation in the fitting procedure; ii) the chiral extrapolation, evaluated by combining the main results with the ones obtained by putting $c_2=0$ in Eq.~(\ref{eq:fit_f});  iii) the discretization effects, calculated by combining the main results with the ones obtained assuming $c_4=0$ in Eq.~(\ref{eq:fit_f}). 
Our results shown in Table~\ref{TABLE} are consistent within the uncertainties with the lattice data of Ref.~\cite{Lubicz:2017syv}. 
\begin{table}[htb!]
\centering
\begin{tabular}{|c||c|c||c|c|}
\hline
 $q^2(GeV^2)$  & $f_+(q^2)\vert_{LQCD}$  &  $f_+(q^2)$     & $f_0(q^2)\vert_{LQCD}$&    $f_0(q^2)$ \\   \hline  
 0.0  & 0.765(31)& 0.724(43)   & 0.765(31)& 0.724(43) \\ \hline
 0.2692 &0.815(31)& 0.790(40)    & 0.792(28)&  0.754(37) \\  \hline
 0.5385  &0.872(31)& 0.866(40)   & 0.820(25)& 0.790(33)  \\  \hline
 0.8077 & 0.937(32)& 0.953(40)  & 0.849(23)& 0.831(31)  \\ \hline
 1.0769  & 1.013(34)& 1.050(40)  & 0.879(21)&  0.876(29) \\ \hline
1.3461  &1.102(38)&1.155(42)&   0.911(19)&0.924(24)\\ \hline
 1.6154  &1.208(44)&1.265(48) &  0.944(19)& 0.965(21) \\ \hline
 1.8846  & 1.336(54)&1.384(58) &  0.979(19)&1.005(23) \\ \hline
\end{tabular}
\caption{\textit{Final results of this work for the vector and scalar form factors extrapolated to the physical pion mass and to the continuum limit (third and fifth columns) evaluated at the eight values of $q^2$ adopted in Ref.~\cite{Lubicz:2017syv}. The errors correspond to the sum in quadrature of the uncertainties related to statistical, chiral extrapolation and discretization effects (see text). For comparison the results of Refs.~\cite{Lubicz:2017syv} are shown in the second and fourth columns. \hspace*{\fill}}}
\label{TABLE}
\end{table}

To conclude we recall the advantages of the present method. The first point is that the determination of the form factors at values of $q^2$ where there isn't  a direct lattice calculation does not assume any functional dependence of the FF on the momentum transfer.  Indeed, the analysis at each bin in $q^2$ is independent of the others.  In addition, the results obtained by using only two points in $q^2$ for each FF and the susceptibilities are of comparable precision with the direct lattice calculations in the full physical range of values of $q^2$ (see Fig.\,\ref{Sovrapposizione}). We are confident that this will remain true for $B\to D^{(*)}$ or $B \to \pi$ decays, where it is much harder, if not impossible, to compute reliably the FFs at small $q^2$. Last but not least, in this analysis we used non perturbative susceptibilities. In the future, we will investigate the use of susceptibilities at non zero momentum.

\section{Conclusions}
\label{sec:conclu}
In this work we have presented an extended study of two- and three-point correlation functions on the lattice, that together with known dispersive techniques~\cite{Okubo:1971jf}-\cite{Lellouch:1995yv} allows to constrain the lattice predictions for the form factors relevant to exclusive semileptonic decays.
The constraints on the form factors have been implemented by using two-point functions computed in a  non-perturbative way.  Contrary to the perturbative calculation of the two-point function, this approach will allow in the future to use the unitarity constraints also at non zero momentum.  We also introduced a straightforward, and simple to implement, treatment  of the uncertainties. This includes the cases where kinematical constraints between form factors are present.

We have then applied the new method to the analysis of the lattice data of the semileptonic $D \to K$ decays obtained in Ref.~\cite{Lubicz:2017syv}.  We have used this example as a training ground for the method and we have shown that it is possible determine the form factors, in a model-independent way, in the  region at low $q^2$ not accessible directly to lattice calculations, as it is the case of exclusive semileptonic $B$-meson decays. This was achieved by comparing the results of the method with the direct calculation of the form factors along the full kinematical range and allowed us to test the validity of the approach. The application to exclusive semileptonic $B$-meson decays will be presented elsewhere.

\section*{Acknowledgments}
We gratefully acknowledge very helpful discussions with G.~D'Agostini and L.~Lellouch. We acknowledge PRACE for awarding us access to Marconi at CINECA, Italy under the grant the PRACE project PRA067.  We also acknowledge use of CPU time provided by CINECA under the specific initiative INFN-LQCD123.
G.M.~and S.S.~thank MIUR (Italy) for partial  support under the  contract  PRIN 2015. 
F.S.~and S.S.~are supported by the Italian Ministry of Research (MIUR) under grant PRIN 20172LNEEZ. F.S.~is supported by INFN under GRANT73/CALAT.
M.D.C.~is supported in part by UK STFC grant ST/P000630/1.

\appendix

\section{Determinants and inequalities}
\label{sec:detine}

In this appendix we give some formulae which are useful to simplify the numerical calculation of the determinants of the matrix $\mathbf{M}$  
and of the minors $\mathbf{M}^{(i,j)}$ discussed in Section~\ref{subsec:exdip}. 
We consider explicitly the case of real matrices only, because in practice this is the standard case.
The generalization to complex matrices is straightforward.   
With respect to Section~\ref{subsec:exdip} we make use of the explicit expressions~(\ref{CI}) of the inner products.

Let us start from the calculation of the determinant of the $N \times N$ matrix corresponding to the inner products $\langle g_{t_i} | g_{t_j} \rangle  = 1 / (1- z_i z_j)$, where $z_i \equiv z(t_i, t_-)$ are assumed to be real numbers satisfying the conditions $z_i \neq z_j$ and $|z_i| < 1$ with $i, j = 1, 2, ... N$. 
Thus, we want to calculate the determinant of the following matrix
\be 
G \equiv \left( 
\begin{tabular}{cccc}
  $\frac{1}{1 - z_1^2}$     & $\frac{1}{1 - z_1 z_2}$ & $...$ & $\frac{1}{1 - z_1 z_N}$ \\[2mm]
 $\frac{1}{1 - z_2 z_1}$  & $\frac{1}{1 - z_2^2}$    & $...$ & $\frac{1}{1 - z_2 z_N}$ \\[2mm]
 $...$                              & $...$                              & $...$ & $...$ \\[2mm]
 $\frac{1}{1 - z_N z_1}$ & $\frac{1}{1 - z_N z_2}$ & $...$ & $\frac{1}{1 - z_N^2}$
\end{tabular}
\right) ~ . ~
\label{eq:DeltaN}
\ee
A simple evaluation by induction shows that  
\be    
G^{N}(z_1, z_2, ... z_N) \equiv {\rm det}[G] = \frac{1}{\prod_{i=1}^N (1 - z_i^2)} ~  \left( \prod_{i < j = 1}^N \frac{z_i - z_j}{1 - z_i z_j} \right)^2 ~ , ~
\ee
where, in the case $N = 1$ it is understood that $\prod_{i < j = 1}^N (...) \to 1$.
 
The matrix of which we want to calculate the determinant is given by Eq.~(\ref{Der}) of Section~\ref{subsec:exdip}, namely it has the form 
\be
\mathbf{M} \equiv \left( 
\begin{tabular}{cccccc}
   $\chi$ & $\phi f$                            & $\phi_1 f_1$                             & $\phi_2 f_2$                           & $...$ & $\phi_N f_N$ \\[2mm] 
   $\phi f$     & $\frac{1}{1 - z_0^2}$     & $\frac{1}{1 - z_0 z_1}$      & $\frac{1}{1 - z_0 z_2}$     & $...$ & $\frac{1}{1 - z_0 z_N}$ \\[2mm]
   $\phi_1 f_1$ & $\frac{1}{1 - z_1 z_0}$  & $\frac{1}{1 - z_1^2}$     & $\frac{1}{1 - z_1 z_2}$ & $...$ & $\frac{1}{1 - z_1 z_N}$ \\[2mm]
   $\phi_2 f_2$ & $\frac{1}{1 - z_2 z_0}$  & $\frac{1}{1 - z_2 z_1}$  & $\frac{1}{1 - z_2^2}$    & $...$ & $\frac{1}{1 - z_2 z_N}$ \\[2mm]
   $... $  & $...$                           & $...$                              & $...$                              & $...$ & $...$ \\[2mm]
   $\phi_N f_N$ & $\frac{1}{1 - z_N z_0}$ & $\frac{1}{1 - z_N z_1}$ & $\frac{1}{1 - z_N z_2}$ & $...$ & $\frac{1}{1 - z_N^2}$
\end{tabular}
\right) ~ , ~
\label{eq:Delta}
\ee
where $\chi$ is the susceptibility that bounds the inner product $\langle \phi f | \phi f \rangle$ and, we remind, $\phi_i f_i$ corresponds to the scalar product $\langle \phi f | g_{t_i} \rangle$ for the known values of the form factor $f_i = f(z_i)$, whereas $\phi f$ is the scalar product $\langle \phi f | g_{t} \rangle$ of the form factor $f(z(t))$ that we want to constrain. 
In order to use a compact notation let us indicate the values of the conformal variable $z$ and of $\phi(z) f(z)$ as $z_0$ and $\phi_0 f_0$, respectively, so that in what follow the index $i$ may run from $0$ to $N$.

A simple evaluation by induction, as before, yields
\bea
    \label{eq:det_Delta}
    \mbox{det[$\mathbf{M}$]} & = & G^{(N+1)}(z_0, z_1, z_2, ... z_N) \left[\chi - \sum_{i=0}^N \phi_i^2 f_i^2 (1 - z_i^2) \left( \prod_{m \neq i = 0}^N \frac{1 - z_i z_m}{z_i - z_m} \right)^2 \right. 
                                                \nonumber \\[2mm]
                                      & + & \left. 2 \sum_{i < j =0}^N \phi_i f_i \phi_j f_j \frac{(1 - z_i^2) (1 - z_j^2) (1 - z_i z_j)}{(z_i - z_j)^2} 
                                               \left(  \prod_{m \neq (i,j) = 0}^N \frac{1 - z_i z_m}{z_i - z_m} \frac{1 - z_j z_m}{z_j - z_m} \right) \right] ~ , ~ \quad
\eea
where
\be
    G^{(N+1)}(z, z_1, z_2, ... z_N) = \frac{1}{1 - z^2} \left( \prod_{i=1}^N \frac{z - z_i}{1 - z z_i} \right)^2 G^{(N)}(z_1, z_2, ... z_N) ~ . ~
\ee

The unitarity bounds for the (unknown) form factor $f_0$ result from the condition
\be
     \mbox{det[$\mathbf{M}$]} = \alpha \phi_0^2 \left[ - f_0^2 + 2 \overline{\beta} f_0 - \overline{\beta}^2 + \overline{\gamma} \right]  \geq 0  ~  ,
\ee
which implies\footnote{The relations of the coefficients $\overline{\beta}$ and $\overline{\gamma}$ with $\beta$ and $\gamma$, defined in Eq.~(\ref{beta1}), are:  $\overline{\beta} = - \beta / (\alpha \phi_0)$ and $\overline{\gamma} = (\beta^2 + \alpha \gamma) / (\alpha \phi_0)^2 = \Delta_1 \Delta_2 / (\alpha \phi_0)^2$.}
\be
    \overline{\beta} - \sqrt{\overline{\gamma}} \leq  f_0 \leq \overline{\beta} + \sqrt{\overline{\gamma}} ~ , ~
    \label{eq:bounds}
\ee 
where (after some algebraic manipulations)
\bea
      \label{eq:alphafinal}
      \alpha & \equiv &G^{(N)}(z_1, z_2, ... z_N) \geq 0 ~ , ~ \\[2mm]
      \label{eq:beta_final}
      \overline{\beta} & = & \frac{1}{\phi_0 d_0} \sum_{j = 1}^N f_j \phi_j d_j \frac{1 - z_j^2}{z_0 - z_j} ~ , ~ \\[2mm] 
      \label{eq:gamma_final}
      \overline{\gamma} & = &  \frac{1}{1 - z_0^2} \frac{1}{\phi_0^2 d_0^2} \left( \chi - \overline{\chi} \right) ~ , ~ \\[2mm]    
       \label{eq:chi0_final}
       \overline{\chi} & = & \sum_{i, j = 1}^N f_i f_j \phi_i d_i \phi_j d_j \frac{(1 - z_i^2) (1 - z_j^2)}{1 - z_i z_j}~,
\eea
with
\bea
   \label{eq:d0}
    d_0 & \equiv & \prod_{m = 1}^N \frac{1 - z_0 z_m}{z_0 - z_m}  ~ , ~ \\[2mm]
    \label{eq:di}
    d_j & \equiv & \prod_{m \neq j = 1}^N \frac{1 - z_j z_m}{z_j - z_m}  ~.    
\eea

Unitarity is satisfied only when $\overline{\gamma} \geq 0$, which implies $\chi \geq \overline{\chi}$.
Note that $d_0$ and $\phi_0$ depend on $z_0$, while the quantities $d_j$ and $\phi_j$ with $j =1, 2, ... N$ do not.
Thus, the values of $\overline{\beta}$ and $\overline{\gamma}$ depend on $z_0$, while the value of $\overline{\chi}$ does not depend on $z_0$ and it depends only on the set of input data. 
Consequently, the unitarity condition $\chi \geq \overline{\chi}$ does not depend on $z_0$.

Note that:
\begin{itemize}
\item When $z_0$ goes toward one of the known values $z_j$, let's say $z _0 \to z_{j^*}$, one has $d_0 \to d_{j^*} (1 - z_{j^*}^2) / (z_0 - z_{j^*}) [1 + {\cal{O}}(z_0 - z_{j^*})]$, so that one gets (as expected)
\bea
     \overline{\beta} & \to & f_{j^*} ~ , ~ \\[2mm]
     \overline{\gamma} & \to & 0 ~ . ~
\eea

\item By expanding the factor $1 / (1 - z_i z_j)$ in Eq.\,(\ref{eq:chi0_final}) for $|z_i| < 1$ one has
\be
    \overline{\chi} = \sum_{k = 0}^\infty \left[ \sum_{i = 1}^N f_i \phi_i d_i (1 - z_i^2) z_i^k \right]^2 ~ , ~
\ee
which implies $\overline{\chi} \geq 0$.
\item Since in terms of the squared 4-momentum transfer $t$ the variable $z_0$ is given by
\be
    z_0 = \frac{\sqrt{t_+ - t} - \sqrt{t_+ - t_-}}{\sqrt{t_+ - t} + \sqrt{t_+ - t_-}}
    \label{eq:zeta}
\ee
the annihilation threshold $t = t_+$ corresponds to $z_0 = -1$, while $t \to - \infty$ corresponds to $z_0 = 1$.
From Eq.\,(\ref{eq:gamma_final}) it follows that unitarity may have no predictive power (i.e.~$\overline{\gamma} \to \infty$) both at the annihilation threshold $t_+$ and for $t \to - \infty$.
\end{itemize}

\section{Simulation details}
\label{sec:simulations}

The gauge ensembles used in this work have been generated by ETMC with $N_f = 2 + 1 + 1$ dynamical quarks, which include in the sea, besides two light mass-degenerate quarks ($m_u = m_d = m_{ud}$), also the strange and the charm quarks with masses close to their physical values~\cite{Baron:2010bv,Baron:2011sf}.
The ensembles are the same adopted to determine the up, down, strange and charm quark masses in Ref.~\cite{Carrasco:2014cwa} and the bottom quark mass in Ref.~\cite{Bussone:2016iua}.

In the ETMC setup the Iwasaki action \cite{Iwasaki:1985we} for the gluons and the Wilson maximally twisted-mass action \cite{Frezzotti:2000nk,Frezzotti:2003xj,Frezzotti:2003ni} for the sea quarks are employed. 
Three values of the inverse bare lattice coupling $\beta$ and different lattice volumes are considered, as it is shown in Table \ref{tab:simudetails}, where the number of configurations analyzed ($N_{cfg}$) corresponds to a separation of $20$ trajectories.

At each lattice spacing different values of the light sea quark mass are considered, and the light valence and sea quark masses are always taken to be degenerate, i.e.~$m_{ud}^{sea} = m_{ud}^{val} = m_{ud}$. 
In order to avoid the mixing of strange and charm quarks in the valence sector we adopt a non-unitary set up in which the valence strange and charm quarks are regularized as Osterwalder-Seiler fermions~\cite{Osterwalder:1977pc}, while the valence up and down quarks have the same action of the sea.
Working at maximal twist such a setup guarantees an automatic ${\cal{O}}(a)$-improvement \cite{Frezzotti:2003ni,Frezzotti:2004wz}.
Quark masses are renormalized through the RC $Z_m = 1 / Z_P$, computed non-perturbatively using the RI$^\prime$-MOM scheme (see Ref.~\cite{Carrasco:2014cwa}).

The lattice scale is determined using the experimental value of $f_{\pi^+}$ so that the values of the lattice spacing are $a = 0.0885(36), ~ 0.0815(30),  ~ 0.0619(18)$ fm at $\beta = 1.90, ~ 1.95$ and $2.10$, respectively, the lattice size goes from $\simeq 2$ to $\simeq 3$ fm.

The physical up/down, strange and charm quark masses are obtained by using the experimental values for $M_\pi$, $M_K$ and $M_{D_s}$, obtaining~\cite{Carrasco:2014cwa} $m_{ud}^{phys} = 3.72 (17)$ MeV, $m_s^{phys} = 99.6 (4.3)$ MeV and $m_c^{phys} = 1.176 (39)$ GeV in the $\overline{\mathrm{MS}}$ scheme at a renormalization scale of 2 GeV.
We have considered three values of the valence quark mass in both the charm and the strange sectors, which are needed to interpolate smoothly to the corresponding physical strange and charm regions.
The valence quark masses are in the following ranges: $3 m_{ud}^{phys} \lesssim m_{ud} \lesssim 12 m_{ud}^{phys}$, $0.7 m_s^{phys} \lesssim m_s \lesssim 1.2 m_s^{phys}$ and $0.7 m_c^{phys} \lesssim m_c \lesssim 1.1 m_c^{phys}$. 

\begin{table}[hbt!]
\begin{center}
\renewcommand{\arraystretch}{1.20}
\begin{tabular}{||c|c|c|c||c|c|c||}
\hline
ensemble & $\beta$ & $V / a^4$ &$N_{cfg}$&$a\mu_{ud}$& $a\mu_s$ & $a\mu_c$ \\
\hline \hline
$A30.32$ & $1.90$ & $32^3\times 64$ & $150$ & $0.0030$ & $\{0.0180,$ & $\{0.21256,$ \\
$A40.32$ & & & $150$ & $0.0040$ & $0.0220,$ & $~0.25000,$ \\
$A50.32$ & & & $150$ & $0.0050$ & $0.0260\}$ & $~~0.29404\}$ \\
\cline{1-1} \cline{3-5}
$A40.24$ & & $24^3\times 48 $ & $150$ & $0.0040$ & & \\
$A60.24$ & & & $150$ & $0.0060$ & & \\
$A80.24$ & & & $150$ & $0.0080$ & & \\
$A100.24$ & & & $150$ & $0.0100$ & & \\
\hline \hline
$B25.32$ & $1.95$ & $32^3\times 64$ & $150$ & $0.0025$ & $\{0.0155,$ & $\{0.18705,$ \\
$B35.32$ & & & $150$ & $0.0035$ & $0.0190,$ & $~0.22000,$ \\
$B55.32$ & & & $150$ & $0.0055$ & $0.0225\}$ & $~~0.25875\}$ \\
$B75.32$ & & & $~75$ & $0.0075$ & & \\
\cline{1-1} \cline{3-5}
$B85.24$ & & $24^{3}\times 48 $ & $150$ & $0.0085$ & & \\
\hline \hline
$D15.48$ & $2.10$ & $48^3\times 96$ & $~90$ & $0.0015$ & $\{0.0123,$ & $\{0.14454,$ \\ 
$D20.48$ & & & $~90$& $0.0020$ & $0.0150,$ & $~0.17000,$ \\
$D30.48$ & & & $~90$& $0.0030$ & $0.0177\}$ & $~~0.19995\}$ \\
\hline   
\end{tabular}
\renewcommand{\arraystretch}{1.0}
\end{center}
\vspace{-0.250cm}
\caption{\it \small Values of the valence-quark bare masses in the light ($a \mu_{ud}$), strange ($a \mu_s$) and charm ($a \mu_c$) regions considered for the $15$ ETMC gauge ensembles with $N_f = 2+1+1$ dynamical quarks (see Ref.~\cite{Carrasco:2014cwa}). $N_{cfg}$ stands for the number of (uncorrelated) gauge configurations used in this work.\hspace*{\fill}}
\label{tab:simudetails}
\end{table}

In Ref.~\cite{Carrasco:2014cwa} eight branches of the analysis were considered. 
They differ in: 
\begin{itemize}
\item the continuum extrapolation adopting for the matching of the lattice scale either the Sommer parameter $r_0$ or the mass of a fictitious P-meson made up of two valence strange(charm)-like quarks; 
\item the chiral extrapolation performed with fitting functions chosen to be either a polynomial expansion or a Chiral Perturbation Theory (ChPT) Ansatz in the light-quark mass;
\item the choice between the methods M1 and M2, which differ by ${\cal{O}}(a^2)$ effects, used to determine the mass RC $Z_m = 1 / Z_P$ in the RI$^\prime$-MOM scheme. 
\end{itemize}
In the present analysis we will make use of the input parameters corresponding to each of the eight branches of Ref.~\cite{Carrasco:2014cwa}.
The central values and the errors of the input parameters, evaluated using bootstrap samplings with ${\cal{O}}(100)$ events, are collected in Tables~\ref{tab:8branches} and \ref{tab:RCs}.
Throughout this work all the results obtained within the above branches are averaged according to Eq.\,(28) of Ref.~\cite{Carrasco:2014cwa}.

\begin{table}[htb!]
\renewcommand{\arraystretch}{1.20}
{\footnotesize
\begin{center}
\begin{tabular}{||c|l ||c|c|c|c||c||} 
\hline 
\multicolumn{1}{||c}{} & \multicolumn{1}{|c||}{$\beta$} & \multicolumn{1}{c|}{ $1^{st}$ } & \multicolumn{1}{c|}{ $2^{nd}$ } & \multicolumn{1}{c|}{ $3^{rd}$ } & \multicolumn{1}{c||}{ $4^{th}$ } & \multicolumn{1}{c||}{ $1^{st} - 4^{th}$ } \\ \hline \hline   
                                  & 1.90 & 2.224(68) & 2.192(75) & 2.269(86) & 2.209(84) & 2.224(84) \\ 
$a^{-1}({\rm GeV})$  & 1.95 & 2.416(63) & 2.381(73) & 2.464(85) & 2.400(83) & 2.415(82) \\ 
                                 & 2.10 & 3.184(59) & 3.137(64) & 3.248(75) & 3.163(75) & 3.183(80) \\ \hline \hline
$m_{ud}^{phys}({\rm GeV})$ & & 0.00372(13) & 0.00386(17) & 0.00365(10) & 0.00375(13) & 0.00375(16) \\ \hline
$m_s^{phys}$({\rm GeV})      & & 0.1014(43)   & 0.1023(39)   & 0.0992(29)   & 0.1007(32)   & 0.1009(38) \\ \hline
$m_c^{phys}$({\rm GeV})      & & 1.183(34)     & 1.193(28)     & 1.177(25)     & 1.219(21)     & 1.193(32) \\ \hline \hline
\end{tabular}
\\ \vspace{0.5cm}
\begin{tabular}{||c|l ||c|c|c|c||c||} 
\hline 
\multicolumn{1}{||c}{} & \multicolumn{1}{|c||}{$\beta$} & \multicolumn{1}{c|}{ $5^{th}$ } & \multicolumn{1}{c|}{ $6^{th}$ } & \multicolumn{1}{c|}{ $7^{th}$ } & \multicolumn{1}{c||}{ $8^{th}$ } & \multicolumn{1}{c||}{ $5^{th} - 8^{th}$ } \\ \hline \hline  
                                  & 1.90 & 2.222(67) & 2.195(75) & 2.279(89) & 2.219(87) & 2.229(86) \\ 
$a^{-1}({\rm GeV})$  & 1.95 & 2.414(61) & 2.384(73) & 2.475(88) & 2.411(86) & 2.421(85) \\ 
                                 & 2.10 & 3.181(57) & 3.142(64) & 3.262(79) & 3.177(78) & 3.191(83) \\ \hline \hline
$m_{ud}^{phys}({\rm GeV})$ & & 0.00362(12) & 0.00377(16) & 0.00354(9) & 0.00363(12) & 0.00364(15) \\ \hline
$m_s^{phys}({\rm GeV})$      & & 0.0989(44)   & 0.0995(39)   & 0.0962(27) & 0.0975(30)  & 0.0980(38) \\ \hline
$m_c^{phys}({\rm GeV})$      & & 1.150(35)     & 1.158(27)     & 1.144(29)   & 1.182(19)    & 1.159(32) \\ \hline \hline   
\end{tabular} 
\end{center}
}
\renewcommand{\arraystretch}{1.0}
\vspace{-0.50cm}
\caption{\it \small The input parameters for the eight branches of the analysis of Ref.~\cite{Carrasco:2014cwa}. The renormalized quark masses are given in the $\overline{\mathrm{MS}}$ scheme at a renormalization scale of 2 GeV. The last columns represent the averages of the previous four columns (according to Eq.\,(28) of Ref.~\cite{Carrasco:2014cwa}). With respect to Ref.~\cite{Carrasco:2014cwa} the table includes an update of the values of the lattice spacing and, consequently, of all the other quantities.\hspace*{\fill}}
\label{tab:8branches}
\end{table} 

\begin{table}[htb!]
\renewcommand{\arraystretch}{1.20}
{\small
\begin{center}
\begin{tabular}{||c|lc|c|c||c|c|c||} 
\hline 
\multicolumn{1}{||c}{} & \multicolumn{3}{||c||}{ $1^{st} - 4^{th}$ branches} & \multicolumn{3}{|c||}{ $5^{th} - 8^{th}$ branches} \\ \cline{2-7}
\multicolumn{1}{||c}{} & \multicolumn{1}{||c|}{ $\beta = 1.90$ } & \multicolumn{1}{c|}{ $\beta = 1.95$ } & \multicolumn{1}{c||}{ $\beta = 2.10$ } 
                                  & \multicolumn{1}{c|}{ $\beta = 1.90$ } & \multicolumn{1}{c|}{ $\beta = 1.95$ } & \multicolumn{1}{c||}{ $\beta = 2.10$ } \\ \hline \hline
\multicolumn{1}{||c}{$Z_V$} & \multicolumn{1}{||c|}{0.5920(4)} & \multicolumn{1}{|c|}{0.6095(3)} & \multicolumn{1}{|c||}{0.6531(2)}
                                             & \multicolumn{1}{|c|}{0.5920(4)} & \multicolumn{1}{|c|}{0.6095(3)} & \multicolumn{1}{c||}{0.6531(2)} \\ \hline
\multicolumn{1}{||c}{$Z_A$} & \multicolumn{1}{||c|}{0.731(8)} & \multicolumn{1}{|c|}{0.737(5)} & \multicolumn{1}{|c||}{0.762(4)}
                                             & \multicolumn{1}{|c|}{0.703(2)} & \multicolumn{1}{|c|}{0.714(2)} & \multicolumn{1}{c||}{0.752(2)} \\ \hline
\multicolumn{1}{||c}{$Z_P$} & \multicolumn{1}{||c|}{0.529(7)} & \multicolumn{1}{|c|}{0.509(3)} & \multicolumn{1}{|c||}{0.516(3)}
                                             & \multicolumn{1}{|c|}{0.573(4)} & \multicolumn{1}{|c|}{0.544(2)} & \multicolumn{1}{c||}{0.542(1)} \\ \hline
\multicolumn{1}{||c}{$Z_S$} & \multicolumn{1}{||c|}{0.747(12)} & \multicolumn{1}{|c|}{0.713(9)} & \multicolumn{1}{|c||}{0.700(6)}
                                             & \multicolumn{1}{|c|}{0.877(3)} & \multicolumn{1}{|c|}{0.822(2)} & \multicolumn{1}{c||}{0.749(3)} \\ \hline
\end{tabular}
\end{center}
}
\renewcommand{\arraystretch}{1.0}
\vspace{-0.50cm}
\caption{\it \small Values of the RCs  using the vector WI  for $Z_V$ and the RI$^\prime$-MOM scheme for  the others bilinear.  The values and uncertainties have been obtained from  the eight branches of the analysis. The scale-dependent RCs $Z_P$ and $Z_S$ are given in the $\overline{\mathrm{MS}}$ scheme at a renormalization scale of 2 GeV. The results  are slightly  different from those of  Ref.~\cite{Carrasco:2014cwa} since they include an update of the values of the lattice spacing and, consequently, of all the other quantities. \hspace*{\fill}}
\label{tab:RCs}
\end{table} 

Besides the RC $Z_P$ we need the RC's of other bilinear quark operators, namely $Z_V$, $Z_A$ and $Z_S$ related respectively to the vector, axial-vector and scalar currents. 
They have been evaluated in the Appendix of Ref.~\cite{Carrasco:2014cwa} in the RI$^\prime$-MOM scheme for $Z_A$ and $Z_S$, while for $Z_V$ we adopt its determination based on the vector WI identity.

\end{document}